\documentclass[twocolumn]{aastex62}
\pdfoutput=1 
\usepackage{graphicx}
\usepackage{amsmath}

\usepackage{color}

\newcommand{\am}[0]{[$\alpha$/M]}
\newcommand{\mh}[0]{[M/H]}
\newcommand{\xh}[1]{[#1/H]}
\newcommand{\xm}[1]{[#1/M]}
\newcommand{\xx}[2]{[#1/#2]}
\newcommand{\xfe}[1]{[#1/Fe]}

\newcommand{\cm}[0]{[C/M]}
\newcommand{\nm}[0]{[N/M]}
\newcommand{\teff}[0]{$T_{\rm eff}$}
\newcommand{\logg}[0]{$\log{g}$}
\newcommand{\vsini}{$v \sin i$}
\newcommand{\water}{H$_2$O}
\newcommand{\eg}{{e.g.}}
\newcommand{\ie}{{i.e.}}

\newcommand{\referee}[1]{\textcolor{black}{#1}}

\begin{document}

\title{APOGEE Data Releases 13 and 14: Data and Analysis}

\author[0000-0002-9771-9622]{Jon A. Holtzman}
\affiliation{New Mexico State University, Las Cruces, NM 88003, USA}
\author{Sten Hasselquist}
\affiliation{New Mexico State University, Las Cruces, NM 88003, USA}
\author[0000-0003-0509-2656]{Matthew Shetrone }
\affiliation{University of Texas at Austin, McDonald Observatory, Fort Davis, TX 79734, USA (shetrone@astro.as.utexas.edu)}
\author{Katia Cunha}
\affiliation{Observat\'orio Nacional, S\~ao Crist\'ov\~ao, Rio de Janeiro, Brazil}
\affiliation{University of Arizona, Tucson, AZ 85719, USA (cunha@email.noao.edu)}
\author{Carlos Allende Prieto}
\affiliation{Instituto de Astrof\'{\i}sica de Canarias, E-38205 La Laguna, Tenerife, Spain}
\affiliation{Departamento de Astrof\'{\i}sica, Universidad de La Laguna, E-38206 La Laguna, Tenerife, Spain (callende@iac.es)}
\author{Borja Anguiano}
\affiliation{ Department of Astronomy, University of Virginia, Charlottesville, VA 22904-4325}
\affiliation{Department of Physics \& Astronomy, Macquarie University, Balaclava Rd, NSW 2109, Australia}
\author{Dmitry Bizyaev }
\affiliation{Apache Point Observatory, P.O. Box 59, Sunspot, NM 88349-0059, USA (dmbiz@apo.nmsu.edu)}
\affiliation{Sternberg Astronomical Institute, Moscow State University, Moscow, Russia}
\author{Jo Bovy }
\affiliation{Department of Astronomy and Astrophysics, University of Toronto, 50 St. George Street, Toronto, ON, M5S 3H4, Canada}
\affiliation{Dunlap Institute for Astronomy and Astrophysics, University of Toronto, 50 St. George Street, Toronto, Ontario, M5S 3H4, Canada}
\author[0000-0003-0174-0564]{Andrew Casey }
\affiliation{School of Physics \& Astronomy, Monash University, Clayton 3800, Victoria, Australia ; Faculty of Information Technology, Monash University, Clayton 3800, Victoria, Australia; andrew.casey@monash.edu}
\author{Bengt Edvardsson }
\affiliation{Department of Physics and Astronomy, Uppsala Astronomical Observatory, Box 515, 751 20 Uppsala, Sweden}
\author{Jennifer A. Johnson}
\affiliation{Department of Astronomy, The Ohio State University, Columbus, OH 43210, USA}
\author[0000-0002-4912-8609]{Henrik J\"onsson}
\affiliation{Lund Observatory, Department of Astronomy and Theoretical Physics, Lund University, Box 43, SE-221 00 Lund, Sweden}
\author{Szabolcs Meszaros}
\affiliation{ELTE E\"otv\"os Lor\'and University, Gothard Astrophysical Observatory, Szombathely, Hungary}
\affiliation{Premium Postdoctoral Fellow of the Hungarian Academy of Sciences}
\author{Verne V. Smith}
\affiliation{National Optical Astronomy Observatories, Tucson, AZ 85719, USA (vsmith@email.noao.edu)}
\author{Jennifer Sobeck}
\affiliation{Department of Astronomy, Box 351580, University of Washington, Seattle, WA 98195, USA}
\author{Olga Zamora}
\affiliation{Instituto de Astrof\'{\i}sica de Canarias, E-38205 La Laguna, Tenerife, Spain}
\affiliation{Departamento de Astrof\'{\i}sica, Universidad de La Laguna, E-38206 La Laguna, Tenerife, Spain (callende@iac.es)}
\author{S. Drew Chojnowski}
\affiliation{New Mexico State University, Las Cruces, NM 88003, USA}
\author{Jose Fernandez-Trincado}
\affiliation{Institut Utinam, CNRS UMR6213, Univ. Bourgogne Franche-Comt\'e, OSU THETA , Observatoire de Besan\c{c}on, BP 1615, 25010 Besan\c{c}on Cedex, France}
\affiliation{Departamento de Astronom\'\i a, Casilla 160-C, Universidad de Concepci\'on, Concepci\'on, Chile}
\author{Anibal Garcia Hernandez}
\affiliation{Instituto de Astrof\'{\i}sica de Canarias, E-38205 La Laguna, Tenerife, Spain}
\affiliation{Departamento de Astrof\'{\i}sica, Universidad de La Laguna, E-38206 La Laguna, Tenerife, Spain (callende@iac.es)}
\author{Steven R. Majewski}
\affiliation{Dept. of Astronomy, University of Virginia, Charlottesville, VA 22904-4325}
\author{Marc Pinsonneault}
\affiliation{Department of Astronomy, The Ohio State University, Columbus, OH 43210, USA}
\author{Diogo Souto}
\affiliation{Observatorio Nacional, Rua General Jose Cristino, 77, 20921-400 S~ao Cristov~ao, Rio de Janeiro, RJ, Brazil}
\author{Guy S. Stringfellow}
\affil{Center for Astrophysics and Space Astronomy, Department of Astrophysical and Planetary Sciences, University of Colorado, Boulder, CO, 80309-0389, USA}
\author{Jamie Tayar}
\affiliation{Department of Astronomy, The Ohio State University, Columbus, OH 43210, USA}
\author{Nicholas Troup}
\affiliation{Department of Physics, Salisbury University, 1101 Camden Ave, Salisbury, MD 21801, USA}
\author{Gail Zasowski}
\affiliation{University of Utah (gail.zasowski@gmail.com)}

\begin{abstract}
Data and analysis methodology used for the SDSS/APOGEE
Data Releases 13 and 14 are described, highlighting differences from the DR12 analysis
presented in \citet{Holtzman2015}.  Some improvement in the
handling of telluric absorption and persistence is demonstrated. 
The derivation and calibration of stellar 
parameters, chemical abundances, and respective uncertainties 
are described, along with the ranges over which calibration was performed.
Some known issues with the public data
related to the calibration of the effective temperatures (DR13),
surface gravity (DR13 and DR14), and C and N abundances for dwarfs (DR13 and
DR14) are highlighted.  We discuss how results
from \referee{a data-driven technique,} The Cannon \referee{\citep{Casey2016},} 
are included in DR14, and compare those with results from
the APOGEE Stellar Parameters and Chemical Abundances Pipeline (ASPCAP).
We describe how using The Cannon in a mode
that restricts the abundance analysis of each element to regions of the
spectrum with known features from that element leads to Cannon abundances
can lead to significantly different results for some elements
than when all regions of the spectrum are used to derive abundances.

\end{abstract}

\section{Introduction}

The fourth phase of the Sloan Digital Sky Survey (SDSS-IV; \citealt{Blanton2017}) includes
APOGEE-2, an extension of the Apache Point Observatory Galactic Evolution
Experiment (APOGEE; \citealt{Majewski2017}). APOGEE-2 continues observations with the APOGEE
spectrograph \citep{Wilson2018} 
using the SDSS 2.5m telescope \citep{Gunn2006} at Apache Point Observatory, and
will extend to observations from the southern hemisphere with a second APOGEE
spectrograph at the 2.5m duPont telescope at the Las Campanas Observatory.
A main goal of the APOGEE surveys is to obtain high-resolution spectra
of red giants
to map out the kinematical and chemical structure of stars across
the entire Milky Way.

SDSS Data Release 12 (DR12, \citealt{Holtzman2015}) made public the
data from the SDSS-III/APOGEE survey (September 2011- July 2014). It presented, for the first time,
chemical abundances of 15 individual elements from the APOGEE spectra.  
The first SDSS-IV data release, DR13, occurred in August 2016; it included
the same APOGEE data released in SDSS-III
DR12, but with revised reduction and analysis.  DR14 \citep{DR14}, released in August
2017, includes a re-reduction and re-analysis of the original APOGEE data, but
also includes the first two years of APOGEE-2 data (September 2014 - July 2016).
While the overall goals of SDSS-IV/APOGEE-2 are mostly an extension of those
of SDSS-III/APOGEE, there were some modifications made to the targeting strategy:
these are described in detail in \citet{Zasowski2017}. The APOGEE-2 data also
include a significant number of observations of stars at high Galactic
latitude taken as ``piggyback''
observations when SDSS/MaNGA primary observations (see \citealt{DR14}) are being made.
Subsequent data releases will include a re-analysis of these data as well as
additional observations, including those taken with the APOGEE-S instrument
at the duPont telescope that started in February 2017.

\begin{figure}
\hspace*{-1cm}\includegraphics[width=0.6 \textwidth]{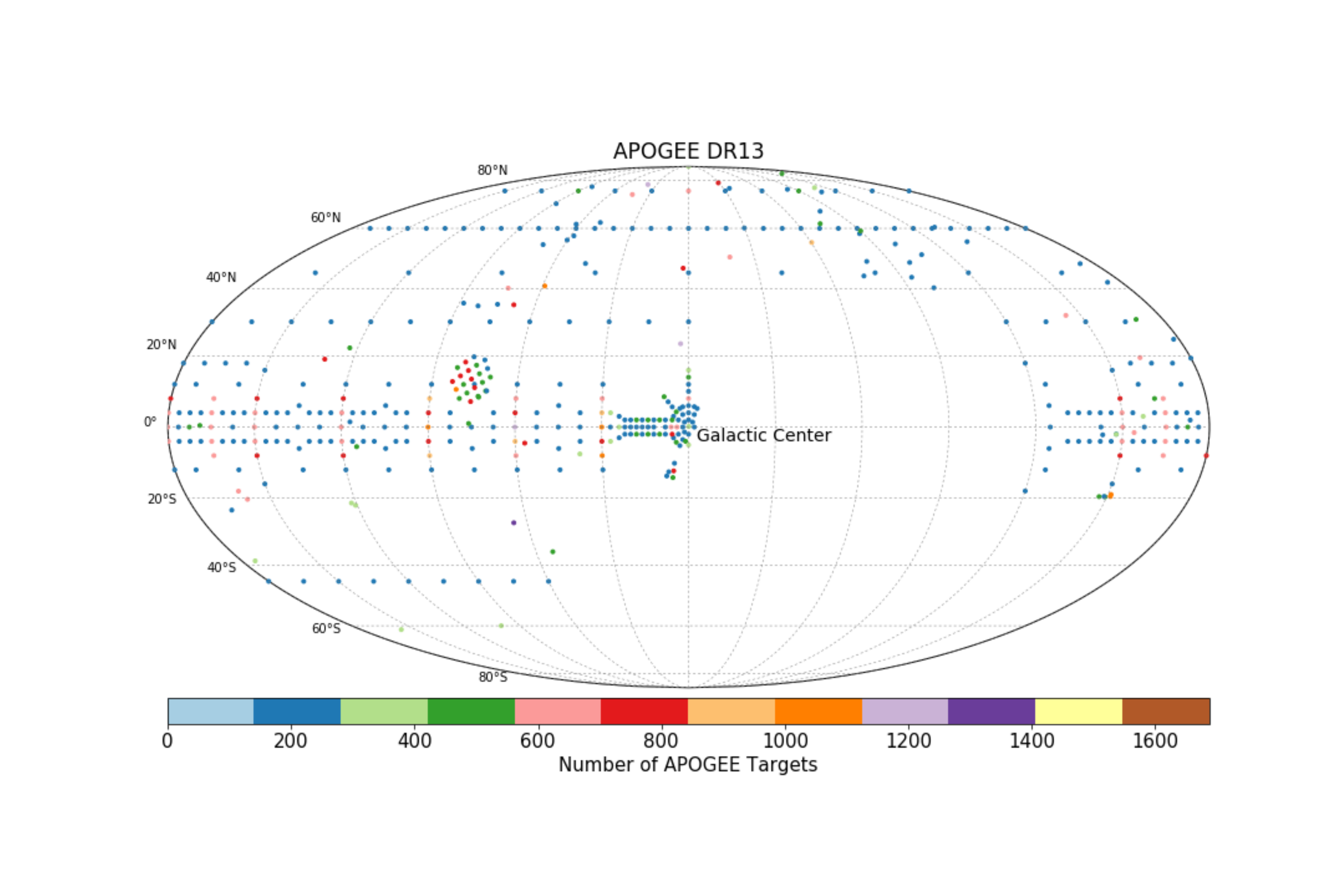}
\caption{Location of fields released in DR13. \referee{Different colors represent
the number of target stars in different fields.}}
\label{fig:dr13_scope}
\end{figure}

\begin{figure}
\includegraphics[width=0.5 \textwidth]{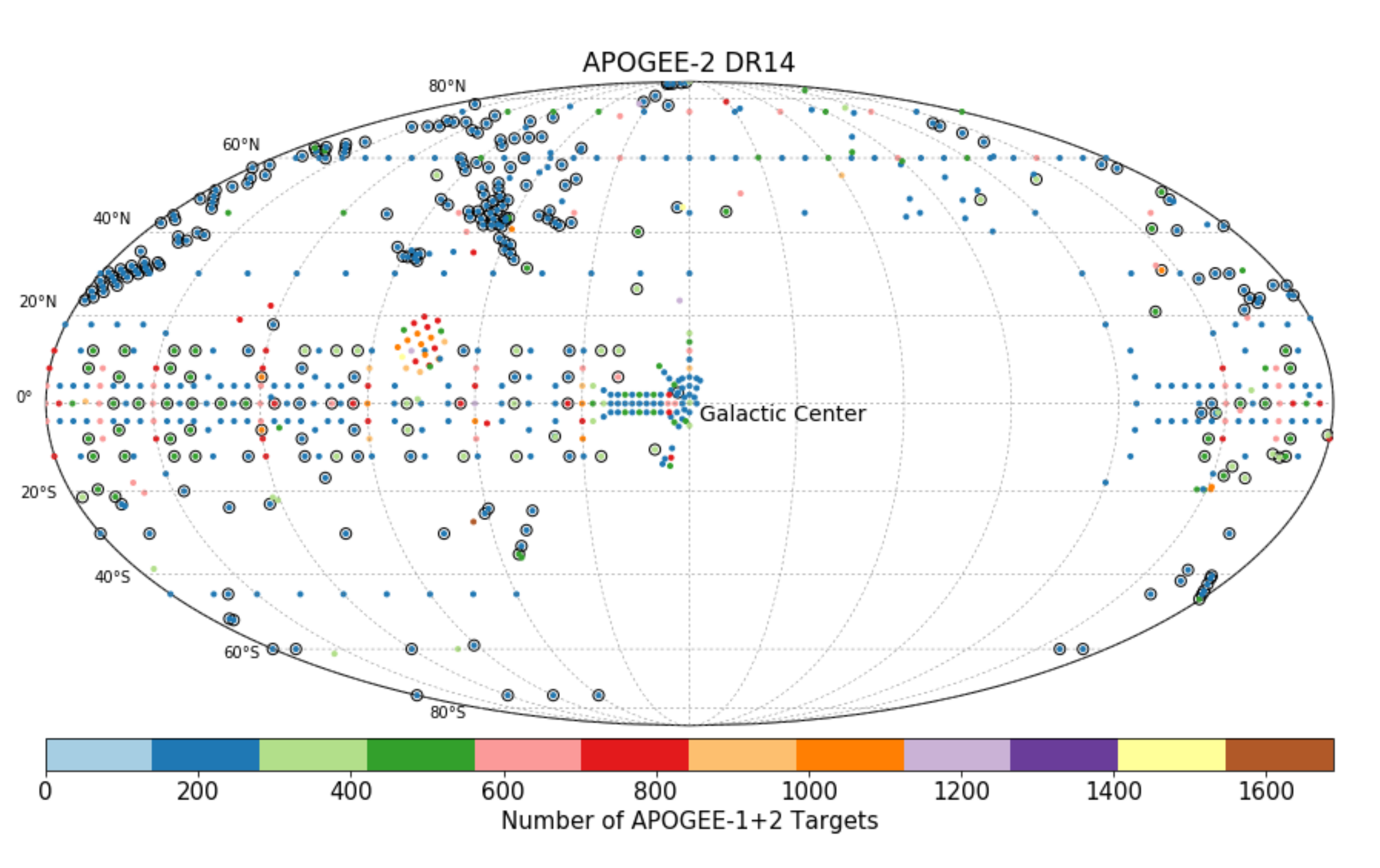}
\caption{Location of fields released in DR14. \referee{Different colors represent
the number of target stars in different fields.}}
\label{fig:dr14_scope}
\end{figure}

Figures \ref{fig:dr13_scope} and \ref{fig:dr14_scope} show the locations
of APOGEE data released in DR13 and DR14, respectively.
For a more general description of the full SDSS DR13, see \citet{DR13}, and
for DR14, see \citet{DR14}.

In this paper, we describe the APOGEE DR13 and DR14 data, focusing
on changes that were made since the DR12 release (as reported
in \citealt{Holtzman2015}).  These include revisions to the APOGEE
data reduction pipeline (\S \ref{sect:reduction}), the APOGEE stellar
parameters and abundances pipeline (ASPCAP, \S \ref{sect:aspcap}), the
calibrations (\S \ref{sect:calibration}), and the data that are released
(\S \ref{sect:data}). We also assess the modifications made to reduce the
impact of persistence (\S \ref{sect:persist}) and describe and analyze
the results from The Cannon  (\S \ref{sect:cannon}).
In addition, we also discuss a few known issues with both DR13 and DR14 that were
discovered after the data releases were frozen and made public.

A companion paper \citep{Jonsson2018} presents assessments of the quality of
the DR13 and DR14 stellar parameters and abundances by comparison with independent
measurements made from optical spectra of a subsample of APOGEE targets.

\section{The APOGEE spectrograph}

All of the data released in DR13 and DR14 were taken with the APOGEE
instrument \citep{Wilson2018} at the Apache Point Observatory (APO), mostly with
the SDSS 2.5m telescope \citep{Gunn2006}, but with a few observations taken
with an APOGEE instrument feed from the NMSU 1.0m telescope.

As a brief review, the APOGEE instrument is a fiber-fed spectrograph,
recording spectra from 300 individual objects in the near-IR (1.51$\mu$m-1.7$\mu$m)
at a resolution of $R\sim 22,500$. The spectra are imaged onto 
three Hawaii-2RG detectors, where each detector records spectra of all
300 objects over roughly a third of the wavelength range, with small
wavelength gaps between detectors.

The APOGEE spectrograph at APO was very stable over the course of the
SDSS-III survey: it was kept under vacuum and cold continuously 
for the entire period
of operation. During the summer of 2014, the instrument was opened for 
routine maintenance and to replace one of the three detectors
(the so-called ``blue" detector that records the shortest wavelength
end of the APOGEE spectra).
This detector was replaced because it exhibited
significant ``superpersistence," as discussed by \citet{Nidever2015}.
We note that the ``green" detector also exhibits some of the same phenomenon at a
lower level, but it was not replaced, since we did not have access to
an additional detector. The new ``blue" detector is of the same
Hawaii-2 format as the old one, and therefore it was a simple detector swap
with no other associated changes required.

Subsequent to the instrument maintenance, it was pumped and cooled,
after which it was refocused. The instrument remained stable under
vacuum and cold from then until the summer of 2017.

\section{Revisions to the data reduction}
\label{sect:reduction}

The APOGEE data reduction pipeline is described in detail in
\citet{Nidever2015}.  A few modifications in the DR13 and DR14
processing have been implemented as compared with DR12. Changes include
an attempt to make some correction for the persistence that occurs over a
portion of the APOGEE detectors, improvements in the LSF characterization,
a small change in the sky subtraction, improvements in telluric correction,
and modified handling of pixels affected by persistence during the
construction of the final combined stellar spectra. Details are given
in the following subsections.

\subsection{Persistence correction}
\label{sect:persistcorr}
As discussed by \citet{Nidever2015}, one of the three original APOGEE detectors
(the short wavelength, or ``blue" detector) suffers from significant
persistence over about a third of its area, and a second detector (the
middle wavelength, or ``green" detector) shows persistence at a lower
level around its periphery. Persistence manifests itself as
elevated counts whose amplitude is related to previous exposure on the
affected pixels. In DR12, no attempt was made to mitigate the effects
of persistence, but data affected by persistence were flagged.
\citet{Holtzman2015} present a discussion suggesting that persistence may not
impact the derivation of stellar parameters too severely, but does
impact the derivation of stellar abundances.

For DR13 and DR14, 
several modifications were made in an attempt to improve the data with 
regard to persistence.
At the individual exposure level, we implemented a correction 
to subtract persistence resulting from previous exposures.
Significant effort was put into parameterizing the amplitude of the
persistence as a function of the previous exposure history. It was found that
this is a complex function that depends not only on the previous
exposure level and elapsed time, but also on the brightness of the previous
source.
A complete characterization proved difficult to obtain
with extant data, but a first order correction was derived that depends
only on the previous exposure level and elapsed time. Specifically,
based on an analysis of illuminated frames followed by a series of
long dark frames, a double-exponential fit for the amplitude of the
persistence was derived for all pixels.

For each science frame, this model was used, along with all of the previous
exposures on a given night, to predict the amplitude of persistence in
each frame. For most of the science frames, the sequence of science
exposures is preceded by two short dark frames (apart from the first plate
of the night). The persistence model was also calculated for these frames,
and a correction factor was derived to make the predicted persistence
better match the observed persistence in these dark frames. This correction
factor was then applied to the predicted persistence for the subsequent
science frames, in an effort to achieve a more reliable correction.
This model was used to try to subtract persistence in both the ``blue"
and ``green" chips. 

Subsequent to the production of the DR13 files, we determined that the
persistence corrections had been calculated without subtracting the true
dark current first. This is not an issue for the ``blue" array since
it has relatively low dark current, but there are a few regions on
the ``green" array that have significant dark current and, as a result,
the persistence correction resulted in an oversubtraction
in these regions.
As a result, we disabled the persistence correction for the ``green" chip
in DR14.

Another, probably more important, persistence amelioration at the visit combination level is 
discussed below (Section \ref{sect:weighting}). An evaluation of the
effectiveness of these improvements on stellar parameter and
abundance determination is presented in Section 
\ref{sect:persist}.

%DR14: turned off green chip correction, dark subtraction in mjdcube

\subsection{The line spread function}
\label{sect:lsf}

The point-spread function of the APOGEE instrument depends on the
location in the spectrograph focal plane, leading to both wavelength
and slit/fiber dependence of the line-spread function (LSF). The LSF
enters the APOGEE analysis in two ways:

\begin{itemize}
\item A fiber-dependent LSF is convolved with an atmospheric
model to correct the observed spectra for the effects of telluric
absorption.

\item The large synthetic library \citep{Zamora2015} used to derive stellar parameters
and abundances in ASPCAP \citep{GarciaPerez2016} is convolved with an
LSF before comparison with the observed spectra.
\end{itemize}
The fiber-by-fiber LSF is derived from observations of night sky lines. 

For DR13 and DR14, several improvements were made with regard to the LSF.
First, it was discovered that one of the OH lines being used for
LSF determination had not been appropriately identified as a doublet,
leading to an incorrect LSF at the long wavelength end of the ``blue"
chip, which happens to be where there is significant CO$_2$
telluric absorption. Second, the functional form of the wavelength
dependence of the LSF characterization was modified to
provide a better LSF in the same wavelength regime.

Modifications related to the LSF in the stellar parameters and
abundances pipeline are discussed in section \S \ref{sect:lsf_grid}.

\subsection{Night sky subtraction}

Emission from the night sky is recorded on a set of ``sky" fibers.
The reduction pipeline attempts a subtraction of the sky emission using the spectra from
sky fibers close in position both in the sky and on the detector.
Most of this emission is in bright OH emission lines, and these are
generally significantly brighter than the underlying spectra of the
objects. As a result, we have not invested significant effort in high
precision sky subtraction, because even perfect sky subtraction would
still result in a spectrum dominated by the Poisson noise of the sky at
the wavelengths of bright OH lines. Instead,
we do a simple, \referee{highly} imperfect subtraction, and flag pixels in the
regions around significant sky lines.

The imperfect sky subtraction leads to reduced spectra that are not
cosmetically appealing, but the regions near bright lines are ignored in
the subsequent analysis. Users of the spectra should be aware of
the poor regions of the spectra around sky lines, which are flagged in the data mask
that accompanies the spectra.

For DR14, we made one small modification to the sky subtraction, namely
we reject spectra from sky fibers that are adjacent to spectra of
very bright stars on the detector, 
since these have the possibility of having inaccurate
measurements of any sky continuum. This happens only rarely because our
fiber management scheme is designed to avoid it, but there are a handful
of observations in which it still occurs.

\subsection{Telluric correction}

In addition to the improvements in the LSF, a few minor additional
modifications were made to the telluric correction routines. These
include masking of the regions affected by hydrogen absorption when
determining the telluric correction factors, and small modifications
in the handling of outliers in the derived correction factors.

\begin{figure}
\includegraphics[width=0.5 \textwidth]{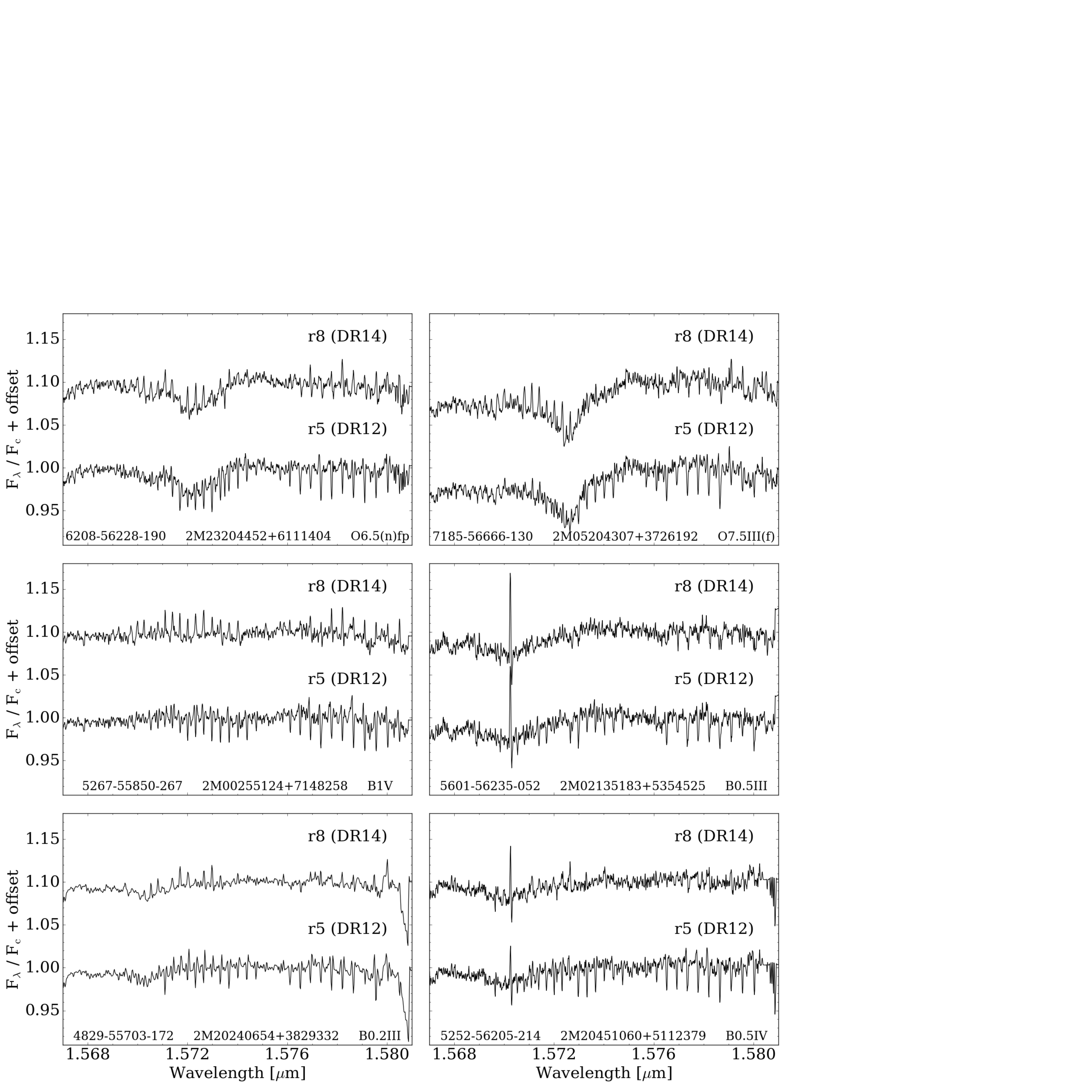}
\caption{A demonstration of the improvement in removal of telluric absorption, comparing DR12 and DR14 visit spectra for a sample of bright OB telluric standard stars. This section of the APOGEE spectra is the most problematic section in terms of telluric features, due to CO$_2$ telluric absorption. The stellar absorption features visible in some of the spectra are a hydrogen Brackett series line at 1.5704 $\mu$m and a He~{\sc ii} line at 1.5723 $\mu$m.}
\label{fig:telluric}
\end{figure}

The LSF modifications (\S \ref{sect:lsf}) had the largest impact on the quality of the
telluric corrections, and the improvements from these can clearly
be seen in essentially all of the hot star spectra and in the
quality of the fits to the cooler star spectra.
Figure \ref{fig:telluric} show some example spectra of hot stars that
should have nearly featureless continua, in a region of the spectrum
with significant CO$_{2}$ absorption. Spectra from both DR12 and DR14
are shown that demonstrate, for the most part, significant improvement
in the telluric absorption correction with the modifications that were
implemented. 

\subsection{Spectra combination}

The majority of APOGEE observations consist of multiple visits so 
radial velocity (RV) variables can be identified while accumulating signal. 
All of the visit spectra (apVisit files), after shifting to zero RV,
are combined into a single final 
spectrum (apStar files) before the ASPCAP analysis is performed.

\subsubsection{Radial velocities}

The combination of spectra from multiple visits requires a measurement of the observed
velocity shift in each visit, which includes a component from the
barycentric correction as well as any RV variation in the object.
In DR12 and DR13, the relative RVs were determined iteratively by cross-correlating
each visit against the combined spectrum, in an effort to avoid any
effects of template mismatch. Once the relative RVs are determined, the
final combined spectrum is cross-correlated against a grid of model
template spectra to determine the absolute RV.

For DR14, several modifications were made to this scheme. First, in
the iterative stage, RVs were determined both by cross-correlation
against the summed observed spectra (as before) and also by cross-correlation
against a best matching template. The final set of RVs are chosen to 
be \referee{those from} whichever of these two methods provides a better result, as quantified
by the scatter in the derived individual visit RVs. 

% \textcolor{red}{Nick to confirm.}

In addition, instead of cross-correlating against a full grid of template
spectra, the template grid is restricted to include only models with 
effective temperatures within 750 K of the effective
temperature implied by the observed $J-K$ color, using the photometric
color-temperature relation of \citet{GHB2009}. This was implemented to
prevent the choice of an inappropriate template, allowing for some
uncertainty in the observed color and extinction, as well as uncertainties
in the color-temperature relation.

The modification to the RV determination resulted in slight improvement
in the precision of the RVs, as judged by the scatter in the individual
visit RVs. As discussed in Nidever et al. (2015), the precision is a
function of \teff , \logg, \mh, and S/N, with a typical value of 100 m/s.

\begin{figure}
\includegraphics[width=0.5 \textwidth]{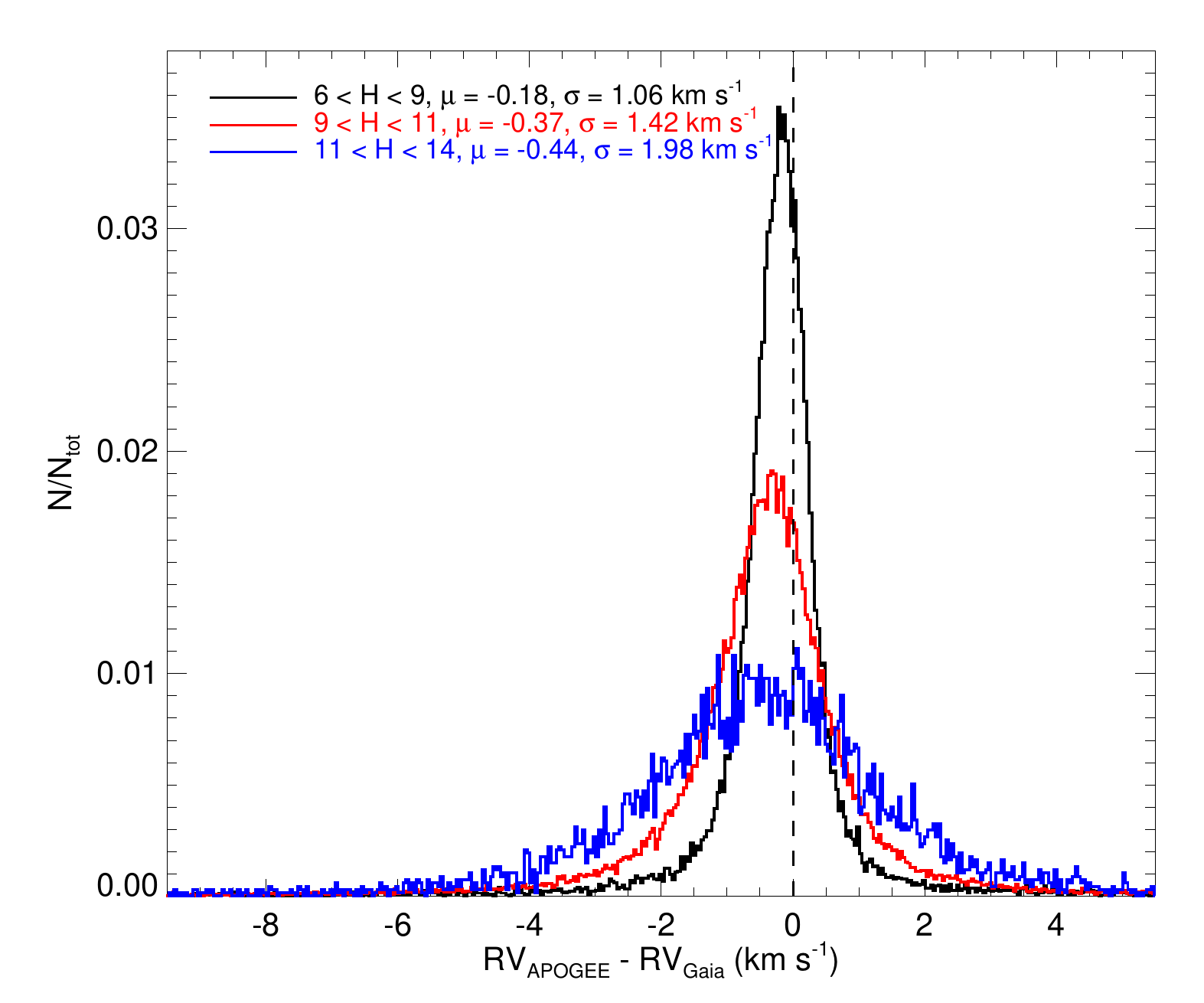}
\caption{Comparison of APOGEE DR14 RVs with those
from GAIA DR2. Stars are binned in three ranges of H-
band magnitude. The scatter is expected to be dominated
by uncertainties in the GAIA RVs, but is a bit larger than
expected from the GAIA uncertainty estimates.}
\label{fig:GAIA_RV}
\end{figure}

Nidever et al. (2015) suggest an accuracy of $\sim$ 0.35
km/s from APOGEE DR12 based on a comparison of
the RVs for 41 stars with literature values. In Figure
\ref{fig:GAIA_RV}, we compare the APOGEE RVs for over 92,000 stars
with RVs from the recent GAIA DR2 (Katz et al. 2018).
We find small median offsets between the APOGEE RVs
and the GAIA RVs that appear to be a function of mag-
nitude, ranging from 0.18 km/s for stars with $6 < H < 9$
to 0.44 km/s for stars with $11 < H < 14$. We note
that the GAIA RVs are expected to have lower precision
than the APOGEE RVs, with precision between 0.2 and
2 km/s, depending on the brightness, \teff, etc., of the
stars, but that the scatter in the comparison is a bit
larger than expected from the combination of the GAIA
and APOGEE RV uncertainties. Katz et al. (2018) dis-
cuss comparison between the GAIA DR2 RVs and sev-
eral catalogs (including APOGEE) and find that other
surveys show comparable offsets, but with different de-
tails, e.g., as a function of magnitude, making it hard to
know what is the correct absolute RV scale at the level
of a few hundred m/s.

%DR14: RVs use trimgrid option, use whichever template (synthetic or combined) is better

\subsubsection{Weighting of individual visits}

\label{sect:weighting}

For DR13 and DR14, the combination of the spectra was modified in an
attempt to reduce the impact of persistence on the
derived stellar parameters and abundances, recognizing that
the persistence correction (discussed above) is imperfect. 
Pixels affected by persistence are given reduced weight in the
spectral combination process, with several consequences:
\begin{itemize}
\item  For stars in which only some visit spectra were recorded in
the regions affected by persistence, the final resulting spectra are dominated by the
unaffected visits, with somewhat larger uncertainties than
would have been achieved by weighting all visits equally; the
improvement in systematic uncertainties resulting from persistence
was judged to be worth the increase in random uncertainties.
\item For stars in which all of the visit spectra were recorded in
regions affected by persistence, the uncertainty in the 
pixels/wavelengths affected by persistence are significantly
inflated. This has the effect of downweighting them in the ASPCAP fits
relative to pixels/wavelengths that are not affected by persistence.
Since the S/N that characterizes the combined spectra is the
median S/N of all of the pixels, the standard S/N reported for these stars
is reduced. To provide a better S/N estimate, we have also 
calculated an alternate S/N, SNREV, that is determined over
a wavelength region in the middle chip that should not have
many pixels that can be affected by persistence. SNREV
is the recommended quantity to use for S/N assessment.
\end{itemize}
An analysis of the effect of the modifications in our treatment of persistence
is presented in Section \ref{sect:persist}, and suggests that
the modifications resulted in significant improvements.

Another change to the weights in the star combination was made to
reduce the impact and inflate the uncertainties of areas of the spectrum
affected by sky lines.  Regions of individual visit spectra in the 
vicinity of sky lines were reduced in weight by a factor of 100
in the combination.  Unfortunately, there was an implementation error in
this for DR13: pixels flagged with PERSIST\_LOW were downweighted
instead of the SIG\_SKYLINE-flagged pixels. This was corrected in DR14.

\section{Improvements to the ASPCAP pipeline}
\label{sect:aspcap}

The ASPCAP pipeline automatically derives stellar parameters and chemical
abundances for the stars observed by APOGEE. The pipeline is described
in \citet{GarciaPerez2016} and \citet{Holtzman2015}. In summary, ASPCAP
determines the best matching syntheses to the observed spectra, 
interpolating in previously computed libraries of synthetic spectra
\citep{Zamora2015}.  A multi-dimensional fit
is first done over the entire spectrum to derive stellar parameters,
and these parameters are then adopted to do single-parameter fits over
limited windows to derive abundances for individual elements.

\referee{When we fit for the stellar parameters, we include a \mh\ and an
\am\ dimension, plus a \cm\ and \nm\ dimension for giants. As
a result, an impact of abundances on stellar parameters only occurs to
the extent to which individual elements depart from the abundance ratios
in the grid (i.e., deviations from solar abundance ratios in non-$\alpha$
elements, and deviations from \am\ for $\alpha$-elements). When these
deviations are small, as they are for most stars, there is not a large
effect on stellar parameters.  In the cases where there are very
atypical abundance ratios, e.g., for second generation stars in globular
clusters, we can erroneous stellar parameters, but this is a
small fraction of stars.}

%The ASPCAP pipeline has evolved over time and efforts have been
%made by the APOGEE team in order to make improvements to both data
%reduction and chemical abundance analysis. In this Section,
%we describe the modifications to the ASPCAP
%pipeline as used for DR13 and DR14.

\subsection{Linelist}
\label{sect:linelist}

Several changes were made to the line list adopted for DR12;
for details on the construction of the DR12 line list, see
\citet{Shetrone2015}.  New lines were
added from NIST and other literature publications, including
hyperfine splitting components for Al and Co. 

%Since the synthetic grids were extended to cooler temperatures (see
%below), lines from $H_2O$ were added to the line list from \citet{Souto2017}.
%Because there are very large number of $H_2O$ lines, we
%restricted their inclusion in the line list to cooler stars: above 4000K,
%no $H_2O$ lines were included, for 3250$ < $\teff$ < $4000K or \mh+\am$<$-1.5,
%only the stronger lines were included, and below 3500, the entire $H_2O$
%line list was used. 

As the synthetic grids were extended to cooler 
temperatures (see below), lines from \water\  were 
added to the line list using the energy levels from
\citet{Barber2006}.  The APOGEE wavelength interval
contains more than 26 million \water\ lines and this large number of
lines makes computation of the entire spectral libraries
with these lines prohibitively expensive in computing time.  
Tests were carried
out that found that \water\ did not contribute 
significant absorption ($\lesssim$1\%) for \teff
$\ge$4000K, so \water\ lines were not included at
these higher effective temperatures.
In the cooler library spectra, 
the \water\ line list was pared down to a
computationally manageable number by including a subset
of the strongest lines.  Tests were done with various
cuts in the line strengths, as defined by 
log (gf$\lambda$) - $\theta$$\chi$$_{\rm lo}$ (where
$\theta$=5040/T and $\chi$$_{\rm lo}$=lower excitation
potential).  Synthetic 
spectra were computed using increasingly smaller numbers
of lines until changes between the complete and reduced
line lists produced differences of less than $\sim$1\%
in flux.  
After this procedure, we ended up using a list containing 443448 lines for
for 3250$ < $\teff$ < $4000K or \mh+\am$<$-1.5, and 
a list with 1891110 lines for \teff $<$ 3250 and \mh+\am$>$-1.5.

Astrophysical gf values for the atomic lines were
derived adopting the same methodology as for DR12
\citep{Shetrone2015}. One change implemented
for DR13/14 was the use of the synthesis code
Turbospectrum (\citealt{Alvarez1998}, \citealt{Plez2012})
to generate synthetic spectra with varying oscillator strengths
and damping values to fit the solar and Arcturus spectra,
respectively.  The use of Turbospectrum provides consistency with 
the calculation of the synthetic libraries described below.
For DR12 the code MOOG (Sneden 1974) was used,
while the computation of the spectral libraries was
done using ASS$\epsilon$T \citep{Koesterke2009}.

When deriving astrophysical $\log gf$ values, the DR13/DR14 line list 
used a different weighting
scheme to combine the results from the Sun and Arcturus than that
described in \citet{Shetrone2015}: the astrophysical solutions were
weighted according to line depth in Arcturus and in the Sun, which
usually gives more weight to the Arcturus solution since the lines are
generally stronger in the cooler, low surface gravity star, Arcturus,
despite it being more metal-poor. When comparing with the center-of-disk
solar spectra, a center-of-disk spectral
synthesis with a microturbulence of 0.7 km s$^{-1}$ was used (correcting
a previous error for the DR12 line list where a full-disk synthesis was used).

The adopted abundances for Arcturus were also updated and modified
slightly, based on new, careful comparisons with the literature, while
retaining the \citet{Asplund2005} abundances for the Sun. \referee{For the
DR12 line list, Arcturus abundances of \xfe{$\alpha$}=0.4 for the $\alpha$ elements,
\xfe{Ca}=0.1, \xfe{Al}=0.3, and \xfe{X}=0. for all others were adopted,
assuming values typical of thick disk stars, but with Al and Ca adjusted
based on measurements by \citet{RamirezCAP2011}. For DR13/14, we 
adopted the abundances from \citet{RamirezCAP2011} exactly.
For the underlying atmospheric model used to do the Arcturus synthesis,
a model with \xh{X}=-0.50 and \xfe{$\alpha$}=0.25 was adopted, i.e.
roughly, but not exactly, consistent with the abundances used in the
synthesis. The adopted stellar parameters and abundances
for Arcturus for both the model atmosphere and synthesis are 
given in Table \ref{tab:arcturus}; we include for reference the
values adopted for the DR12 synthesis.}

%\textcolor{red}{references or appendix?}.

\begin{table}
\caption{\referee{Adopted Arcturus parameters/abundances}}
\begin{center}
\begin{tabular}{llll}
\hline
Quantity& DR13/14 model & DR12  & DR13/DR14 \\
        & atmosphere & synthesis & synthesis \\
\hline
Teff & 4286  &  4286  & 4286  \\ 
\logg & 1.66 &  1.66  & 1.66  \\
\mh &  -0.50 & -0.52  & -0.52 \\
C   &  7.89  &  7.96  &  7.96 \\
N   &  7.28  &  7.64  &  7.66 \\
O   &  8.41  &  8.64  &  8.62 \\
Na  &  5.67  &  5.65  &  5.86 \\
Mg  &  7.28  &  7.41  &  7.38 \\
Al  &  5.87  &  6.15  &  6.25 \\
Si  &  7.26  &  7.39  &  7.32 \\
P   &  4.86  &  4.84  &  4.91 \\ 
S   &  6.89  &  7.02  &  6.97 \\  
K   &  4.58  &  4.80  &  4.76 \\ 
Ca  &  6.06  &  5.89  &  5.88 \\ 
Sc  &  2.55  &  2.53  &  2.72 \\ 
Ti  &  4.65  &  4.78  &  4.63 \\ 
V   &  3.50  &  3.48  &  3.64 \\ 
Cr  &  5.14  &  5.12  &  5.07 \\ 
Mn  &  4.89  &  4.87  &  4.75 \\
Fe  &  6.95  &  6.93  &  6.93 \\
Co  &  4.42  &  4.40  &  4.44 \\
Ni  &  5.73  &  5.71  &  5.74 \\
Cu  &  3.71  &  3.69  &  3.64 \\
\hline
\end{tabular}
\end{center}
\label{tab:arcturus}
\end{table}

\subsection{Synthetic spectral grids}

The ASPCAP pipeline determines stellar parameters and abundances
by finding the best match between observed spectra and a large grid
of synthetic spectra, using the FERRE\footnote{Available at http://github.com/callendeprieto/ferre} 
code \citep{AllendePrieto2006} to determine
the best match. The synthetic spectral grid is multi-dimensional,
since the main features in the near-IR portion of the spectra can depend
on multiple quantities, including \teff, \logg, \mh, \am, \cm, \nm, microturbulent
velocity, macroturbulent velocity, and stellar rotation.

Several modifications were made for the spectral grids used in DR13 and DR14, and
these are discussed below.

\subsubsection{Inclusion of a cool grid}
\label{sect:coolgrid}
For DR12, we used grids covering two temperature ranges: the
GK grid (3500-6000 K), and the F grid (5500-8000K). For DR13/DR14,
we added cooler (M) grids for giants and dwarfs. However, we could not 
compute the M grids using Kurucz model atmospheres as we
did for the warmer grids, since Kurucz atmospheres are not
available below 3500 K. In addition, very cool giants are
expected to have large radii where sphericity effects are
likely to be important. As a result, we used a set
of self-consistent MARCS model atmospheres \citep{Gustafsson2008} computed
specifically for the APOGEE project by B. Edvarsson for the M grid.
While it would be preferable to use a homogeneous set of model atmospheres
in all calculations,
Kurucz atmospheres are not available at cooler temperatures, and MARCS
atmospheres were only available in a coarser grid at warmer temperatures.
For subsequent data releases, we are planning on using a homogeneous
grid, with a spacing in stellar parameters similar to the Kurucz grid,
based on a new grid of MARCS model atmospheres.

The M grid of model atmospheres covers the range between 2500-4000 K in steps of 100K,
\logg\  from -0.5 to 5 in steps of 0.5, \mh\ from -2.5 to
0.5 in steps of 0.5 dex, and \am\ and \cm\ from -1.0 to 1.0
in steps of 0.5 dex. Unfortunately, a significant fraction
($\sim$ 20\%) of the model atmospheric structures could not
be computed because of convergence issues; while many of these failures
were on a grid edge in one or more dimensions, there are some
that fall within the grid. To complete the rectangular grid
needed for the FERRE analysis, we filled in the missing models
with the ``nearest" adjacent models, where the following
metric was adopted:

\begin{multline}
dist = 0.7 |\Delta \textrm{\mh} | + 0.4 X |\Delta \textrm{\am}|  + \\
      0.17 X |\Delta \textrm{\cm}| + 0.62 |\Delta \textrm{\teff}|/100. + \\    
      1.5 |\Delta \textrm{\logg}|
\end{multline}
where $X$ was taken to be unity for models where \cm-\am\ had the
same sign as \cm-\am\ of the missing model, and $X=4$ for those where \cm-\am\
had an opposite sign.  This prescription was developed in an effort
to choose those neighboring model atmospheres that are expected to be
most similar to the missing ones: 
a change in \cm\ is probably less significant than a change in \am,
which is less significant than a change in \teff, etc. After adopting
the nearest model atmosphere, the appropriate abundances for the 
grid location were used in the synthesis.

\begin{figure}
\includegraphics[width=0.5 \textwidth]{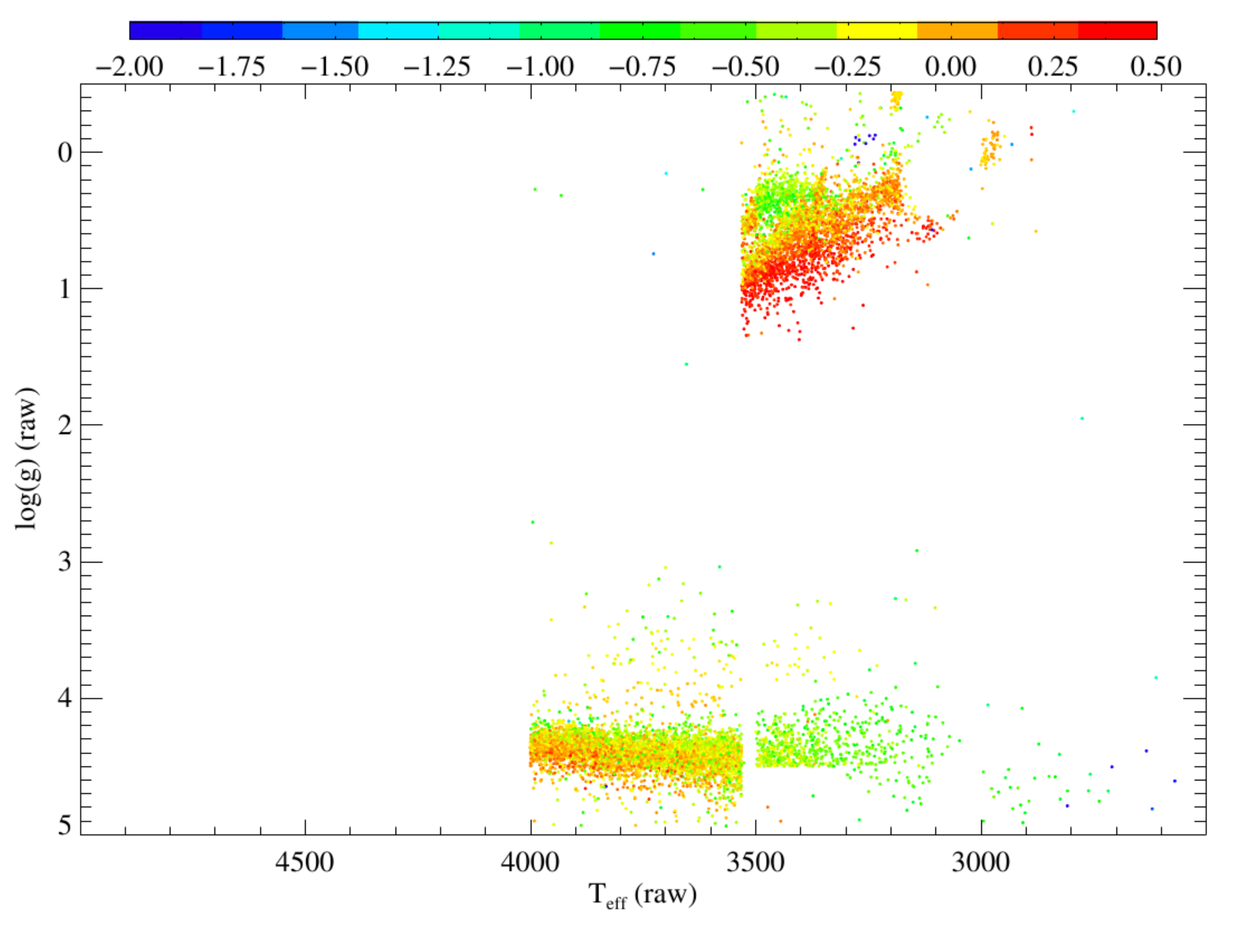}
\caption{Location in an HR diagram of DR14 stars that have parameters within
one grid point (ATMOS\_HOLE\_BAD bit set) of the best fit solution in any of the 
dimensions of the fit. \referee{Color indicates the overall metallicity of
the stars.}}
\label{fig:holes}
\end{figure}

In practice, the presence of the missing models is most significant for
the coolest giants and for cooler dwarfs.
Figure \ref{fig:holes} shows the location in an HR diagram of stars
in DR14 for which there is a missing model within one grid point in any of the
dimensions in the relevant grids. More locations are affected that have
a missing model within two grid points, \ie, are still relevant for the 
cubic interpolation that is used.

While DR13 and DR14 do include results from the M grid, there are a 
number of issues that warrant caution in the use of resulting
parameters and chemical abundances for M dwarfs and giants:
\begin{itemize}
\item The MARCS grids are coarser than the Kurucz grids,
with a spacing of 0.5 in \mh, \am, as opposed to 0.25 
for the Kurucz grids.
\item Inaccuracies may occur because of the missing models that have been filled
as described above.  This is
especially severe given the coarseness of the MARCS grid, which
results in only five points in the \am\ and \cm\ dimensions, since 
the cubic interpolation used by FERRE considers four points in each
dimension of the grid. For DR14, we implemented a
flag to warn when the interpolation includes a point for
which a neighboring model atmosphere was used in the solution
for a given star.
\item In the transition region between the cool and warm
grids (\teff $\sim$ 3500 K) for giants, there are 
significant discontinuities 
in the sets of results from the Kurucz and MARCS models
(see \ref{sect:atmos}).
Some discontinuity is expected because of the different
geometries of the two sets of models.
\end{itemize}

\subsubsection{Spectral synthesis}

For DR13/DR14, we adopted the publicly available Turbospectrum code
\citep{Plez2012}, rather than the ASS$\epsilon$T
\citep{Koesterke2009} synthesis code that was 
used for DR12 and previous versions. On top of the self-consistency 
mentioned in \S \ref{sect:linelist}, an additional
motivation for this change included the public availability and the related
ability to run the synthesis on the SDSS computer cluster with different input
parameters, reducing the need for transfer of large data sets.

For previous releases, a single set of grids, covering different
temperature ranges, were used to get parameters and abundances for both
giants and dwarfs. For DR13, we have split the grids to allow giants
and dwarfs to be treated differently, in several respects:

\begin{itemize}
\item For the giant grids, we adopt carbon isotopic ratios more
appropriate for low surface gravity stars ($^{12}C/^{13}C=15$); 
for the dwarf grids, we use solar isotopic ratios ($^{12}C/^{13}C=90$).
For DR12, solar isotopic ratios were used for all grids.

%\textcolor{red}{need reference for isotopic ratio}

\item For the dwarf grids, we allowed for the non-negligible
stellar rotation that is observed in a substantial fraction of dwarfs; previous
data releases only used grids without rotation. To allow for the
storage required by this extra
dimension, we eliminated the \cm\ and \nm\ dimensions and
set \cm$=$\nm$=0$ at the parameter fitting stage (but see below);
\referee{the need for these dimensions in dwarfs is reduced because
the mixing that leads to significant variations in \cm\ 
and \nm\ in giants is not present in dwarfs.}
\end{itemize}

Spectra were calculated at a fixed wavelength spacing of 0.05 \AA, but
were subsequently resampled (after LSF convolution) to a uniform spacing
in $\log \lambda$, with $d\log \lambda = 6.e-6$ (corresponding to 1.8 km/s/pixel),
and split into three sections corresponding to the wavelengths covered by
the three APOGEE detectors.
For the DR13/DR14 libraries, we also expanded the wavelength range stored for each detector
slightly compared with the DR12 libraries, to allow for the inclusion of a few additional lines.
However, data for all stars may not be available in the expanded region 
because of different radial velocity shifts.  To ensure that the same wavelengths are used 
for the determination of stellar parameters in all stars, the trimmed wavelength
range that was used for the DR12 libraries was adopted in the global parameter
fit via an input mask, and the extended spectra were only used for the fits
for individual element abundances.
Table \ref{tab:wavelengths} gives the wavelength ranges that were used. 

\begin{table}
\begin{center}
\caption{Wavelengths stored for synthetic grids}
\begin{tabular}{lllll}
\hline
Detector & $\log\lambda_0$ & npixels & d$\log \lambda$ & wavelength range\\
\hline
\multicolumn{2}{l}{DR12} & & \\
blue & 4.180932 & 2920 & 6.e-6 &  15168.13 - 15792.32\\
green& 4.200888 & 2400 & 6.e-6 &  15881.37 - 16416.55\\
red  & 4.217472 & 1894 & 6.e-6 &  16499.55 - 16936.75\\
\multicolumn{2}{l}{DR13/DR14} & & \\
blue & 4.180476 & 3028 & 6.e-6 &  15152.21 - 15799.31\\
green& 4.200510 & 2495 & 6.e-6 &  15867.55 - 16423.81\\
red  & 4.217064 &1991 & 6.e-6 &  16484.05 - 16943.53\\
\hline
\end{tabular}
\end{center}
\label{tab:wavelengths}
\end{table}

\subsubsection{Separate grids for different LSFs}
\label{sect:lsf_grid}

Subsequent to DR12, we recognized that the variable LSF across
the detector results in small systematic differences in stellar parameters
and abundances depending on where spectra were recorded on the
instrument detectors, when analyzed using a single synthetic grid calculated
with an average LSF (see also \citealt{Ness2017}).

To partially account for the fact that the LSF varies with location on the
detector, for DR13/DR14 we constructed and used five different versions
of each grid, one for an average LSF, and four others for four different
fiber ranges that capture the main variations of the LSF with fiber. We
determined the different fiber ranges using a clustering analysis
of the LSF's FWHM at three representative wavelengths: $15450\,\AA$,
$16130\,\AA$, and $16740\,\AA$ in the blue, green, and red detectors,
respectively. We clustered the FWHMs at these three wavelengths
using a Gaussian mixture (using the XD code; \citealt{Bovy11a}) and
we determined with a cross-validation test set of 60 fibers that four
fiber groupings suffices to capture the variation of the LSF with fiber
at these wavelengths. These fiber ranges are: 1-38, 39-150, 151-250,
251-300. Within each of these, LSFs from five fibers were averaged and
used to create each spectral grid.

While this change in LSF strategy should be an improvement over the single
average LSF used for the DR12 grids, it is still an
approximation because:
\begin{itemize}
\item The LSF varies continuously, even within the
adopted ranges.
\item Most stars are obtained on multiple
visits where different fibers are used in different
visits. Generally, the fibers from different visits lie
relatively near one another on the slit, but this is not always
the case; we adopt the LSF grid appropriate to
the mean fiber position of the fibers in which observations were
obtained.  However, the typical dispersion in FWHM of a single star over different visits is only 
$\approx$ 20\% of the full dispersion of FWHM over all 300 fibers.
\end{itemize}

%\textcolor{red}{Could demonstrate change with plot of
%derived vmacro vs metallicity, color coded by fiber
%number, for fits with combo LSF and fiber-specific LSFs,
%but would need to recreate.}

\subsubsection{Spectral synthesis grid management}

Because of the separation of giant/dwarf grids, the
addition of the cool grids, and the multiple LSF grids,
the DR13 and DR14 analyses involves a large number of grids:
F dwarf, GK dwarf and giant, and M dwarf and giant, 
with five different LSFs for each,
leading to 25 total grids.  To avoid having to run FERRE on all stars through
all of the temperature grids, we perform initial coarse
characterization fits with the F dwarf grid, the
GK giant grid, and the M giant grid. In these fits, 
\cm\ and \nm\ are set to solar, and an average LSF
is used for all stars. Based on the derived \teff\
and \logg\ from this coarse characterization, full
fits are done in the different grids according to:
\begin{itemize}
\item stars with \teff $>$ 5000K are fit with the F grid
\item stars with 3000 $<$ \teff $<$ 6000K and \logg $<$ 4 are
fit with the GK giant grid
\item stars with 3000 $<$ \teff $<$ 6000K and \logg $>$ 3.5 are
fit with the GK dwarf grid
\item stars with \teff $<$ 4000K and \logg $<$ 4 are
fit with the M giant grid
\item stars with \teff $<$ 4000K and \logg $>$ 3.5 are
fit with the M dwarf grid
\end{itemize}

For stars that are fit with multiple grids (note that there
is overlap in the ranges), we adopt the fit with the lowest
$\chi^2$. We avoid mixing MARCS and Kururcz results
above 3500K by severely penalizing the MARCS grid above 3500K (by increasing the
$\chi^2$ by a factor of ten). We also penalize fits within
1/8 of a grid step of any grid edge (by increasing $\chi^2$ by 25\%). 
The parameters from the adopted fit are
used to derive the stellar abundances.

As described in \S \ref{sect:data}, we record and release information on which
was the best grid for each star, and also release the parameters from all
grids that were used for a given star.

\subsection{Pseudo-continuum normalization}

Before the observed spectra can be compared with model spectra, they
must be continuum-normalized to remove instrument/reduction signatures
(\eg, fiber-dependent and wavelength-dependent sensitivities)
and continuum slope introduced by interstellar absorption. To ensure a
correct comparison, both observed and model spectra are normalized in
the same way. In DR13 and prior data releases, a polynomial fit with an
iterative asymmetric rejection scheme (preferentially rejecting low pixels 
to high pixels) was used to do this normalization,
in an effort for the normalization continuum to more closely approximate the
true continuum, \ie\ by rejecting absorption lines in the continuum fit.

However, the asymmetric rejection causes the derived continuum to be a function
of S/N, especially at lower S/N, because pixels with larger statistical
fluctuations that are below the continuum are rejected in the fit, biasing
the fit high. This is apparent, \eg, in metal-poor
stars that have weaker absorption features, in which statistical fluctuations
in the continuum in lower S/N spectra may be rejected. To remove this bias,
DR14 adopts a continuum normalization that is just a straight 4th order 
polynomial fit to the spectrum, with no iterative rejection. To avoid 
contamination of the fit from bad pixels, \eg, those with imperfect sky
subtraction, pixels marked as bad or in the vicinity of sky lines in the 
observed spectra are masked in the fit.
It is not possible to use the same masks for the model
as for the observed spectra because sky features appear at different rest wavelengths
in stars with different radial velocities, so no pixels are masked in the fits
to the model spectra. Since the fit is low order, however, applying masks
to the model spectra would have little effect.

Because of the new normalization scheme, the pseudo-continuum normalized spectra
in DR14 have a noticeably different mean level than for spectra in
previous data releases; in DR14 pseudo-normalized spectra, there is a significant
number of pixels with values above unity.

\subsection{Fitting for stellar parameters}

The details of the spectral fitting procedure differed slightly between DR13 and DR14, as
described in the following subsections. One new feature, discussed below, for both DR13 and DR14,
(as compared with DR12) is that a relation for the macroturbulent velocity as a function
of other stellar parameters was derived and adopted (\eg,
\citealt{Massarotti2008}); for DR12, a single value was used for all stars.

\subsubsection{DR13}

For red giants, the best matching spectral syntheses are obtained from fits in 6D 
(\teff, \logg, \mh, \am, \cm, and \nm), adopting relations for microturbulent 
and macroturbulent velocities that are a function of the other parameters. These 
relations were derived as follows.

\begin{figure}
\includegraphics[width=0.5 \textwidth]{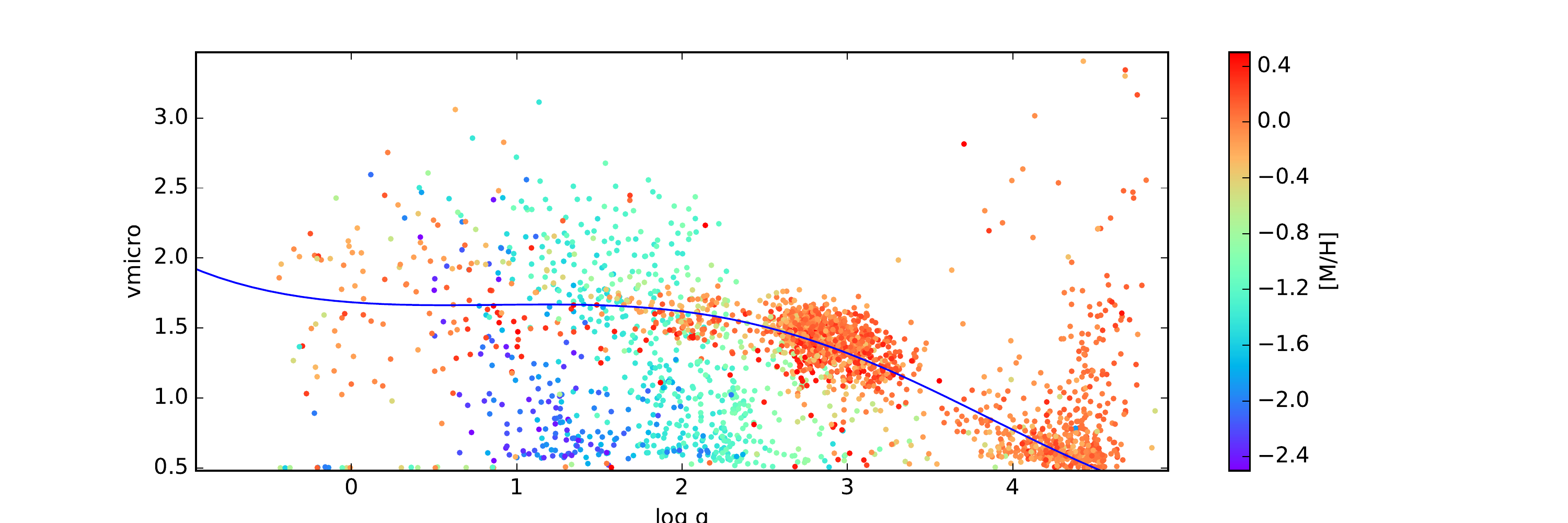}
\caption{Derived microturbulent velocity as a function of surface
gravity for DR13 calibration sample (points), and adopted relation for
microturbulent velocity (line). \referee{Color of the points indicate the
metallicities of the stars.}}
\label{fig:dr13_vmicro}
\end{figure}

First, a 7D fit (\teff, \logg, microturbulent velocity, \mh, \am, \cm, and \nm), 
adopting a fixed macroturbulent velocity of 4 km/s,  was performed on a stellar calibration
subsample that was chosen to include stars from across the HR diagram.
Figure \ref{fig:dr13_vmicro} shows the derived microturbulent velocities as
a function of surface gravity, color-coded by metallicity. At lower metallicity,
the weaker lines are expected to be less sensitive to microturbulent velocity,
perhaps explaining the large scatter seen in these stars. From these results,
a cubic fit to surface gravity was derived using stars with \mh$>-1$ and \logg $<3.8$,
giving the following relation:

\begin{equation}
v_{micro}=10.^{\left( 0.226 - 0.0228 \log g + 0.0297 (\log g)^2 - 0.0113 (\log g)^3\right)}
\end{equation}
Figure \ref{fig:dr13_vmicro} shows this adopted fit.

% calib.vmicro_dr13() using dr6/stars.protodr13/l30a/l30a

\begin{figure}
\includegraphics[width=0.5 \textwidth]{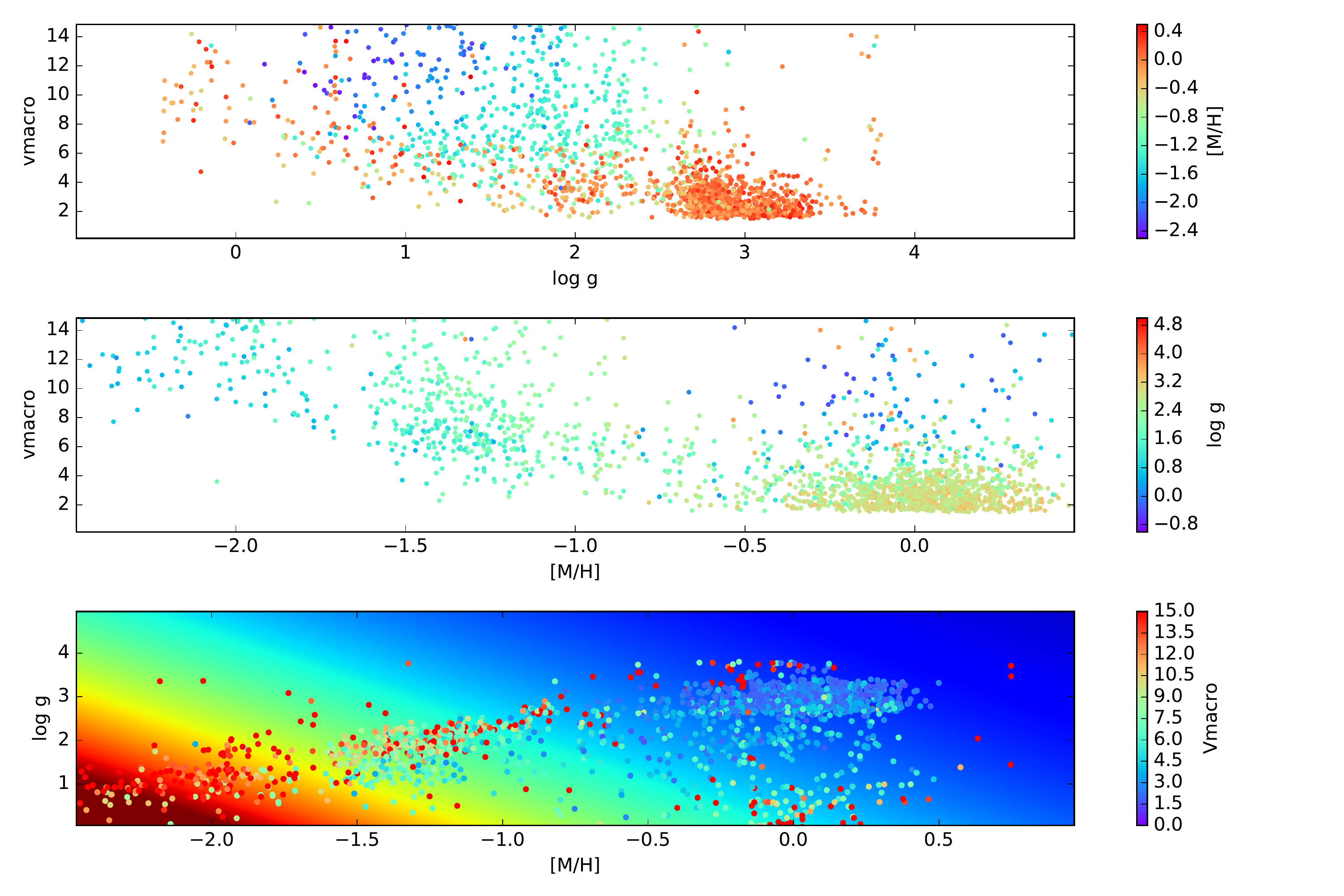}
\caption{Derived macroturbulent velocity for DR13 calibration
sample. Top panel shows $v_T$ as a function of surface
gravity, middle panel as a function of metallicity, and bottom
panel shows adopted 2D fit. \referee{Colors of the points encode the
metallicity, surface gravity, and macroturbulent velocity, 
respectively, as indicated by the color bars.}}
\label{fig:dr13_vmacro}
\end{figure}

Adopting this relation, we perform another 7D fit, now
adding macroturbulent velocity as a free parameter. Figure
\ref{fig:dr13_vmacro} shows derived macroturbulent velocity as
a function of both \mh and \logg. Based on these, we
adopted a 2-D fit depending on both quantities:

\begin{equation}
v_{macro}=10.^{\left(0.741 -0.0998 \log g - 0.225 [M/H]\right)}
\end{equation}
Figure \ref{fig:dr13_vmacro} shows the derived macrotubulent
velocities and the adopted fit. Since the derived relation
is a strong function of metallicity, we cap the maximum
macroturbulent velocity at 15 km/s, to avoid extrapolation.

Finally, a 6D grid is generated, adopting the relations for
both microturbulent and macroturbulent velocities, and
this grid is used to derive stellar parameters and abundances for the
entire sample.

For dwarfs, an additional dimension to account for stellar rotation
is needed, so the same
methodology would require a 7D grid. The effects of stellar rotation
and macroturbulence are essentially indistinguishable at the
spectral resolution of the APOGEE spectra. To reduce the dimensionality
of the fit, we take advantage of the fact that 
dwarfs do not show the same range of \cm\ and \nm\ variations
as is seen in red giants. As a result, we derived stellar parameters
setting \cm$=$\nm$=0$ and using a 6D grid (\teff, \logg,
microturbulent velocity, \mh, \am, and \vsini) for dwarfs.
However, this choice leads to significant problems in the derivation
of C and N abundances in dwarfs, as discussed below.

%Because the
%carbon and nitrogen is not expected to depart from solar abundance
%ratios as much as in giants, we have dropped the \cm and \nm
%dimensions in the initial fit for stellar parameters, (but we still
%solve for carbon and nitrogen abundances subsequently using spectral
%windows!). This is done so that we can add a rotation
%dimension to the grid, the lack of which was known
%to be a significant deficiency of the DR12 analysis.
%We also retain microturbulent velocity as a grid dimension,
%leading to the 6D grid used for the dwarfs: \teff, \logg,
%microturbulent velocity, \mh, \am, and rotational velocity.

\subsubsection{DR14}

\begin{figure}
\includegraphics[width=0.5 \textwidth]{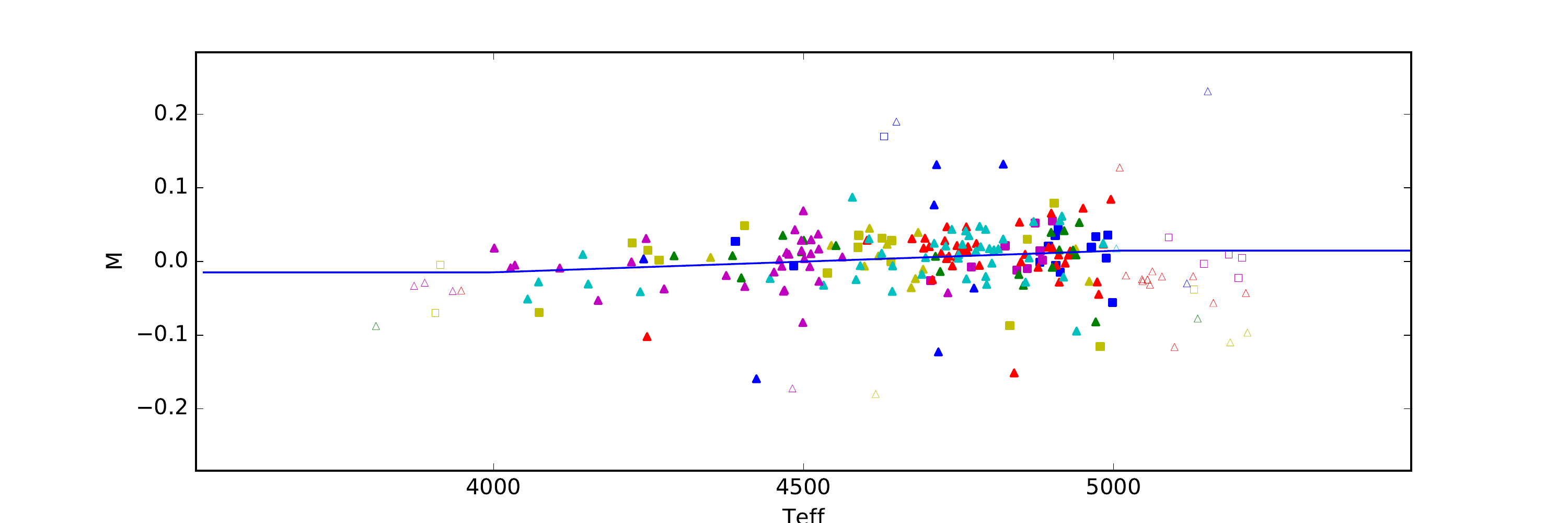}
\includegraphics[width=0.5 \textwidth]{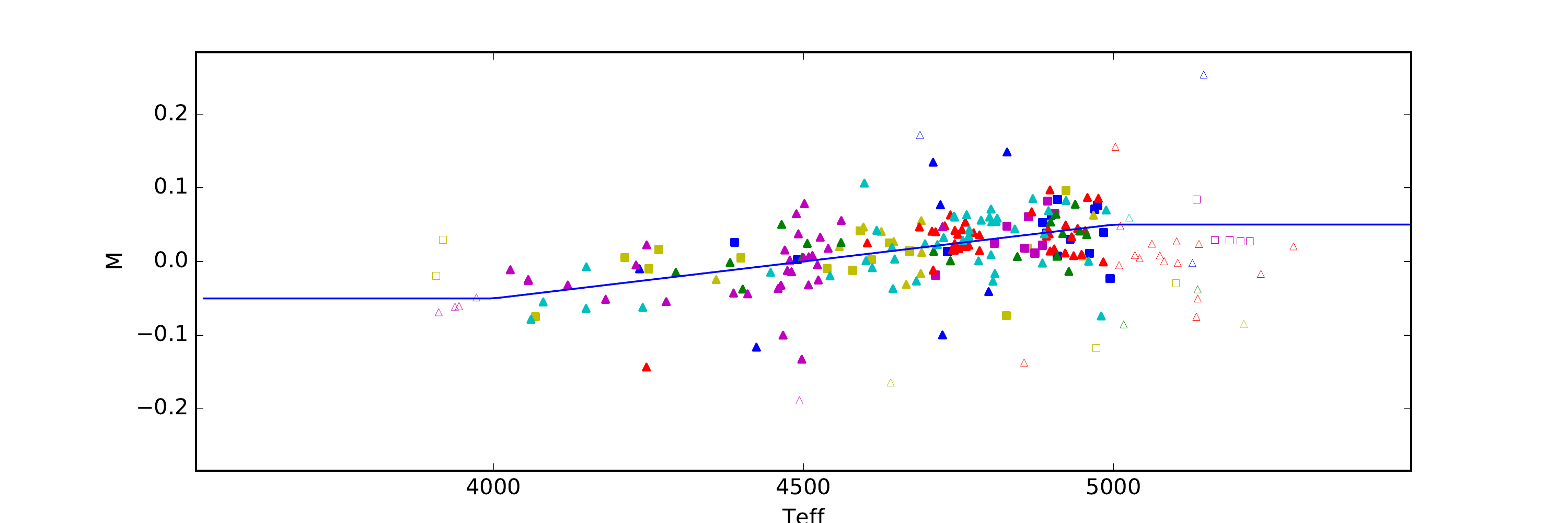}
\caption{Derived raw \mh\ vs \teff\ for stars in clusters, allowing for
microturbulent velocity to float (upper) vs fixing with a relation based
on \logg\  (bottom). There is significantly less trend with temperature when
microturbulent velocity is left as a free parameter. \referee{Different point colors
and symbols are used to distinguish different clusters.}}
\label{fig:dr14_vmicro}
\end{figure}

For DR14, a similar scheme was adopted, with one significant change
for giants: microturbulent velocity was left as a free parameter in the
final fits, \ie, 7D fits in \teff, \logg, microturbulent velocity, \mh,
\am, \cm, and \nm\ were done. While this increased the computational
analysis time significantly (by roughly a factor of five), it was felt
to be warranted for two reasons. First, from the initial calibration
sample fit, there was a non-negligible range in derived microturbulent
velocity in some regions of \teff-\logg-\mh\ parameter space. Second,
the derived metallicities of cluster stars showed less of a trend with
effective temperature when the microturbulent velocity was allowed to
float as a free parameter than when a fixed relation was adopted; see
Figure \ref{fig:dr14_vmicro}.

\begin{figure}
\includegraphics[width=0.5 \textwidth]{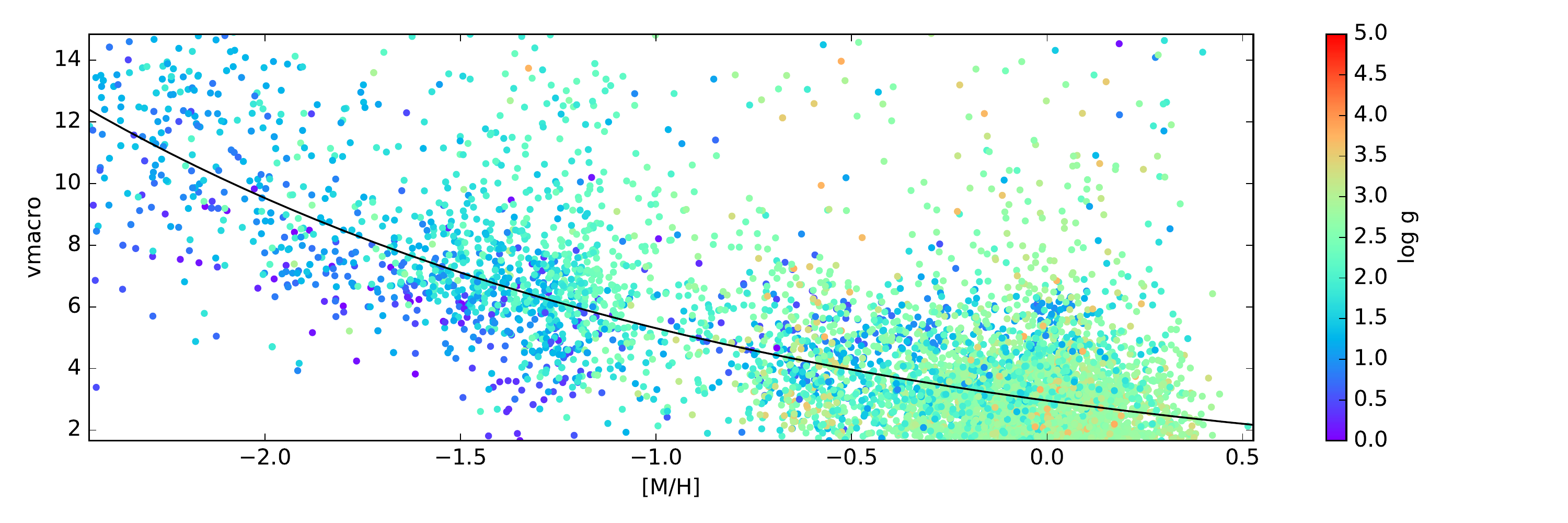}
\caption{Derived macroturbulent velocity relation as a function of \mh\ 
for DR14 calibration sample\referee{; point color encodes the surface gravity,
as indicated by the color bar}. Line shows the adopted relation.}
\label{fig:dr14_vmacro}
\end{figure}

Although the basic methodology was the same as for DR13, DR14 adopted
a different prescription for macroturbulent velocity, with
a dependence only on metallicity (see Figure \ref{fig:dr14_vmacro}):

\begin{equation}
v_{macro}=10.^{\left( 0.471 - 0.254 [M/H]\right)}
\end{equation}

For dwarfs, the DR14 methodology was the same as the DR13
methodology.

\subsection{Fitting for stellar abundances}

After the stellar atmospheric 
parameters are determined for each star, individual elemental
abundances are determined using spectral windows located on features of each element,
varying the \mh, \am, \cm, or \nm\ dimension of the grid, depending on
the element, with other dimensions fixed, \ie, the \am\  dimension is used
to derive abundances for the different $\alpha$-elements (O, Mg, Si, S, Ca, Ti)
and the \mh\  dimension is used for all other elements (except C and N
in giants, see below).
We note that, as in previous releases, the
stellar parameter dimensions are fixed to their fitted
values, \ie, before any calibrations (discussed below) are applied. 
The rationale for this is that these values provide the best match of synthetic spectrum, hence would do the best job in removing blends from lines of interest.
In addition,
differences between the uncalibrated spectroscopic parameters
and independent estimates may be absorbing inaccuracies in some 
of the assumptions in the models, such as the 1D LTE methodology.

Carbon and nitrogen abundances in giants were determined by varying
the C and N dimensions in the grid, which provides an accurate modeling
of their abundances. However, as discussed above, the grids for dwarfs
do not have separate C and N dimensions. As a result the metallicity
dimension was used to solve for C and N in dwarfs.  However, it was
subsequently realized that this procedure is fundamentally flawed because
the C and N abundances come largely from molecular lines. For these lines,
the metallicity dimension changes abundances of all constituent species
simultaneously, so leads to incorrect results for C and N. So while
C and N abundances should be reliable in giants, they are not in dwarfs,
and should not be used in the latter.

\subsubsection{New elements and revised element windows}

DR13/DR14 includes detailed abundances for additional elements that
generally have weaker lines than the 15 elements presented in the
previous data releases: the newly added elements include P, Co, Cr,
Cu, Ge, Rb, Nd \citep{Hasselquist2016}, as well as carbon abundances from 
atomic lines of C I.  We also
note that although an abundance labelled Y (yttrium) is provided,
the dominant feature in the windows used for this is actually a
Yb (ytterbium) line, so no meaningful abundance is provided.
As discussed below, the lines from many of these elements are weak
and sensitive to blending, and abundances from them should be used with 
caution, if at all. 
\referee{The current methodology does not produce meaningful results for
elements that are derived from a feature that
is blended with a line from another element that varies in the same
dimension in the grid; we are considering how best to ameliorate this
for future analysis. }

The procedure used to determine the windows for each element was
slightly modified from that used for DR12. The windows were determined by 
finding regions of the spectrum that are sensitive to variations in the 
abundance of each element (at the stellar parameters of Arcturus), 
while at the same time being less sensitive to the variation of other
elements within the same grid dimension, and weighting pixels
according to these considerations. Within these windows, higher
weight was given to features where a model Arcturus spectrum better
matches the observed Arcturus spectrum.

The window determination also uses the mean residuals from fits to the full
APOGEE sample.  In DR12, those pixels with residuals larger than a given
threshold were removed by assigning them zero weight.
This results in some windows with peculiar shapes; the windows for 
Al provide a good example. To avoid this, we adopted for DR13/DR14 the same
procedure used to identify pixels not well reproduced
in the Arcturus spectrum, but instead of simply completely removing 
those pixels with larger differences, they are scaled from the pixel
with the largest residuals, which is assigned a zero value, to the
pixels at the threshold, which are not downweighted. Therefore,
pixels with residuals near the threshold have weights close
to one (\ie, the weights of the pixel are only slightly
lowered) and only the pixel with the largest residuals is completely
removed.

\section{Effect of model atmospheres}
\label{sect:atmos}

As described in section \ref{sect:coolgrid}, stellar atmospheres from 
Kurucz were used for the bulk of the sample, but MARCS stellar atmopheres
were used for stars with \teff$<$3500 K. However, there was an overlap region
between $3500<$\teff$<4000$ in which stars were fit by both grids.

\begin{figure}
\includegraphics[width=0.5 \textwidth]{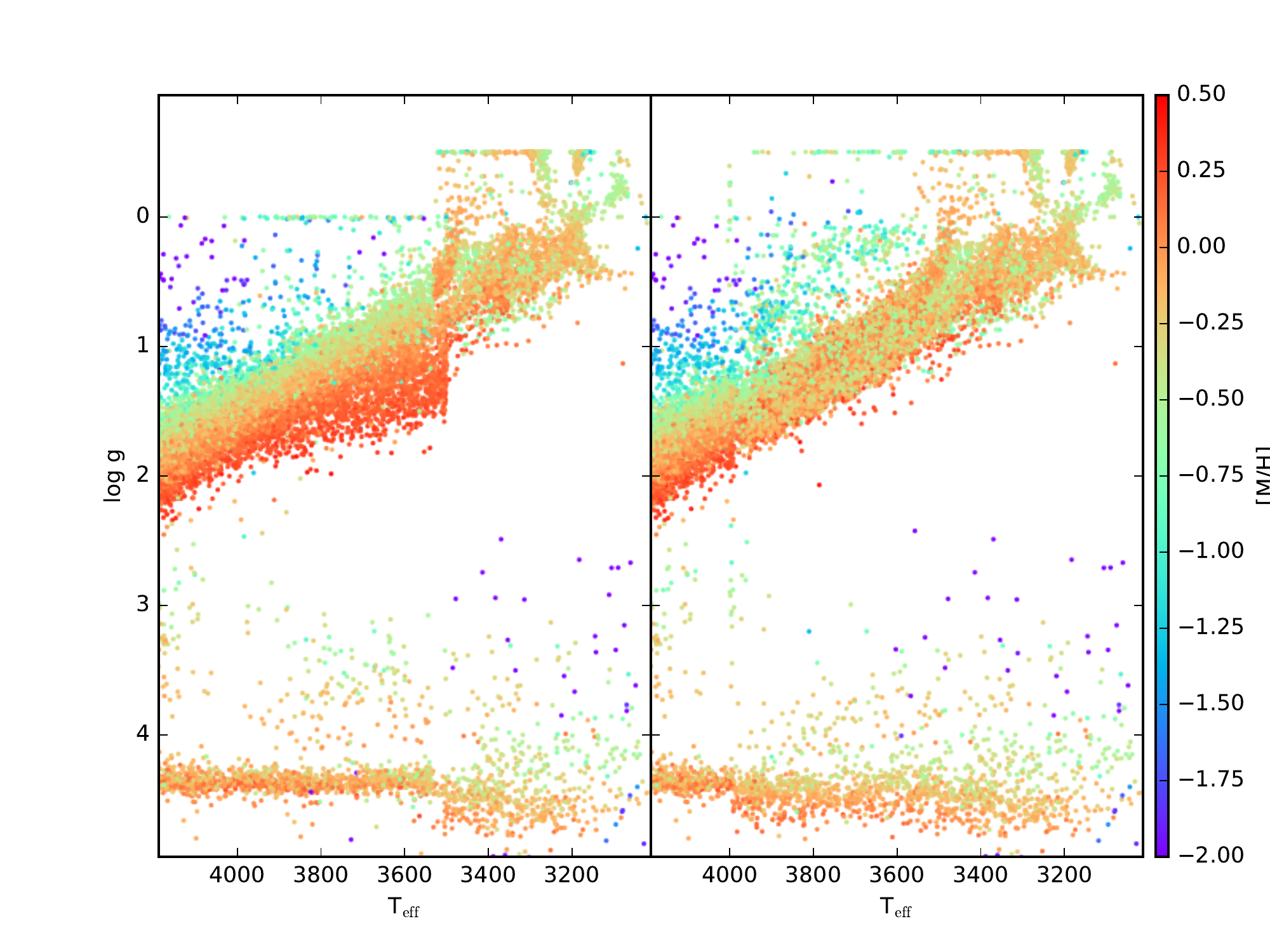}
\caption{Raw spectroscopic HR diagrams at $3500<$\teff$<4000$ K 
derived using Kurucz atmospheres (left) and MARCS atmospheres (right).
\referee{Point color is used to represent the derived \mh.}
As discussed in the text, results from the Kurucz atmospheres were
adopted in this temperature range for DR13 and DR14.}
\label{fig:kurucz_marcs}
\end{figure}

Figure \ref{fig:kurucz_marcs} shows the uncalibrated spectroscopic 
HR diagrams in this effective temperature range derived using the
Kurucz (left panel) and MARCS (right panel) atmospheres. The transition between
the two atmospheres is easily seen at 3500K in the left panel and at
4000 K in the right panel.

For DR13 and DR14, we have adopted the results from the Kurucz grid in
this temperature regime. Several factors contributed to this choice.
First, the MARCS grid available at the time was significantly coarser
than the Kurucz grid, so this led to a preference for the Kurucz grid,
especially for regions in \teff\ and \logg\ where sphericity effects
are not expected to be very significant (i.e., warmer than 4000-4500 K). 
Given this choice, we felt that it
was better to use a  homogeneous grid extending to 3500 K rather than
to mix results from different sets of atmospheres. On top of this,
results from the Kurucz grid seem to show a cleaner trend of stellar
parameters with metallicity at $3500<$\teff$<4000$ K 
on the upper giant branch \referee{than the results from the MARCS grid}
(seen as the mix of colors in the upper giant branch in the right panel
of Figure \ref{fig:kurucz_marcs}).

Clearly, this leaves a significant discontinuity at 3500 K, which leads
to our recommendation that results below 3500 K be used with extreme
caution, and recognition of the fact that there may be 
systematic uncertainties for stars with $3500<$\teff$<4000$ K.

Subsequent to this analysis, a finer grid of MARCS models has been calculated
by one of us (BE), which should allow a homogeneous analysis over the full
effective temperature range, so future data releases will likely use these
atmospheres.

\section{Calibrations and uncertainties}
\label{sect:calibration}

In this section, we describe the calibrations that have been derived
for the values of stellar parameters and elemental abundances from the
spectral fits.  We also describe how we derive empirical uncertainties
in these quantities. To avoid over-complication and since DR14
supercedes DR13, we present here the DR14 calibrations in detail and
only qualitatively describe the DR13 calibrations; additional details
on DR13 are presented in an Appendix.

\subsection{Effective temperature calibration}

In DR12, the spectroscopic temperatures were compared with photometric
temperatures from \citet{GHB2009} and a single zero-point correction
was applied to provide a ``calibrated" effective temperature.

\subsubsection{DR13}

Subsequent inspection of the different photometric
temperature calibrations from the literature
highlighted that different photometric scales differ 
by an amount comparable to the offset applied to DR12. 
Given the uncertainty in photometric temperature scales,
it was decided that no external calibration would be applied 
to the effective temperatures in DR13.

After the DR13 release was frozen, however, it became apparent that there
are trends in the comparison of DR13 spectroscopic temperatures with
photometric temperatures, in particular, as a function of metallicity. As
a result, we have suggested a ``post-calibration" correction to effective
temperatures, as described in Appendix \ref{appendix:DR13_teff}.

\subsubsection{DR14}

\begin{figure}
\includegraphics[width=0.5 \textwidth]{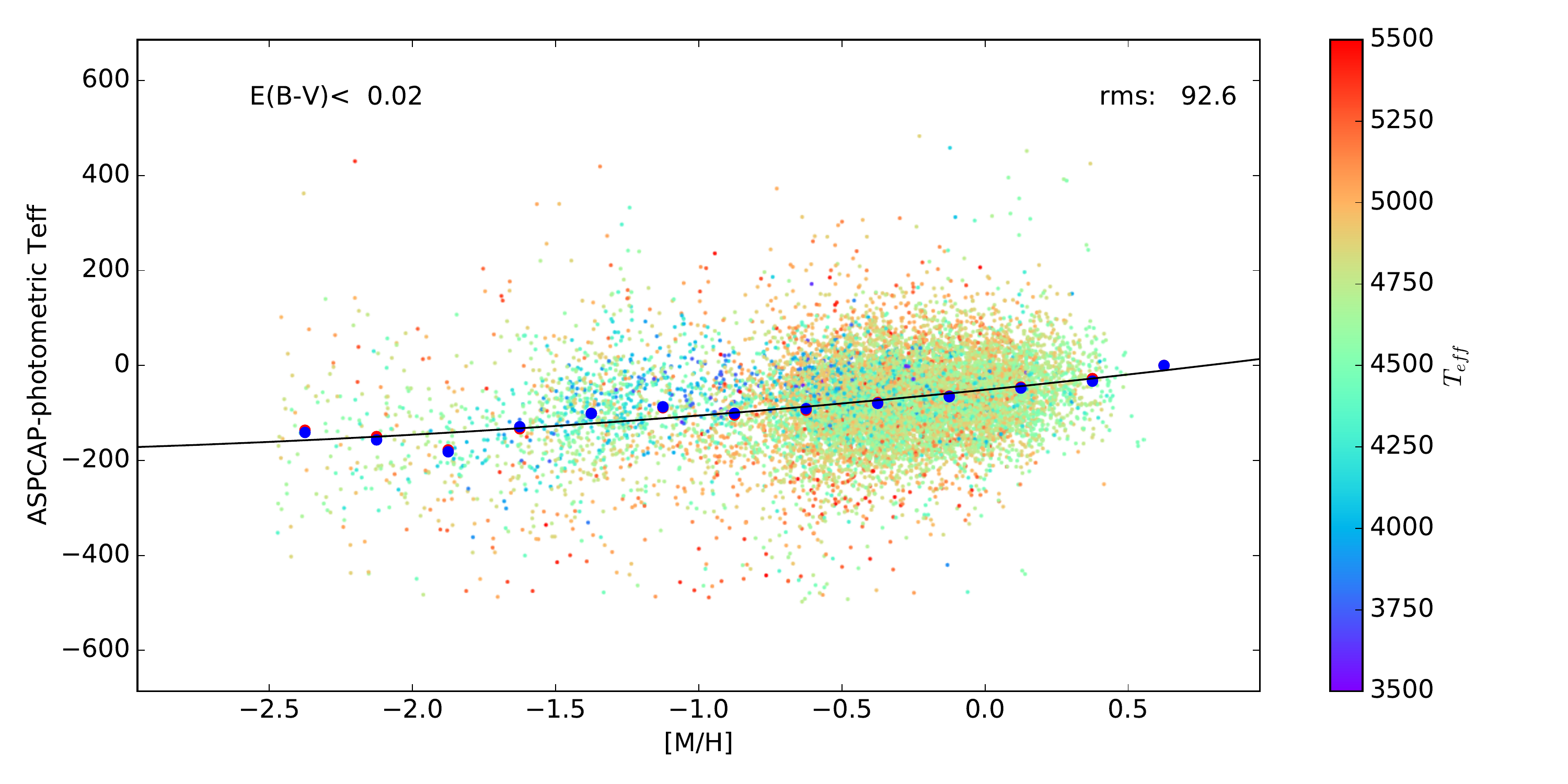}
\includegraphics[width=0.5 \textwidth]{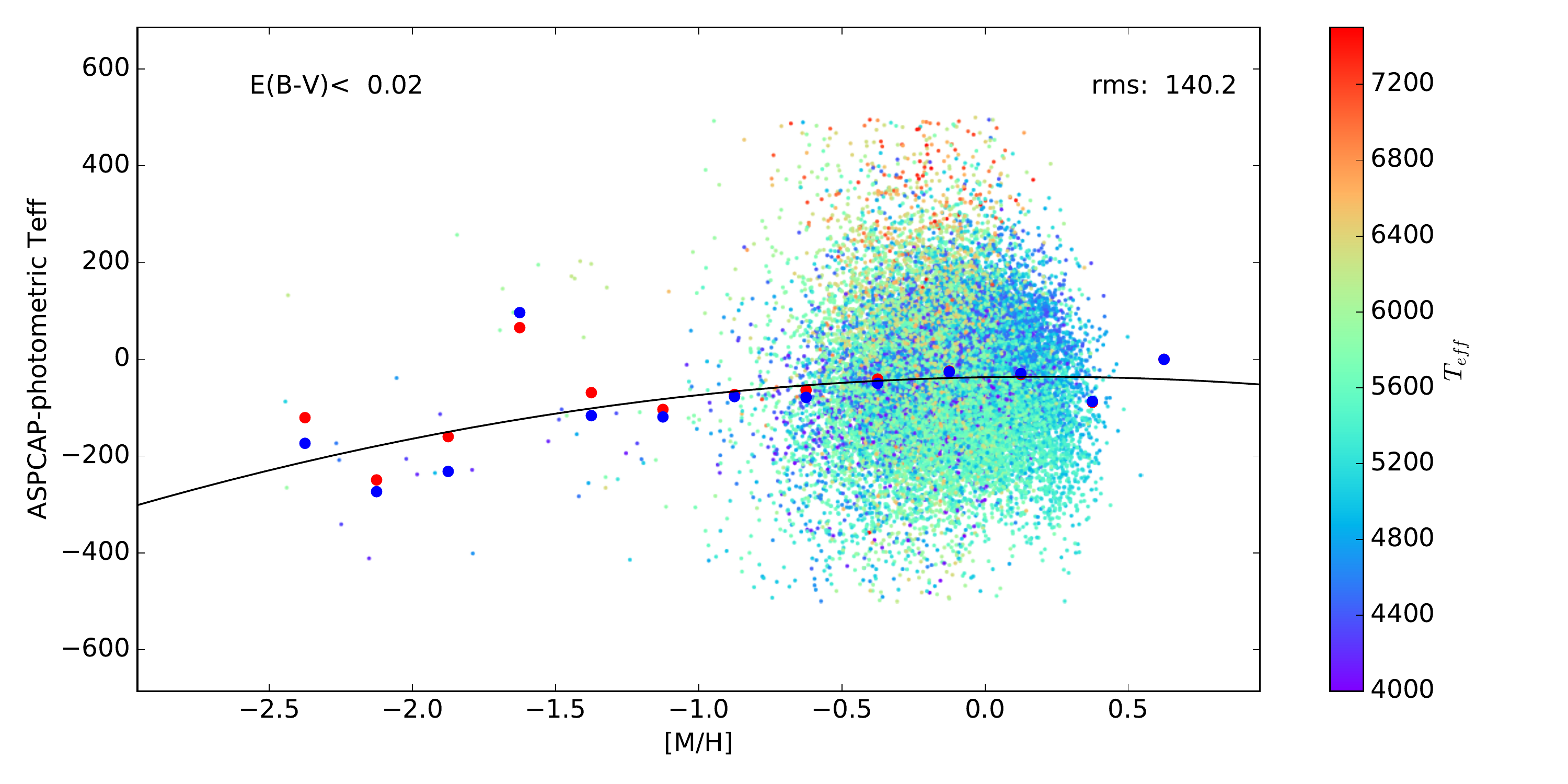}
\caption{Relation between raw DR14 ASPCAP \teff\ and photometric \teff\  from \citet{GHB2009}
as a function of metallicity (color-coded by \teff) for giants (top) and dwarfs (bottom). 
\referee{Large red} and blue points show mean and median difference
in bins of metallicity. Curves show derived and adopted DR14 calibration.}
\label{fig:teffcomp_dr14}
\end{figure}

Because of the issues discovered with DR13, a calibration relation
for \teff\  was adopted for DR14.  
Figure \ref{fig:teffcomp_dr14} shows the difference between DR14 raw
ASPCAP derived effective temperatures and photometric temperatures
for a low reddening ($E(B-V)<0.02$) sample. A trend with metallicity exists,
but it is not so large as in DR13; the improvement comes from a
revised handling of the normalization of the spectrum. It should be
kept in mind that the photometric effective temperature scales may
also have uncertainties. 

We adopted an effective temperature calibration that is a quadratic function of metallicity:
\begin{multline}
\label{eqn:dr14_teff}
T_{eff}(ASPCAP)-T_{eff}(GHB) = A_{Teff} + \\
 B_{Teff} [M/H] + C_{Teff} [M/H]^2
\end{multline}
with separate parameters for giants and dwarfs, as given in
Table \ref{tab:dr14_tecal}, where we define giants to have:
\begin{equation}
\log g < 2./1300 (T_{eff}-3500) + 2
\label{eqn:giantsep}
\end{equation}
The adopted calibration relations are shown in \referee{Figure} \ref{fig:teffcomp_dr14}. These
relations were derived from giants with $3750<T_{eff}<5500$ and from dwarfs
with $4000<T_{eff}<7500$. They were applied to all stars with $T_{eff}>3532$,
pinning the applied correction to the lower and upper ends of the range from
which the relations were derived outside of that range.

\begin{table}
\caption{Parameters for DR14 \teff\ calibration.}
\begin{center}
\begin{tabular}{llll}
\hline
Sample &$A_{Teff}$ & $B_{Teff}$ & $C_{Teff}$ \\
\hline
giants & -51.59  &   61.48  &   7.176\\
dwarfs & -36.38  &   13.16  &  -26.10\\
\hline
\end{tabular}
\end{center}
\label{tab:dr14_tecal}
\end{table}

While our relation is derived from a comparison with photometric
effective temperatures, we note that \citet{Jonsson2018}
provide an independent indication that even the calibrated
ASPCAP temperatures have a small remaining metallicity-dependent error with
respect to optical spectroscopic \teff, highlighting the challenge
of achieving a true effective temperature scale.

For an estimate of the uncertainties in \teff, we calculate the scatter
between the photometric and spectroscopic effective temperatures in
bins of \teff, \mh, and S/N, and a linear surface fit was performed to
derive coefficients that approximate the observed scatter:

%\begin{equation}
\begin{multline}
\ln(\sigma_{Teff}) = A_{\sigma Te} + B_{\sigma Te} (T_{eff} - 4500) + \\ C_{\sigma Te} [M/H] + D_{\sigma Te} (S/N - 100)
\label{eqn:dr14_teff_error}
\end{multline}
%\end{equation}
where we fit to the logarithm to ensure that the uncertainty never reaches
negative values (\ie, outside of the range of the calibration data). Again, a separate fit
was performed for red giants and dwarfs, with parameters as given in Table \ref{tab:dr14_teff_error}.

\begin{table}
\caption{Parameters for DR14 \teff\ uncertainties.}
\begin{tabular}{lllll}
\hline
Sample & $A_{\sigma Te}$ & $B_{\sigma Te}$ & $C_{\sigma Te}$ & $D_{\sigma Te}$ \\
\hline
giants & 4.361 &0.000604 &-0.00196  &-0.0659\\
dwarfs & 4.583 &0.000290 &-0.00130  & -0.243\\
\hline
\end{tabular}
\label{tab:dr14_teff_error}
\end{table}

\begin{figure}
\includegraphics[width=0.5 \textwidth]{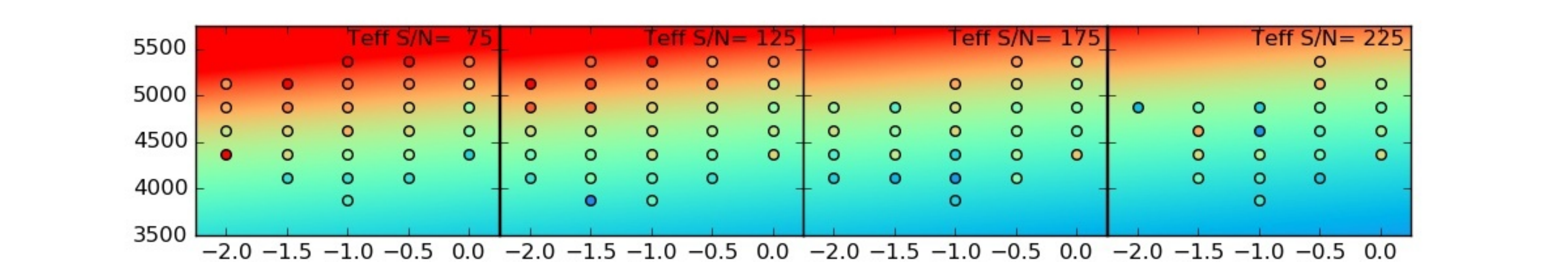}
\caption{Measured scatter in \teff\ in bins of \mh, \teff, 
and S/N (different panels): color of points give observed scatter. 
Background color shows derived fit that was used to determine uncertainties
for individual stars.}
\label{fig:dr14_teff_error}
\end{figure}

The fit surfaces are shown in Figure \ref{fig:dr14_teff_error}, which demonstrate that the \teff\ uncertainties
are largely a function of \teff, with larger uncertainties at higher temperatures.

\subsubsection{Raw and calibrated effective temperatures}

Both the raw and calibrated temperatures are included in the data release,
although, as noted above, the calibrated and uncalibrated effective
temperatures are identical in DR13.  Raw quantities can be found in the
FPARAM array, while calibrated quantities are found in TEFF and in the
PARAM array.

\subsection{Surface gravities}

Spectroscopic surface gravities are challenging to derive accurately. 
We have taken the approach of calibrating the derived surface gravites
using observations of stars in the Kepler field, for which high
precision measurements of surface gravity are available from
asteroseismic analysis \citep{Pinsonneault2014}; we adopted asteroseismic
values from version 3.6.0 of the APOKASC catalog.

For DR12, \citet{Holtzman2015} discuss offsets between the ASPCAP values of
\logg\  when compared to those derived from asteroseismology;
the differences between asteroseismic and spectroscopic
surface gravities was found to be different for red giants and red clump
stars. For DR14, the offsets between the ASPCAP and asteroseismic surface
gravities have been reduced slightly, which results from the improved
treatment of the LSF and macroturbulence.  However, differences still
remain, including the offset between red giant (RGB) stars and red clump
(RC) stars that deserve further study (see, \eg, \citealt{Masseron2017}).

The surface gravities for dwarf stars are more problematic. The 
spectroscopic surface gravities for dwarfs generally seem to be
too low, especially for cooler stars, based on expectations from 
stellar isochrones. In addition, we do not have a significant number 
of asteroseismic calibrator for dwarfs. As a result, we do not provide
calibrated surface gravities for dwarf stars at all, and defer improvements
on this issue to subsequent analysis.

The following subsections describe the surface gravity calibrations applied
to giants for DR13 and DR14.

\subsubsection{DR13}

For DR13, we adopted a \logg\  correction for red giants 
that depends both on surface
gravity and metallicity.  We note that this differs from DR12, where
only a surface gravity dependence was found and calibrated. As with DR12,
separate corrections were derived for RGB and RC stars.  Due to limited
availability of asteroseismic data when the calibration was frozen,
we chose to clip the metallicity correction to \mh$>$-1.5. Subsequent
analysis of addditional data demonstrate that this clipping was incorrect,
so we recommend a ``post-calibration" correction to the DR13 surface
gravity for low metallicity stars.

Appendix \ref{appendix:DR13_logg} provides the details of the DR13
surface gravity calibrations and recommended correction, including 
the criteria used to distinguish RGB and RC stars.

\subsubsection{DR14}

\begin{figure}
\includegraphics[width=0.5 \textwidth]{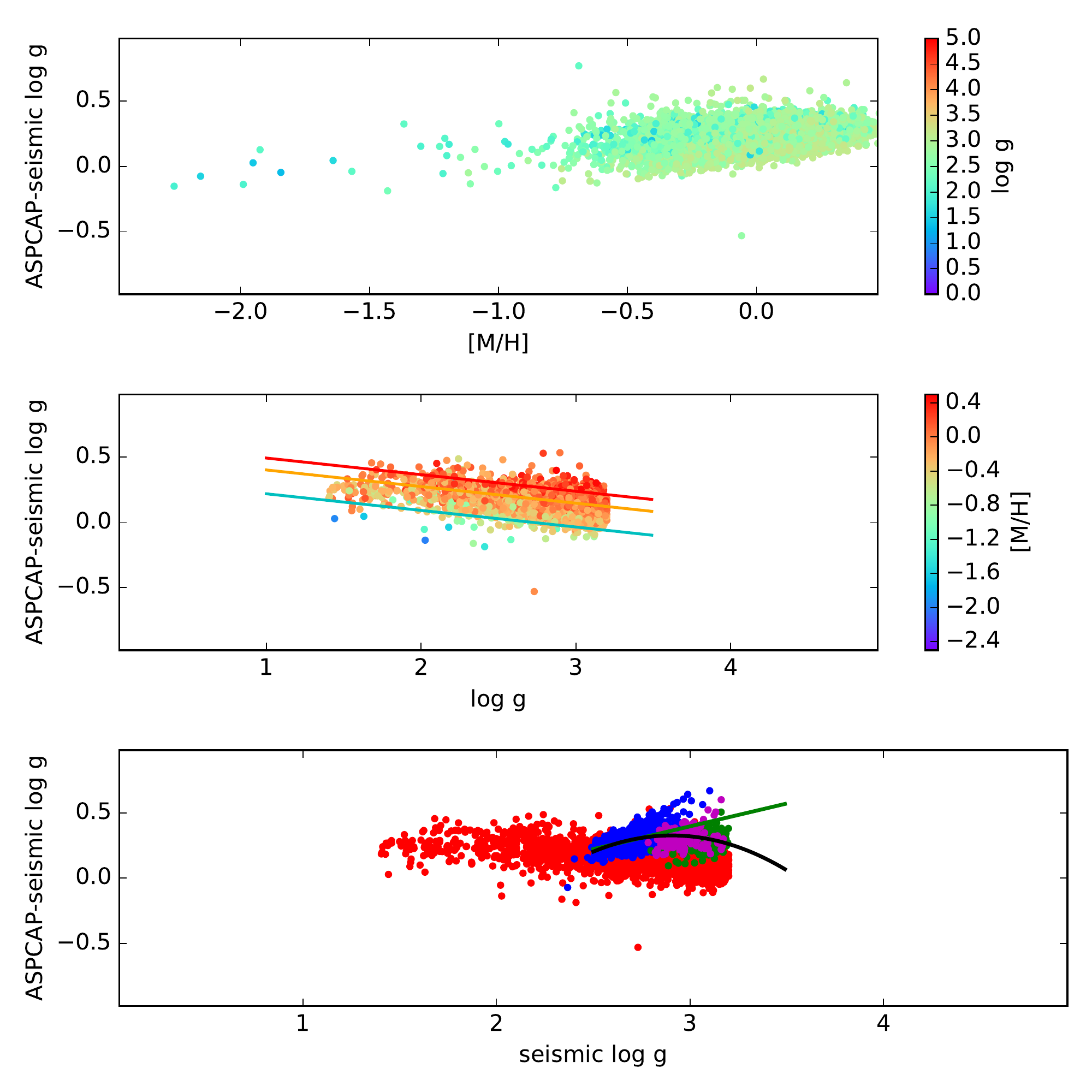}
\caption{Comparison of raw ASPCAP surface gravity with asteroseismic
surface gravity. Top panel shows difference between spectroscopic
and asteroseismic surface gravity as a function of metallicity,
color-coded by gravity. Middle panel shows difference as a function
of surface gravity, color-coded by metallicity, for red giants
only; lines show derived RGB calibration relation for metallicities
0.5, 0, and -1. Bottom panel shows different as a function of 
surface gravity, with RGB sample in red, RC  sample in blue, 2CL
sample in green, and 2CL/RC sample in magenta (see text for
description); the green line shows the RC calibration relation used for DR14,
while the black line show the recommended RC calibration.}
\label{fig:dr14_loggcomp}
\end{figure}

Figure \ref{fig:dr14_loggcomp} shows the difference between raw DR14 ASPCAP
and asteroseismic surface gravities for red giant stars.
The top panel of Figure \ref{fig:dr14_loggcomp} demonstrates a
trend with metallicity, the middle panel demonstrates a trend
with surface gravity (RGB stars only), and the bottom panels
shows the difference between RGB stars (red) and RC stars (blue).
From these data, we derived
separate calibration relations for RGB and RC stars:

\begin{multline}
 \log g = \log g (raw) - (0.528 - 0.127 \log g + \\
            0.183 [M/H])
\end{multline}
for RGB stars, and, for RC stars:
\begin{multline}
 \log g = \log g (raw) - (-0.643 + 0.346 \log g + \\
            0.0147 [M/H])
\end{multline}

Unfortunately, we discovered after the data release that the
RC stars denoted as such in the APOKASC catalog do not include
so-called secondary clump stars, which are denoted as 2CL, with
transition objects denoted as RC/2CL; these are shown in the bottom
panel of Figure \ref{fig:dr14_loggcomp} as green and magenta
points, respectively. As a result, the calibration relation
derived and applied for RC stars (green line) is not valid for the
RC/2CL and 2CL stars; rather than increasing, the correction 
should decrease for the higher surface gravity core helium burning 
stars.  We discuss how to do so below.

To apply a separate calibration for RGB and RC stars requires
some way of distinguishing them for stars without asteroseismic
analysis.  We revisited the methodology used in DR12 for RC/RGB separation using
the asteroseismic sample (which provides an RC/RGB classification
from seismology). The basic idea is to use a ridgeline in the
\teff-\logg\  plane that is a function of metallicity, and supplement
this with the measurements of the \xx{C}{N} ratio, since this
is expected (and observed) to further separate the RGB and RC.
We define a difference, $\Delta t$, between ASPCAP \teff\  and
a metallicity-dependent ridgeline:

\begin{multline}
$$\Delta t=T_{eff}(raw)-(4444.14+ 554.311 (\log g (raw)-2.5) - \\
307.962 [M/H] (raw))$$
\end{multline}

Using these definitions, we then classify stars as RC if 
$2.38 <$\logg$<3.5$ and

\begin{equation}
\textrm{\xx{C}{N}}> -0.08 - 0.5\textrm{\mh} - 0.0039 \Delta t
\label{eqn:rgbrc}
\end{equation}
where \xx{C}{N} = \xm{C}(raw) - \xm{N}(raw). 

%RGB stars those where 
%\logg $> -1$ and \logg$< 3.8$  
%and \teff $>$ 3500 and \teff$<$ 5500 and
%(\logg$<$2.38  or \logg$>$3.5 or

\begin{figure}
\includegraphics[width=0.5 \textwidth]{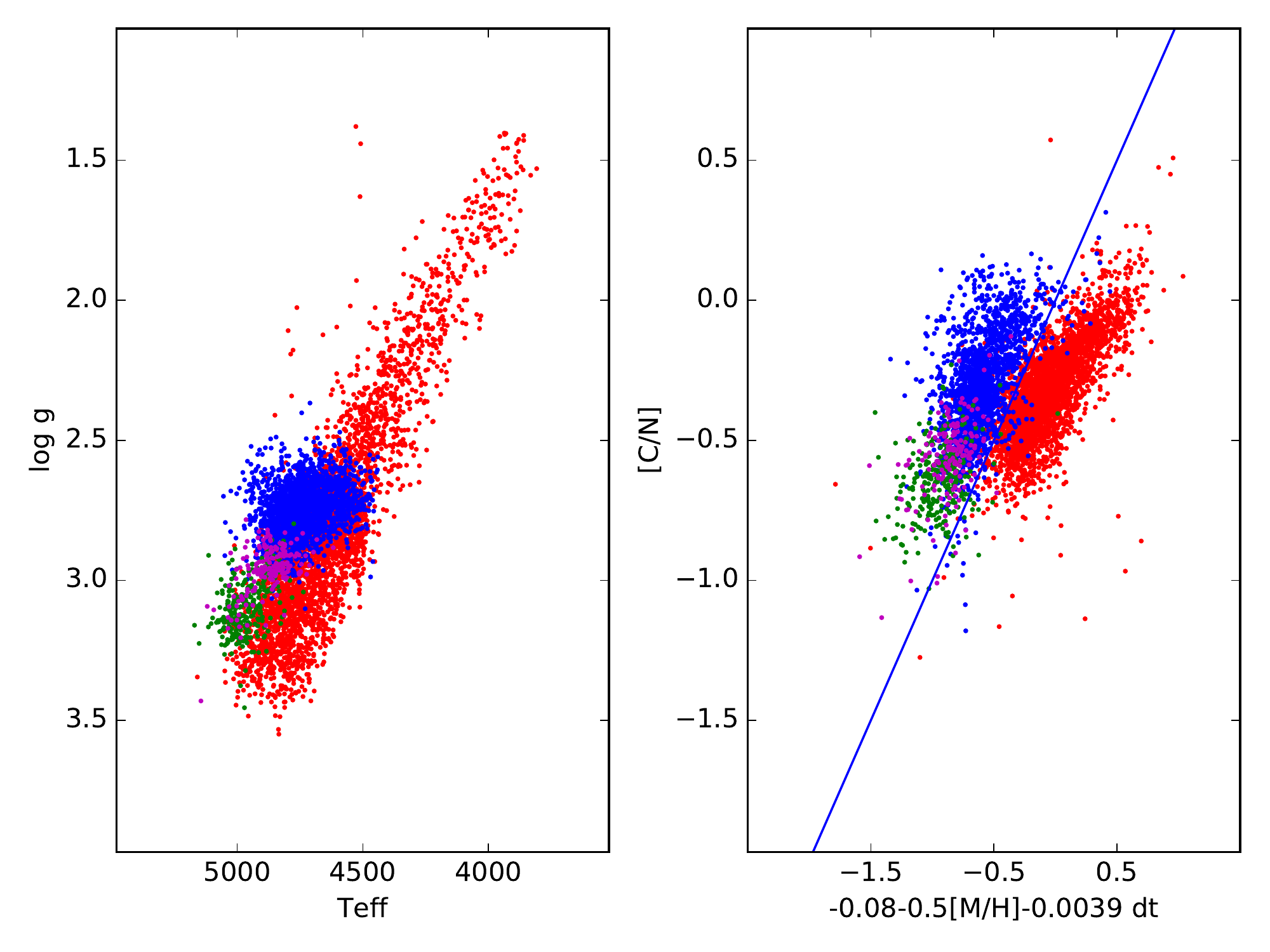}
\caption{Left panel show a spectroscopic HR diagram for stars from
the APOKASC catalog, with stars marked by evolutionary state: RGB (red),
RC (blue), RC/2CL (magenta), 2CL (green). Right panel shows the 
separation of RGB/RC using the relation discussed in the text.}
\label{fig:dr14_rcrgb}
\end{figure}

Figure \ref{fig:dr14_rcrgb} shows the location of the RGB, RC, 2CL/RC,
and 2CL stars in a spectroscopic HR diagram (left); the right panels
show \xx{C}{N} as a function of the right hand side of Equation
\ref{eqn:rgbrc}.  The line shows the relation used to separate RGB and RC.
We note that while this relation was derived to separate RGB and RC
stars in the asteroseismic sample, it is not perfect, with about a 5\%
failure rate for each category. On top of this, different chemistry
in different regions of the Galaxy could lead to different \xx{C}{N}
ratios that might affect the validity of this separation. The effect
of misclassifications would be the difference between the RGB and RC
surface gravity calibration relations.

Because of the issue with the secondary red clump calibration, users
wanting the best estimates of surface gravities for these stars should
redo the \logg\  calibration for RC stars (those satisfying Equation
\ref{eqn:rgbrc}), instead using:
\begin{equation}
 \log g = \log g (raw) - (-6.05 + 4.39 \log g - 0.7556 (\log g)^2
\end{equation}
In practice, this is only important for stars with \logg $>\sim 2.9$,
but this still amounts to $\sim$ 13000 stars.

\subsection{Elemental abundances}

Individual raw elemental abundances are derived using fixed stellar
parameters from the full spectrum fit, by performing fits to windows
in the spectrum containing lines from the element.  As mentioned above,
we use the raw parameters rather than the calibrated ones when deriving
abundances, because these are the ones that provide the best global
fit to the spectrum, and ensure a consistent continuum fit. However,
\citet{Jonsson2018} suggest that this may lead to poorer abundances,
as compared with independent abundance analysis from the literature,
especially for elements whose abundances are strongly dependent on
effective temperature. Analysis has shown that this is the
case for titanium, but future analysis will investigate this in greater detail.

As was the case for DR12, we find that the derived abundances of
some of the individual elements show a small dependence on effective temperature
for stars within a given star cluster. Assuming that abundances are
homogeneous within clusters (e.g., \citealt{Bovy2016}, \citealt{DeSilva2006,
DeSilva2007}), such a dependence might result from effects that are not
well characterized by our models, such as NLTE effects, or effects from
blending that are a function of temperature. Under the assumption of
homogeneity, we
apply a small internal calibration to the derived abundances as a function
of \teff, using a set of star clusters to derive this calibration.
We have chosen to employ only clusters with metallicity greater than
\mh$>$ -1, because this is representative of the vast majority of
the APOGEE stars; this restricts the sample to mostly open clusters.
Unlike DR12, in DR13 and DR14 we fit for a \teff\  dependence of \xm{X}
rather than the \xh{X}\ that was used for DR12 because the scatter
in the \xm{X} vs \teff\ relations is smaller. As in DR12, no internal
calibration is made for C or N because these cannot be assumed to be
homogeneous within clusters along the giant branch because of mixing.

The observed cluster stars cover a more limited range of
effective temperature than that of the full sample. In an
effort to remove effective temperature dependences over a larger range of
the APOGEE data, we inspected \xm{X} vs \teff\  diagrams for
each element for a subsample of the APOGEE data with $70<l<110$, 
which limits the stars to be at Galactocentric radii not
dramatically different from the solar radius,
under the assumption that there should not be
a temperature dependence of abundance ratios within this sample.
We note that this sample was not used to derive the effective
temperature dependence of the calibration relations, but only
to inspect what different calibration relations from the
cluster fits (different orders and \teff\  limits for the fit)
had an impact on the \xm{X} locii; we adopted the calibration that
most effectively removed any effective temperature trends.

Separate calibration relations were derived for giants
and for dwarfs, where  equation \ref{eqn:giantsep} was
used to classify stars as giants or dwarfs. For most elements,
no calibration was applied to stars cooler than 3500 K because of
the lack of calibrators and because the extrapolation of the calibration
did not appear to work well judging from the solar circle sample,
although we do provide calibrated values for a few elements for which 
the extrapolation looked reasonable.

In addition to internally calibrating the abundances as a function of
\teff, we also adopted a zeropoint shift to force the mean abundance
ratios of all observed stars with -0.1$<$\mh$<$0.1, $-5<|b|<5$, and
$70<l<110$ to be zero, \ie, we forced the mean abundance ratio of
stars near the solar circle within $\pm 0.1$ of the solar abundance to
have solar abundance ratios. This is motivated by studies of the solar
neighborhood (e.g., \citealt{Bensby2014}) that suggest that
most stars in the solar neighborhood have solar abundance ratios
at solar abundance. Such an assumption
could be questioned, but given the internal calibration with \teff,
we have to adopt some \xm{X} zeropoint (DR12 simply adopted the raw
abundance at \teff$=$ 4500 K). The need for zeropoint corrections
might result from issues with the astrophysical $\log gf$ values, e.g., 
from incorrect assumed abundances for Arcturus and/or the Sun.

\subsubsection{DR13}

Calibration relations for DR13 elemental abundances are presented in
Appendix \ref{appendix:dr13_abundances}.  We note that abundances for
Nd and Y show large scatter in clusters for both dwarf and giants,
potentially indicating that windows used in determining the chemical
abundances are not ideal. These elements are derived from weak, blended
features in the spectra that are likely not present in stars with \teff\
$\gtrsim$ 4200 K (see \eg, \citealt{Hasselquist2016}).  Moreover,
it was later discovered that Y abundances were actually derived from
a spectral line that was actually due to a transition of Yb (see \eg,
\citealt{Hawkins2016}). While we do provide raw values for Nd and Y,
these values do not represent the abundance of these elements and should
not be used.

Rb, Cu, and Ge show strong temperature trends, and while calibrated
abundances are provided, they should \referee{probably not be used}.  A detailed
analysis of the two Rb I lines used for the DR13/DR14 abundances reveals
that both lines are almost certainly affected significantly by blends,
with the bluer Rb I line blended with a CN line and the redder Rb I line
blended with an unclassified, high-excitation Fe I line.  Inspection
of the single germanium line in the APOGEE spectral window (Ge I at
$\lambda$=16759.76\AA$_{\rm Air}$) finds that this line is very weak.

\subsubsection{DR14}

Figures \ref{fig:dr14_giants_intcal} and \ref{fig:dr14_dwarfs_intcal}
show the internal calibration relations for red giants and dwarfs for DR14,
as derived from cluster stars.   \referee{Figure \ref{fig:dr14_clusters} lists the clusters
that were used and the symbols that represent them in 
Figures \ref{fig:dr14_giants_intcal} and \ref{fig:dr14_dwarfs_intcal}.}
In general, the trends with \teff\  are
small, with the largest slopes occuring in giants for Na, Ti, Cr, Mn, and
Co.

\begin{figure}
\includegraphics[width=0.5 \textwidth]{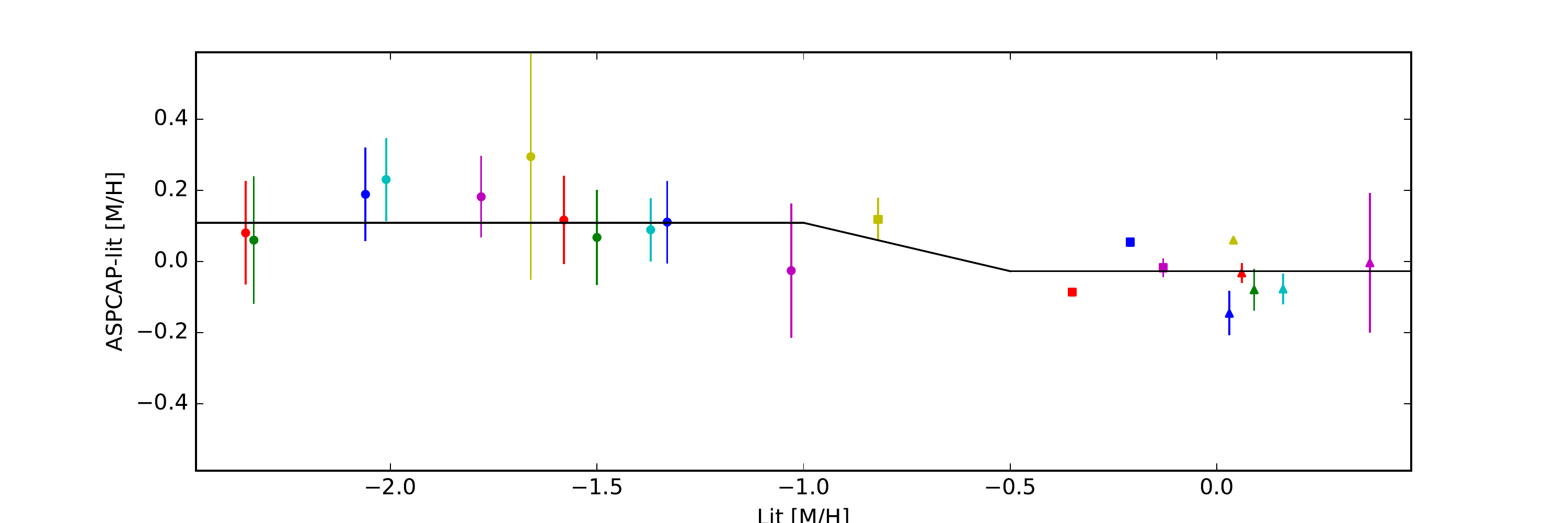}
\caption{Mean \mh\ of cluster stars compared with cluster metallicities
from the literature, and derived \mh\ calibration.
\referee{Different point colors and symbols are used to distinguish different clusters.}}
\label{fig:mhcal}
\end{figure}

\begin{figure}
\includegraphics[width=0.5 \textwidth]{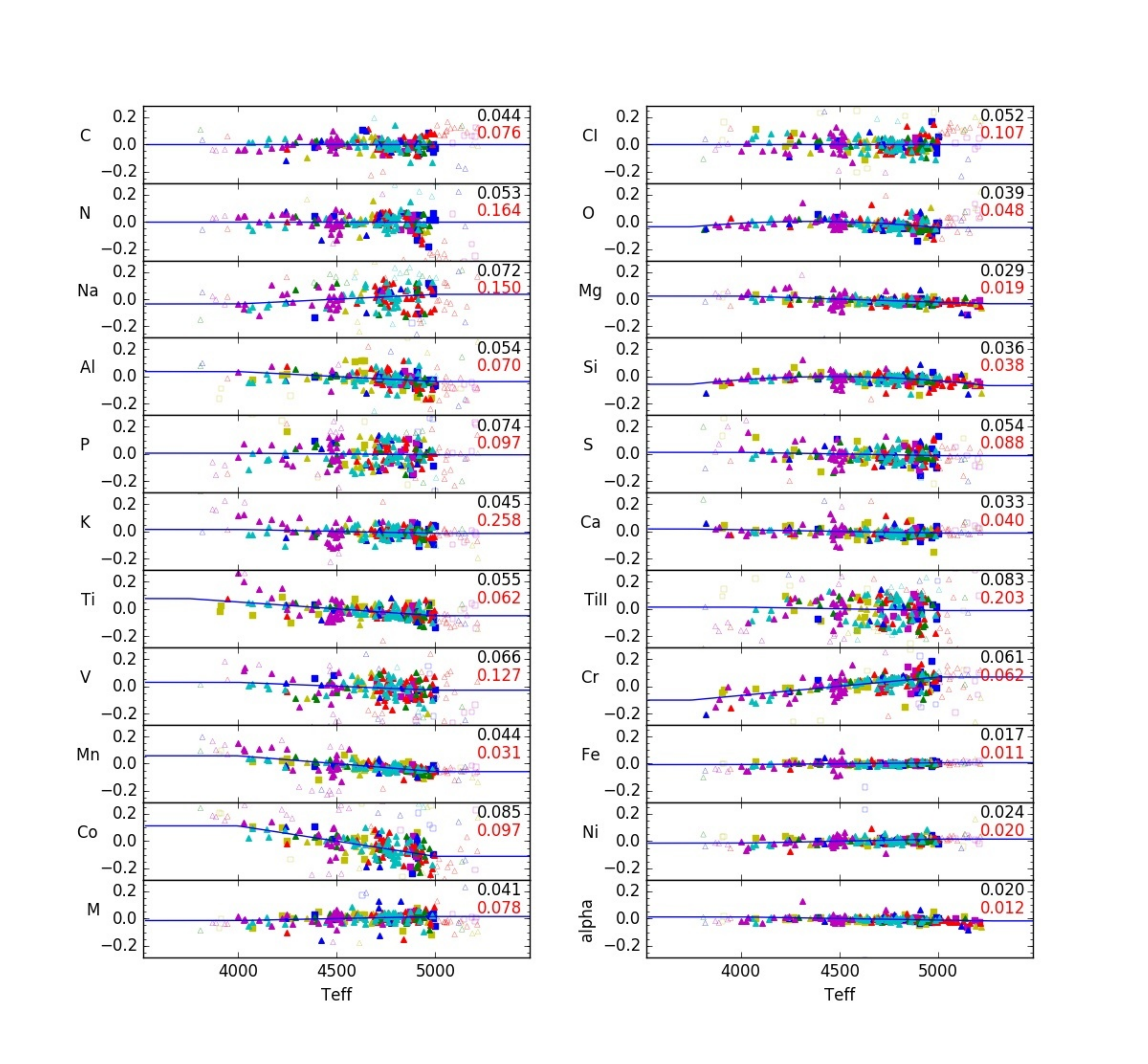}
\caption{DR14 internal abundance calibrations for giants applied as a 
function of \teff. Each panel shows the results for abundances in calibration
clusters along with the adopted calibration relation (line). \referee{Different colors
and point types are used to distinguish different clusters as denoted
in Figure \ref{fig:dr14_clusters}.} Number in the
upper right gives residual scatter around the calibration relation; red number
is scatter for M67 stars only.}
\label{fig:dr14_giants_intcal}
\end{figure}

\begin{figure}
\includegraphics[width=0.5 \textwidth]{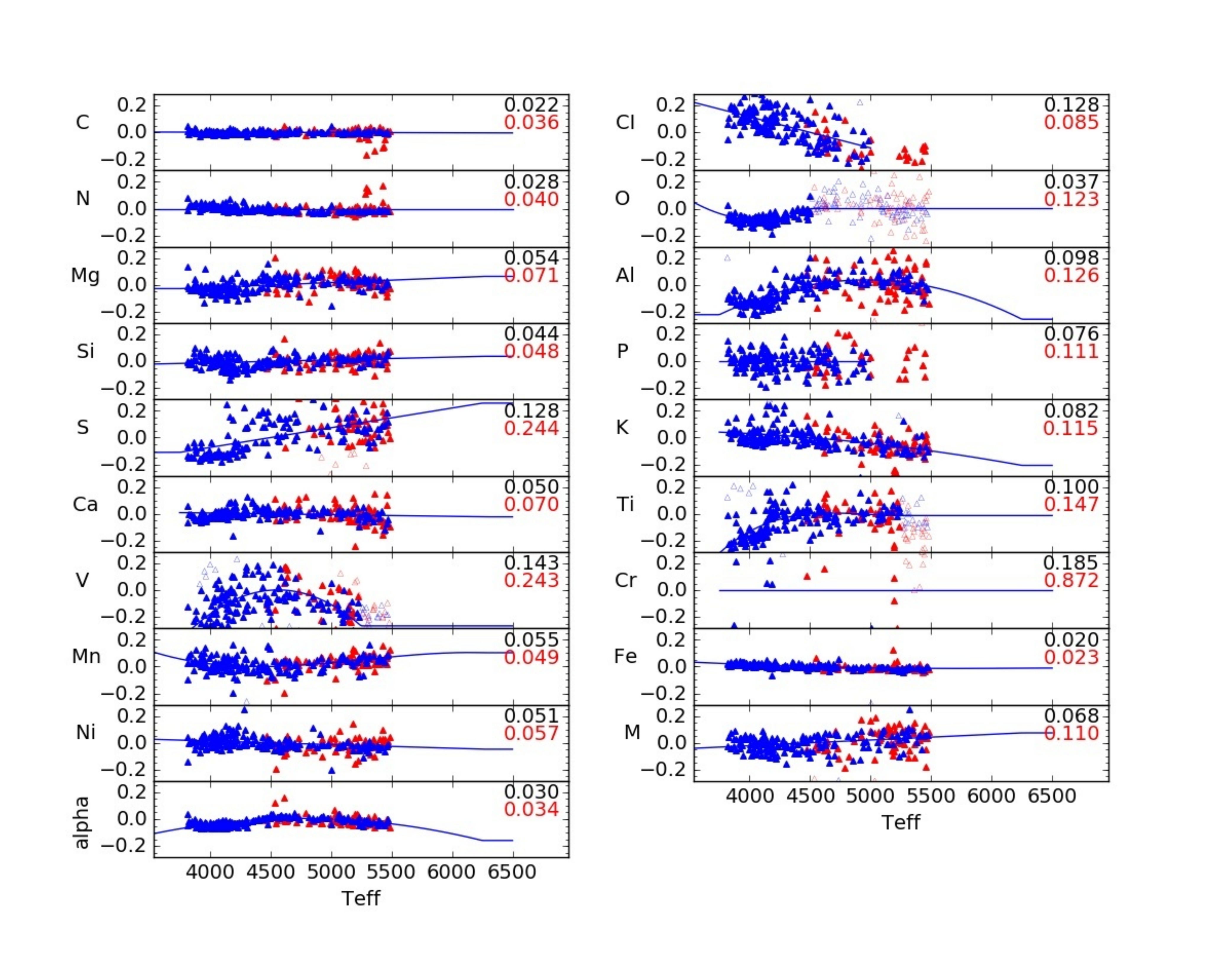}
\caption{As Figure \ref{fig:dr14_giants_intcal}, but for dwarfs.}
\label{fig:dr14_dwarfs_intcal}
\end{figure}

\begin{figure}
\includegraphics[width=0.5 \textwidth]{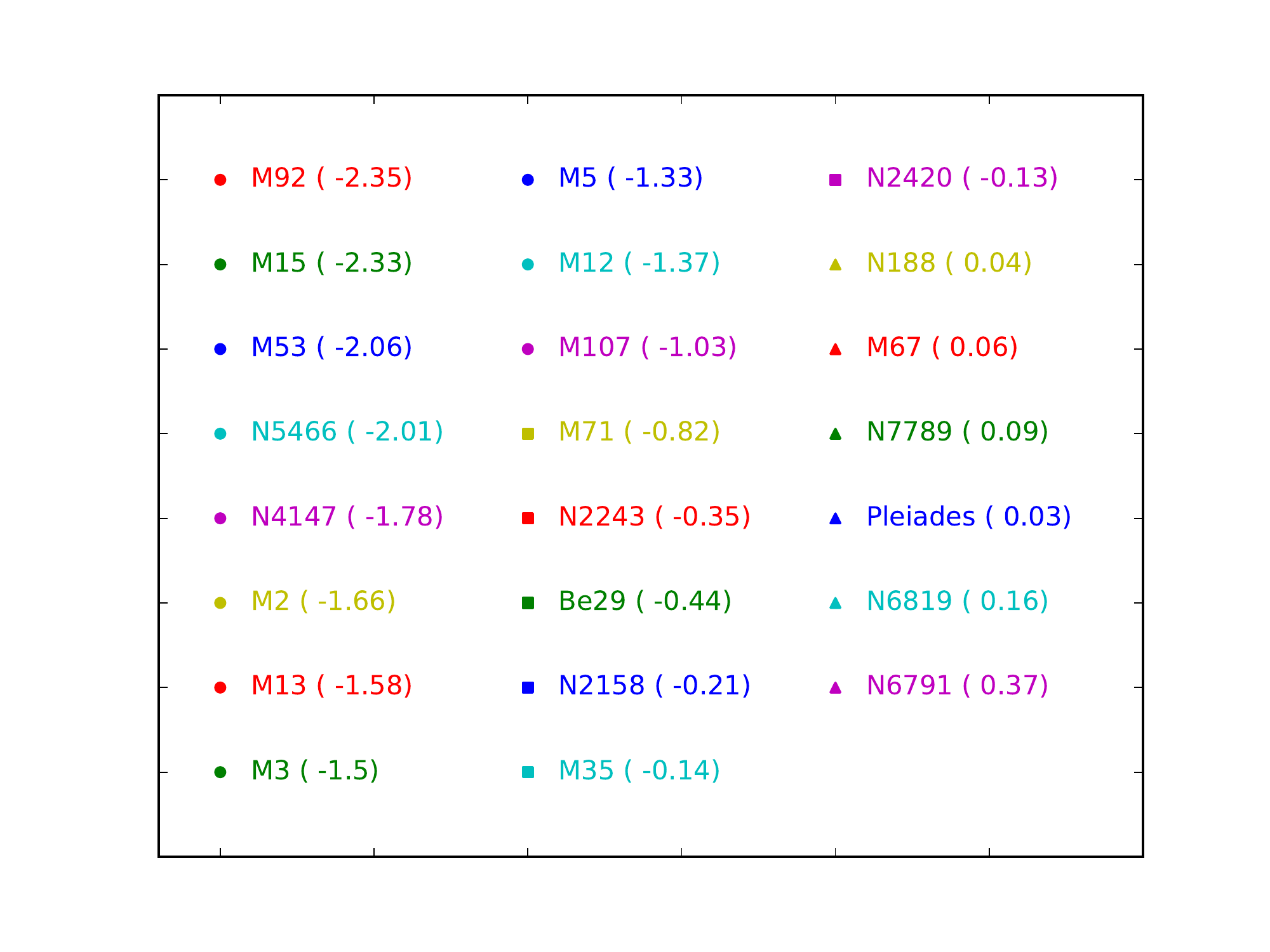}
\caption{\referee{List of clusters used for internal calibration, with
associated symbols used in Figures \ref{fig:dr14_giants_intcal} and
\ref{fig:dr14_dwarfs_intcal}.}}
\label{fig:dr14_clusters}
\end{figure}

Figure \ref{fig:mhcal} compares the mean derived \mh\ from cluster stars
with cluster metallicities from the literature. While different studies do
not always agree on cluster metallicities, the raw ASPCAP
\mh\ for metal poor clusters is $\sim$ 0.15 dex higher than most literature
values. As a result, we applied a simple calibration of a constant offset
at \mh$ < $-1, with a linear ramp to zero correction at \mh$ > $-0.5.
Note that while we applied this external correction to \mh, we did not
apply it to \xh{Fe}.

Table \ref{tab:dr14_offsets} show the zeropoint offsets applied to each
of the individual element abundances based on the solar circle sample,
for both giants and dwarfs and for
DR13 and DR14. In general these zeropoint shifts are modest ($<0.1$ dex), 
with the exceptions of Na, Al, Si, and V. Future line list work will
attempt to understand whether there are plausible reasons for these that
can be addressed.

While raw abundance measurements using the Rb, Cu, Ge, Y (Yb!), and Nd
windows were made for DR14, no calibrated abundances of these elements
are presented because of the challenges/problems involved with measuring
the relevant lines (discussed above), and we do not recommend 
the use of these abundances with the current analysis.

Similarly, a more in depth analysis
of the Na windows have revealed that the Na features are not measurable
at \mh\ $<$ -1.0 so, as a result, we do not provide calibrated abundances
of Na for metal-poor stars.

\begin{table}
\caption{Elemental abundance zeropoint offsets}
\begin{center}
\begin{tabular}{lrrrr}
\hline
Element&\multicolumn{2}{c}{DR14}& \multicolumn{2}{c}{DR13} \\
  & giants & dwarfs & giants & dwarfs \\
\hline
C & 0.000 & 0.000 & 0.000 & -0.019 \\
CI & 0.000 & -0.038 & 0.000 & -0.026 \\
N & 0.000 & -0.003 & 0.000 & -0.010 \\
O & 0.035 & 0.020 & 0.060 & 0.068 \\
Na & 0.103 & $\cdots$ & 0.186 & 0.096 \\
Mg & 0.022 & -0.035 & 0.045 & -0.003 \\
Al & 0.208 & 0.053 & 0.108 & 0.043 \\
Si & 0.127 & -0.034 & 0.107 & -0.023 \\
P & 0.003 & 0.000 & -0.008 & 0.000 \\
S & 0.003 & -0.074 & -0.092 & -0.017 \\
K & -0.046 & 0.001 & -0.026 & -0.029 \\
Ca & -0.027 & 0.045 & -0.021 & 0.023 \\
Ti & 0.016 & 0.049 & -0.014 & -0.002 \\
TiII & 0.090 & $\cdots$ & 0.166 & 0.000 \\
V & 0.142 & 0.186 & 0.110 & 0.002 \\
Cr & -0.137 & -0.066 & -0.057 & -0.044 \\
Mn & 0.012 & -0.106 & 0.041 & -0.077 \\
Fe & 0.003 & 0.023 & -0.005 & 0.016 \\
Co & -0.061 & $\cdots$ & 0.003 & 0.000 \\
Ni & -0.005 & 0.047 & -0.001 & 0.030 \\
Cu &$\cdots$ & $\cdots$ & 0.452 & 0.026 \\
Ge &$\cdots$ & $\cdots$ & 0.354 & 0.000 \\
%Rb & -0.319 & $\cdots$ & -0.105 & -0.217 \\
%Y &$\cdots$ & $\cdots$ & 0.000 & 0.000 \\
%M & 0.109 & -0.098 & 0.000 & 0.000 \\
$\alpha$ & 0.038 & 0.027 & 0.056 & -0.004 \\
\hline
\end{tabular}
\end{center}
\label{tab:dr14_offsets}
\end{table}

%\subsection{Abundance precision}

Estimated uncertainties in abundances were derived using the same
methodology as for DR12.  We use the abundance derivations in both
open and globular cluster stars (removing known second generation stars
from the latter) with the underlying assumption that 
individual element abundances are uniform in all cluster members (apart
from C and N).  
In the selected cluster
sample, we measure the element abundance scatter in \xm{X} 
in bins of effective temperature,
metallicity, and signal-to-noise (S/N). For each individual element,
we fit these values with a simple functional form:

\begin{equation}
log \sigmaσ = A + B (T_{eff} -4500) + C (S/N -100) + D [M/H] 
\end{equation}
where $\sigma$σ is the scatter among cluster stars relative to the 
mean derived abundance. The $S/N$ used in this relation is capped
at $S/N=200$.  Note that in the above relation, the fit to
log  ensures that the derived relation will always yield a positive
uncertainty. The values for the coefficients (A, B, C, D) associated with
each element for giants are given in Table \ref{tab:dr14_abun_err}. 

\referee{The abundance precision for \xh{Fe} is underestimated by this methodology
since the scatter is computed from \xm{Fe}, but the measurement of \mh\ 
is strongly dominated by Fe lines, so \xm{Fe} will show very little scatter.}

\referee{
As another estimate of uncertainty, we also include in Table \ref{tab:dr14_abun_err}
the ``global'' uncertainties, which represent the total scatter around
the effective temperature fits shown in Figures \ref{fig:dr14_giants_intcal}
and \ref{fig:dr14_dwarfs_intcal}. For Fe, we calculate this in \xh{Fe} to
avoid the problem mentioned above. However, these global uncertainties do not
capture the dependence on \teff, \mh, and S/N.}

\begin{table}
	\caption{Parameters for DR14 abundance uncertainties.}
	\begin{tabular}{l c c c c c r}
\hline
		\textbf{El} & \textbf{A} & \textbf{B} & \textbf{C} & \textbf{D} & \textbf{$\sigma$\footnote{\teff = 4500, \mh=0, S/N=100}}	& $\sigma_{global}$\\ 
\hline
		C & -3.488 & 9.42E-04 & -1.93E-03 & -0.685 & 0.030 & ---\\
		C I & -3.010 & 4.24E-04 & -2.82E-03 & -0.567 & 0.049 & ---\\
		N & -3.138 & 8.24E-04 & -1.20E-03 & -0.632 & 0.043 & ---\\
		O & -3.454 & 8.48E-04 & -3.15E-03 & -0.649 & 0.031 & 0.039\\
		Na & -2.413 & 4.62E-04 & -2.84E-03 & -0.188 & 0.089 & 0.132\\
		Mg & -3.826 & -7.13E-05 & -2.50E-03 & -0.693 & 0.021 & 0.039\\
		Al & -2.974 & 6.91E-04 & -2.00E-03 & -0.345 & 0.051 & 0.081\\
		Si & -3.643 & 3.17E-04 & -1.60E-03 & -0.473 & 0.026 & 0.037\\
		P & -2.233 & 3.10E-04 & -2.59E-03 & -0.149 & 0.10 & 0.130\\
		S & -2.704 & 1.12E-04 & -3.68E-03 & -0.453 & 0.066 & 0.062\\
		K & -2.966 & 2.52E-04 & -5.55E-03 & -0.467 & 0.051 & 0.061\\
		Ca & -3.510 & 2.02E-04 & -5.21E-03 & -0.634 & 0.029 & 0.038\\
		Ti & -3.243 & 5.48E-04 & -2.68E-03 & -0.508 & 0.039 & 0.064\\
		Ti II & -2.386 & 4.63E-04 & -1.49E-03 & -0.188 & 0.092 & 0.147\\
		V & -2.626 & 6.87E-04 & -2.50E-03 & -0.381 & 0.072 & 0.117\\
		Cr & -3.100 & 4.30E-04 & -4.13E-03 & -0.626 & 0.045 & 0.071\\
		Mn & -3.424 & 3.30E-04 & -4.60E-03 & -0.582 & 0.032 & 0.054\\
		Fe & -4.757 & -1.80E-04 & -8.32E-04 & -0.443 & 0.009 & 0.047\\
		Co & -2.469 & 7.21E-04 & -4.16E-03 & -0.065 & 0.084 & 0.141\\
		Ni & -3.779 & 2.84E-04 & -5.71E-03 & -0.659 & 0.022 & 0.024\\
		Rb & -2.434 & -4.91E-05 & -8.50E-04 & 0.071 & 0.087 & ---\\
		M & -3.667 & 5.80E-04 & 3.98E-04 & -0.568 & 0.025 & 0.035\\
		$\alpha$ & -4.284 & 2.10E-05 & -1.45E-03 & -0.793 & 0.013 & 0.014\\
\hline
	\end{tabular}
	\label{tab:dr14_abun_err}
\end{table}

%
%Dwarf sequence in [alpha/Fe]?
%
%Giant sequence differences from DR12?

\subsection{Derived parameters and abundances}

\subsubsection{Comparison of DR12, DR13, and DR14}

\begin{figure*}
\includegraphics[width=0.9 \textwidth]{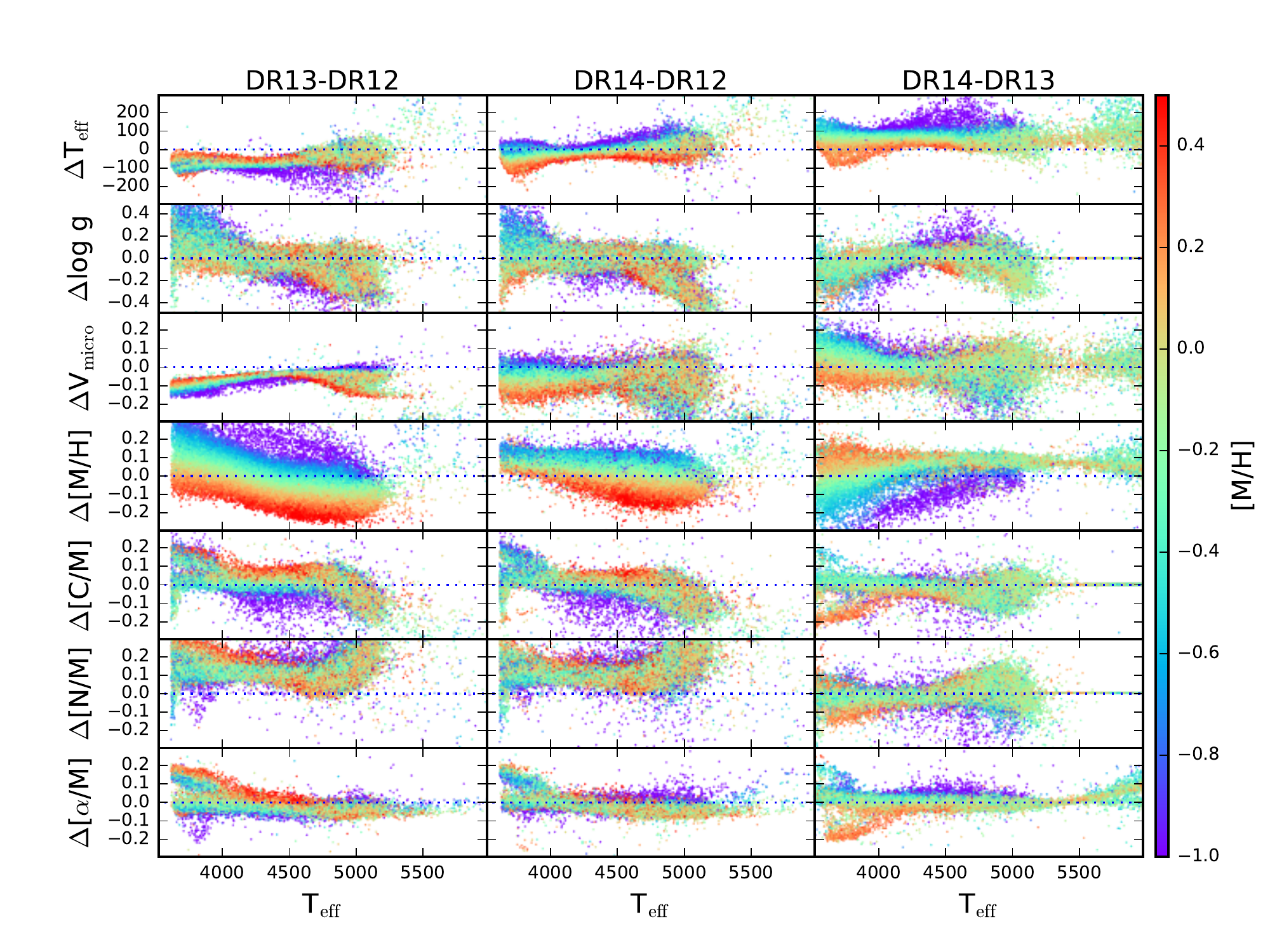}
\caption{Comparison of calibrated stellar parameters from DR12, DR13, and
DR14 for stars that overlap. Points are color-coded by \mh.}
\label{fig:drcomp_cal}
\end{figure*}

\begin{figure*}
\includegraphics[width=0.9 \textwidth]{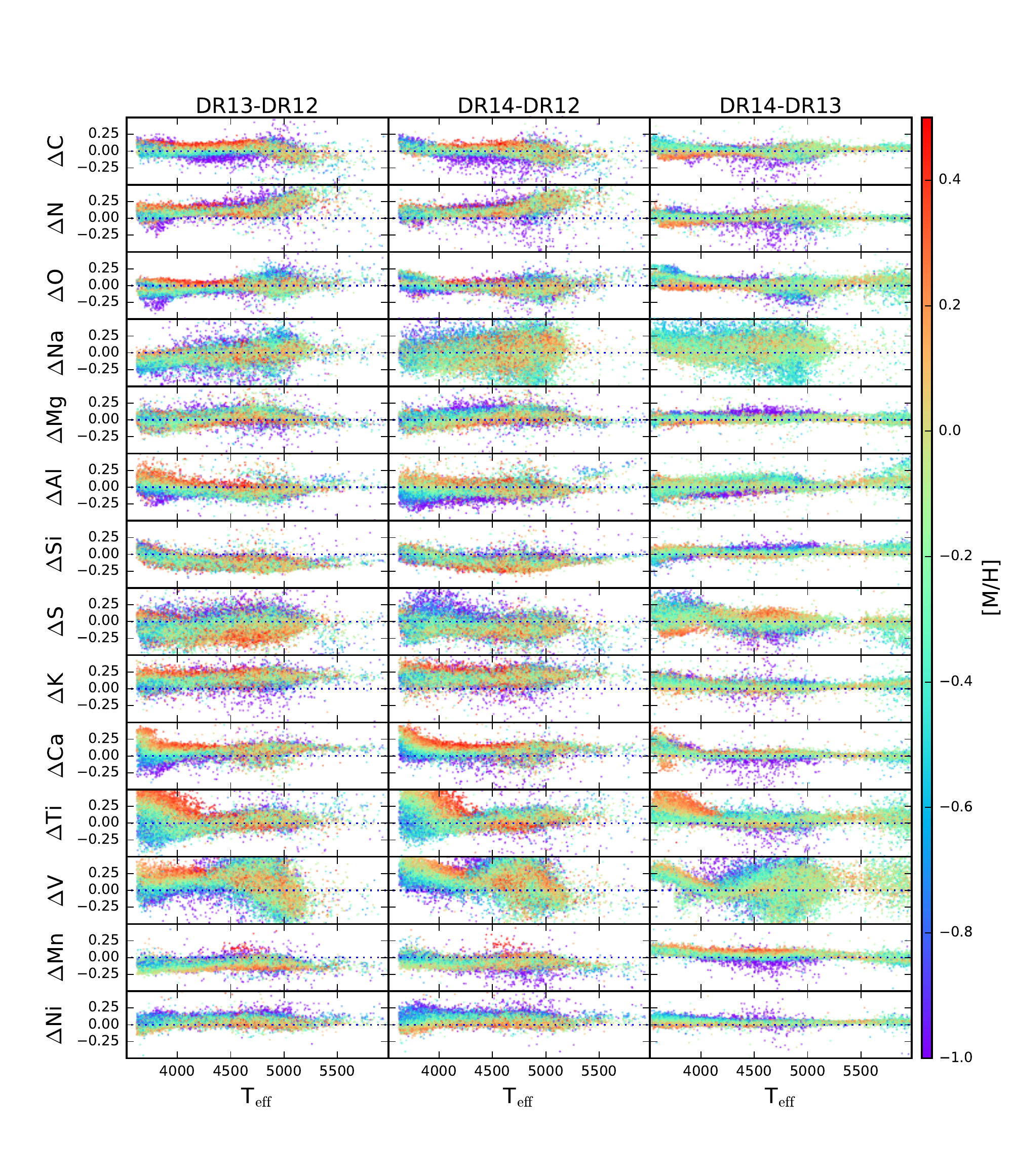}
\caption{Comparison of calibrated elemental abundances from DR12, DR13, and
DR14 for stars that overlap. Points are color-coded by \mh.}
\label{fig:drcomp_elem}
\end{figure*}

To illuminate the differences that result from modifications in the analysis,
Figure \ref{fig:drcomp_cal} compares the calibrated
stellar parameters between DR12, DR13, and DR14, for high
S/N ($S/N>150$) stars that overlap between the
releases. Figure \ref{fig:drcomp_elem} compares the calibrated elemental
abundances.

%\begin{figure*}
%\includegraphics[width=0.9 \textwidth]{drcomp_uncal.pdf}
%\caption{Comparison of uncalibrated stellar parameters from DR12, DR13, and
%DR14 for stars that overlap. Points are color-coded by \mh.}
%\label{fig:drcomp_uncal}
%\end{figure*}

As described in previous sections, differences include changes
in the line list (from DR12 to DR13/DR14), changes in the handling of
microturbulence, changes in continuum normalization, and changes
in calibration.
It is apparent that changes in the analysis and calibration make a
difference in the derived results; we hope and believe that the changes
are in the direction to make things more accurate. In general, there
is a larger difference between DR12 and DR13 than there is between
DR13 and DR14, which is not surprising given the larger changes in the
analysis between DR12 and DR13, e.g., different line list and synthesis.
Differences in \teff, \logg, microturbulent velocity, and \mh\ arise
from different calibration choices as much as from analysis differences.

\subsubsection{Comparison with independent measurements}

To provide independent assessment of the accuracy of the DR13 and DR14
parameters and abundances, we compiled a set of independent measurements
of stellar parameters and abundances derived from optical spectra and analysis
for a subset of APOGEE stars.  These are presented and discussed in
\citet{Jonsson2018}, a companion paper to this work.  Here we present
a brief summary, and we direct the reader to \citet{Jonsson2018} for a
complete description.

\citet{Jonsson2018} find the same trend of APOGEE/ ASPCAP \teff\
with metallicity as discussed in comparison to the photometric \teff\
above. As expected, this effect is much smaller for DR14 when a calibration
was applied to take care of this problem. However, the optical
spectroscopic effective temperatures suggest that even the calibrated DR14
\teff\  is still about 100 K too high for high-metallicity stars. The
other (calibrated) stellar parameters show no trends or systematic
offsets for giants.

For most of the abundances -- C, Na, Mg, Al, Si, S, Ca, Cr, Mn, Ni --
the DR14 ASPCAP analysis have systematic differences to the comparisons
samples of less than 0.05 dex (median), and random differences of less
than 0.15 dex (standard deviation). Magnesium is the most accurate
alpha-element, showing a very clear thin/thick disk separation, and
nickel is the most accurate iron-peak element.

Some abundances -- N, O, K, Ti I, V, Co -- have differences with the 
optical abundances that are correlated with stellar parameters. Given
the systematic trends of \teff\ with metallicity, some of these abundances
might be improved if the calibrated \teff\  were used instead of the
uncalibrated \teff, especially for elements where the derived
abundances are a strong function of the adopted \teff\, such as Ti I.

Some elemental abundances -- P, Cu, Ge, Rb, Nd, Yb -- are not evaluated
in \citet{Jonsson2018} due to either lack of comparison samples with
overlapping stars/element or due to the APOGEE-analysis of the elements
being unreliable in the present analysis.

Future data releases will consider these issues, which will hopefully
lead to better stellar parameters as well as abundances.

\section{Persistence}
\label{sect:persist}

As discussed in \S \ref{sect:persistcorr} and \ref{sect:weighting}, 
several modifications were made to the pipeline to reduce
the impact of persistence on the data. Persistence is mostly relevant for the data
taken before Fall 2014, when the ``blue" chip was replaced (although
there is evidence for some persistence in parts of the ``green" chip as well).

\begin{figure}
\includegraphics[width=0.5 \textwidth]{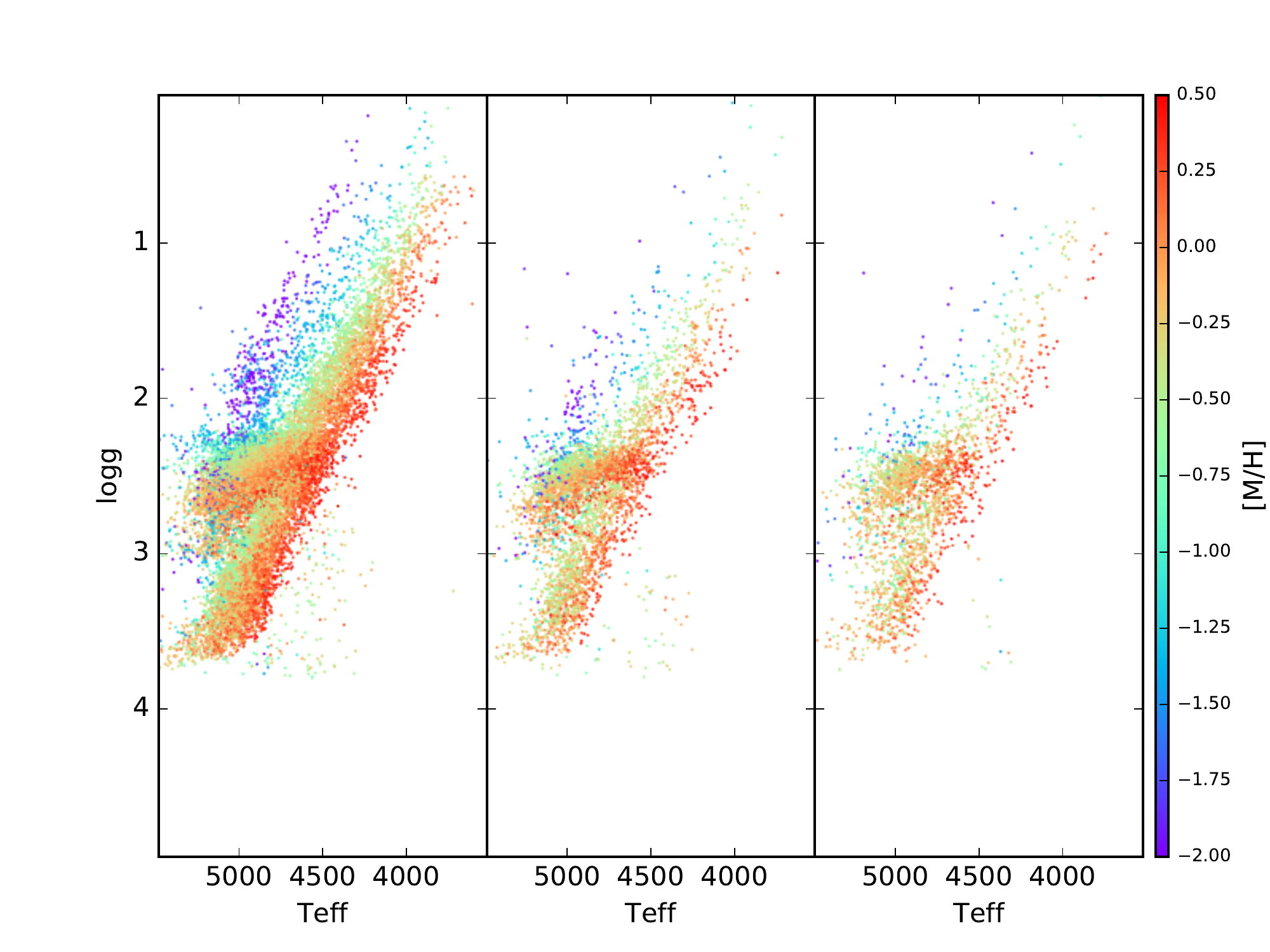}
\caption{HR diagrams for 3 subsamples split by persistence behavior:
left panel has stars that are not flagged as having significant number
of persistence-affected pixels in \textit{any} of the visits, middle panel has
stars that are flagged with significant persistence-affect pixels in 
\textit{some}
visits, and right panel has stars that are flagged with significant
persistence-affected pixels in \textit{all} visits. 
\referee{Points are color-coded by \mh.}}
\label{fig:persist_f_hr}
\end{figure}

\begin{figure}
\includegraphics[width=0.5 \textwidth]{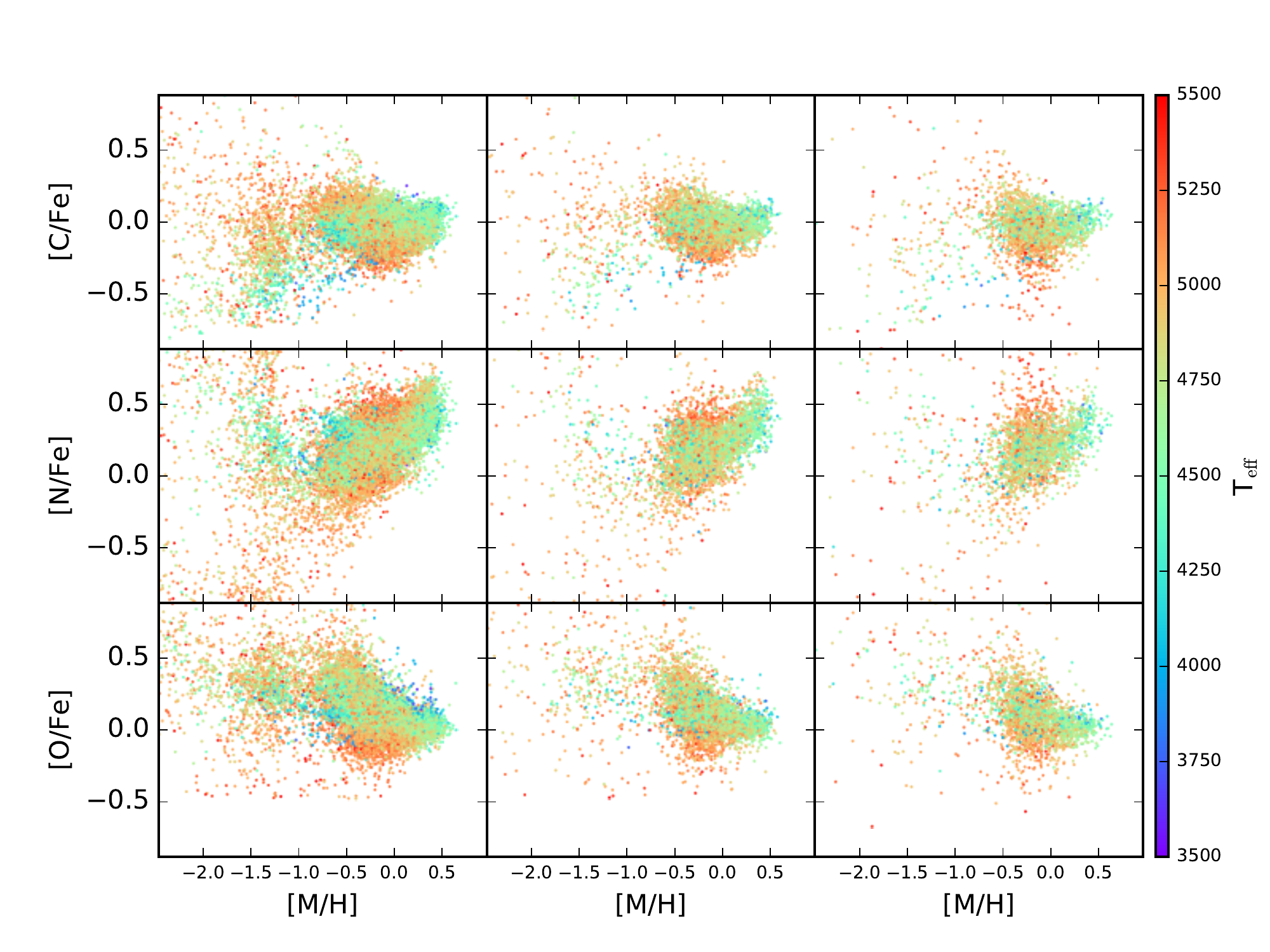}
\caption{\referee{CNO abundances relative to Fe for the same subsamples presented in Figure \ref{fig:persist_f_hr}.
Points are color-coded by \teff.}}
\label{fig:persist_f_cno}
\end{figure}

\begin{figure}
\includegraphics[width=0.5 \textwidth]{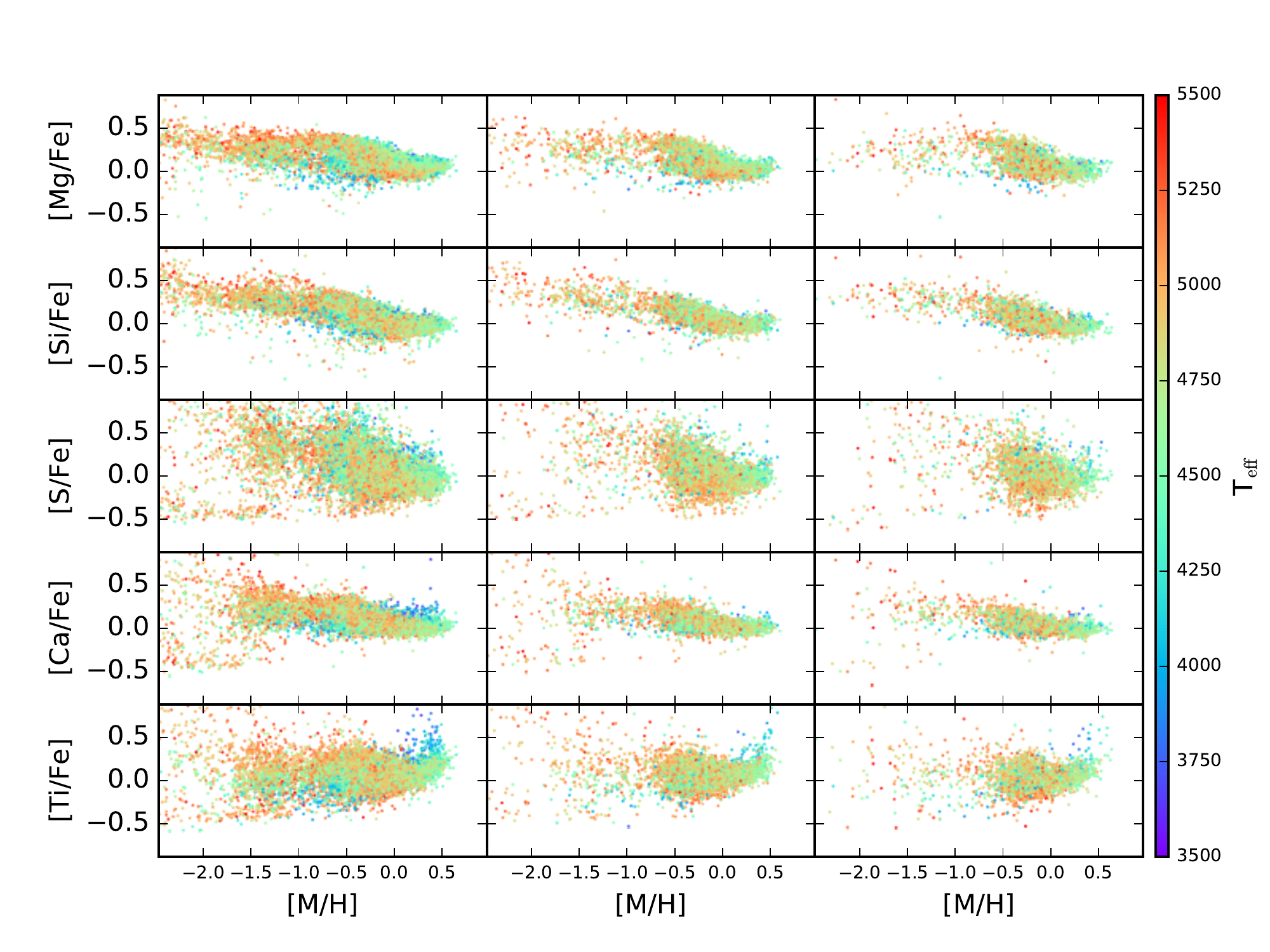}
\caption{Same as Figure \ref{fig:persist_f_hr}, but for $\alpha$-element
abundances relative to Fe.}
\label{fig:persist_f_alpha}
\end{figure}

\begin{figure}
\includegraphics[width=0.5 \textwidth]{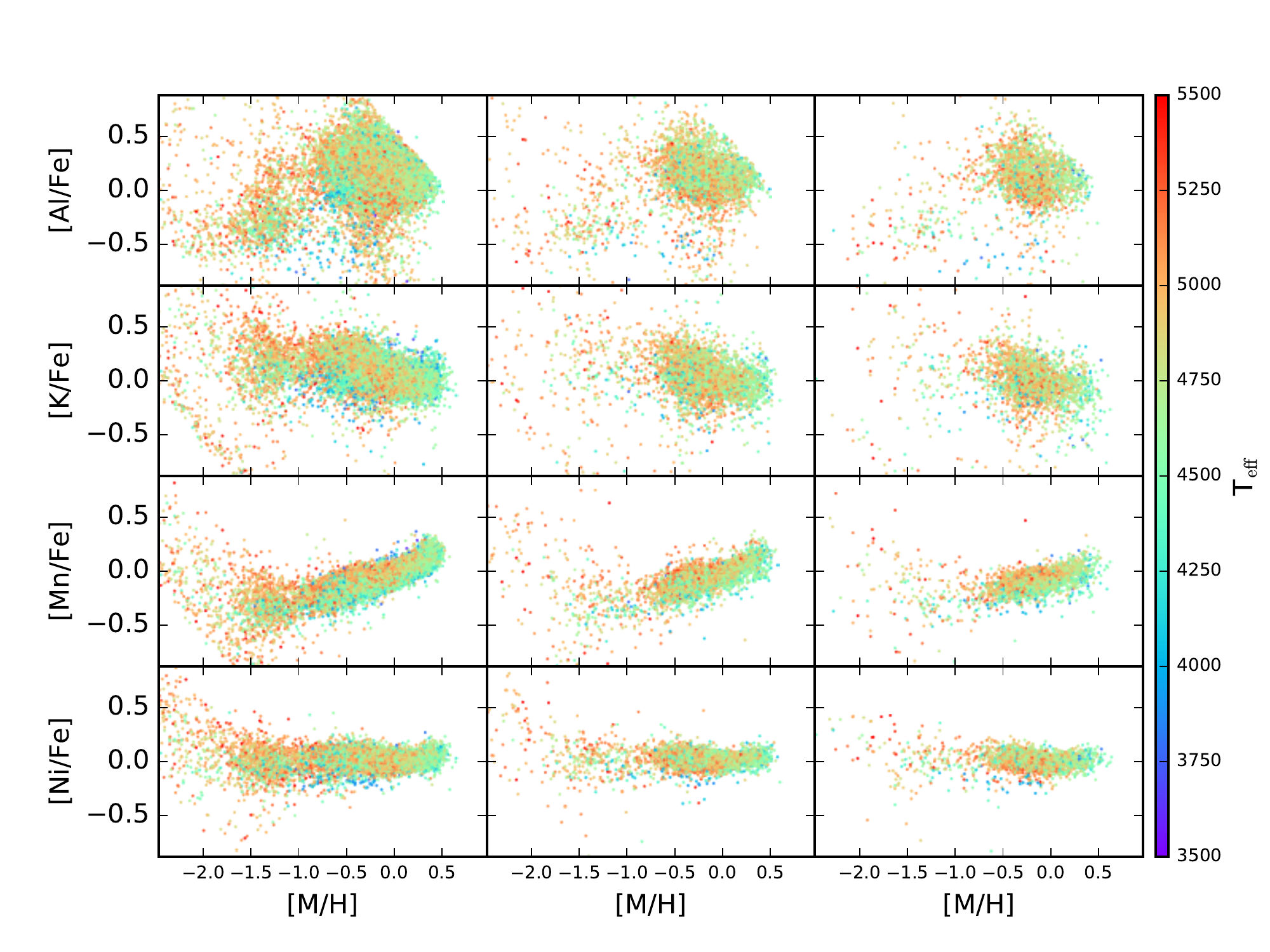}
\caption{Same as Figure \ref{fig:persist_f_hr}, but for other element
abundances relative to Fe.}
\label{fig:persist_f_fe}
\end{figure}

To assess whether these modifications had an impact on the resulting
stellar parameters and abundances, we repeat the same comparison of 
parameters and abundances for stars unaffected by persistence with stars
most affected by persistence that was performed in \citet{Holtzman2015}.
Persistence is most likely to affect faint stars, so we are probing
maximal effects by investigating a subsample with $H>12$.
Figures \ref{fig:persist_f_hr}, \ref{fig:persist_f_cno}, \ref{fig:persist_f_alpha}, \ref{fig:persist_f_fe} 
show locii of parameters and abundances for
three samples: stars with no persistence flags set (left panel), stars
with PERSIST\_HIGH flag set for some, but not all visits (middle panel), 
and stars with PERSIST\_HIGH flag set in all visits. As described above,
one of the significant changes was to reduce the weight of pixels
affected by persistence in the visit combination. Given this procedure, we expect
significant improvement for stars in which only some of the visits had
persistence, as these will contribute little to the combined spectra; even
for stars with persistence in all visits, we expect improvements because
the reduced weights mean that persistence-affected pixels will carry
less weight in the ASPCAP fits than other pixels.

The results presented in Figure \ref{fig:persist_f_hr}--\ref{fig:persist_f_fe} suggest 
that there is little difference in the derived 
parameters and abundances between the different
subsets, suggesting that we have significantly reduced the effect of
persistence on derived parameters and abundances. Interested readers can
compare these plots with Figures 15-18 in \citet{Holtzman2015}.

\citet{Jahandar2017} have recently presented data for a few stars that
clearly demonstrate very poor results in APOGEE DR12 that result from
persistence; they claim that some of the poor results persist in DR13.
In general, however, we find no evidence that persistence strongly affects
results of a significant fraction of the APOGEE sample in DR14.
The low metallicity (\xh{Fe} = -0.6), warm temperatures (\teff = 4800 K),
and faintness (H $>$ 14) of the targets in \citet{Jahandar2017} lead
to multiple issues, including problems in the RV determination
and ASPCAP pipeline.  For a boutique
analysis, correction of the persistence and careful RV combination 
(as was done by \citet{Jahandar2017})
will make a significant difference.

While we believe that significant improvements in persistence handling
have been made in DR14, users are reminded that the presence of a significant number
of persistence-affected pixels in the spectrum of an object is flagged
in the stellar catalog files/tables with the STARFLAG bitmask (significant fraction pixels affected in any of the
component visit spectra) and  the ANDFLAG bitmask (significant fraction in
\textit{all} of the component visit spectra). In the spectra themselves,
pixels known to be affected by persistence are flagged in the PIXMASK.

\section{Results from The Cannon}
\label{sect:cannon}

DR14 includes, for the first time, an alternate set of stellar parameters
and abundances as derived from a data driven method called The Cannon
\citep{Ness2015,Casey2016}. This technique parameterizes the
spectral fluxes as a function of a set of externally-determined stellar
parameters and abundances; in principle these could be any physical
quantities, so are generically referred to as labels. The method uses
a training set of stellar spectra to determine the coefficients of
the parameterization that best match the training set spectra, and these are then
applied to a broader data set to derive labels for a larger data set. The
method has the power of exploiting all of the information that may be
present in the stellar spectra.
In general, it has been claimed that the method
produces higher precision than the ASPCAP results (\eg, \citealt{Ness2017}). This is plausible
because the method can respond to individual features in the spectra that we may not model
well with the ASPCAP analysis, \eg, lines with imperfect atomic data,
lines missing from the line list, lines that are not well modeled with
the 1D LTE approach used by ASPCAP, etc.  For more details, refer to the 
papers listed above.

For DR14, Cannon results have been determined using the Cannon-2 
code \citep{Casey2016}, except that we used a different prescription
for the uncertainties in the input spectra. Specifically, we adopted the
same uncertainties used in the ASPCAP pipeline. In particular, these
uncertainties use better knowledge of the sky spectra to mask broader
regions around sky lines that are often imperfectly subtracted.

We initially ran The Cannon after training a model on the ASPCAP stellar
parameters and abundances and found that it appeared to give higher
precision, based on the tightness of locii in, \eg, plots of \xm{X} as
a function of \mh. However, we subsequently used the model to generate spectra, varying
individual elemental abundances one at a time, and found that the resulting
spectra showed variations where no identified lines of the element in
question were found and, in fact, in some cases, where identified lines
of other elements were found. 
%\textcolor{red}{(example?)}. 
As a result,
it appeared that The Cannon was training on features of multiple
elements that may be well correlated in the bulk of the training set.
This is in some sense not surprising, and in some situations might
actually be desired: if we were interested in a single label that has
a complicated — and not fully described — relationship with other
labels (e.g., age), then we might \emph{want} The Cannon to be building
on all information available. If the training set is representative of
the test sets, then correlated information can be useful.

This behavior, however, may jeopardize
the ability of the model to derive abundances for stars that might
have slightly different abundance patterns; in fact, the ability to
distinguish these is a key goal of the APOGEE project. To prevent this, for
the labels that are associated with individual elemental abundances, we
use ``censoring" in The Cannon parameterization, which means that we
only allow pixels that we expect, based on our line list, to be affected
by the abundance of the element. This minimizes the potential issue of
having correlations of abundances of different elements in the training
set imposing such correlations on the full data set. In practice, we implemented
this by only allowing The Cannon to use pixels that have non-zero weight
in the windows used by FERRE for the ASPCAP abundances, although we note that
these windows may not perfectly include \textit{all} wavelengths affected by the 
abundance of any given element. 

\begin{figure}
\includegraphics[width=0.5 \textwidth]{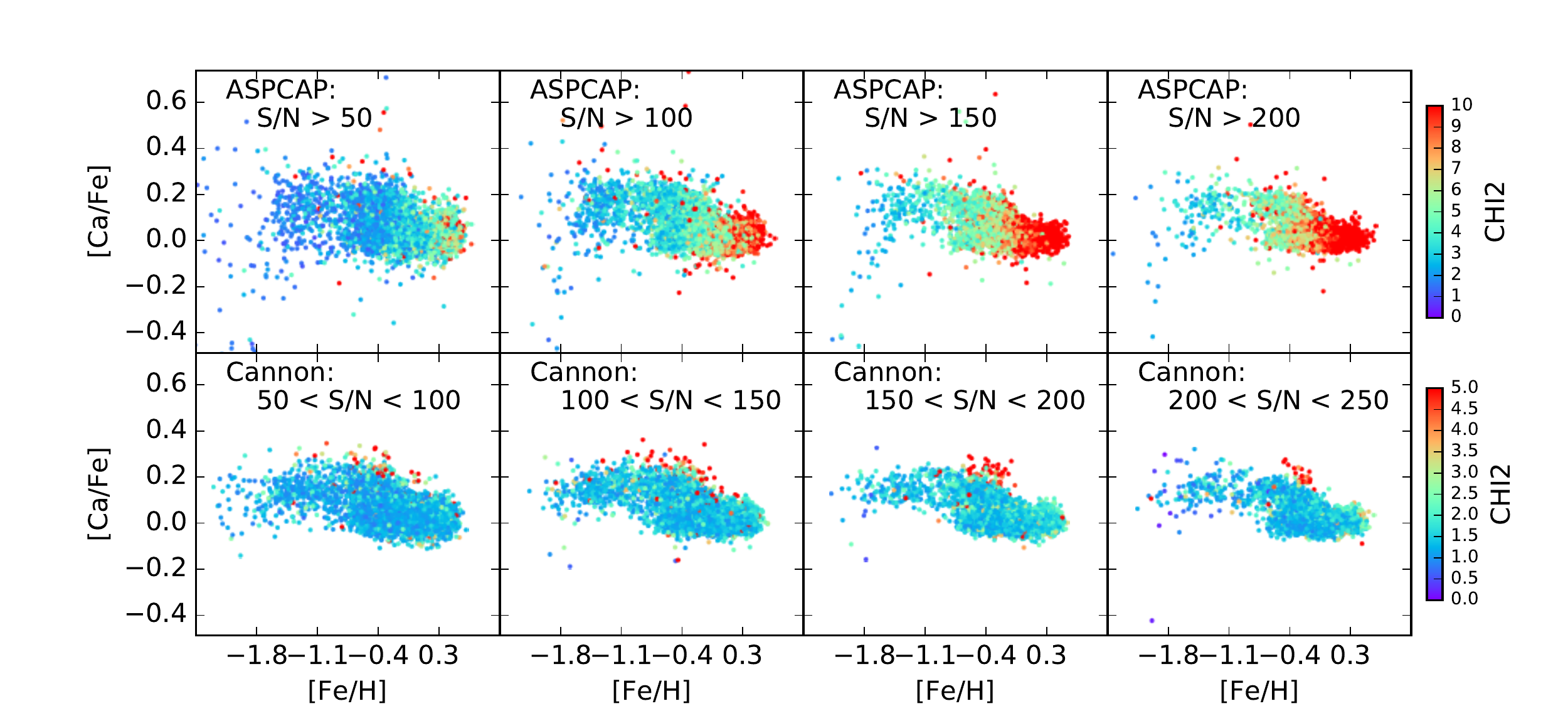}
\caption{Uncensored Cannon results for Ca compared with ASPCAP results
for stars with 4000 $<$ \teff $<$ 4500 at a range of S/N. Points
are color-coded by the $\chi^2$ of the fit.}
\label{fig:cannon_Ca}
\end{figure}

\begin{figure}
\includegraphics[width=0.5 \textwidth]{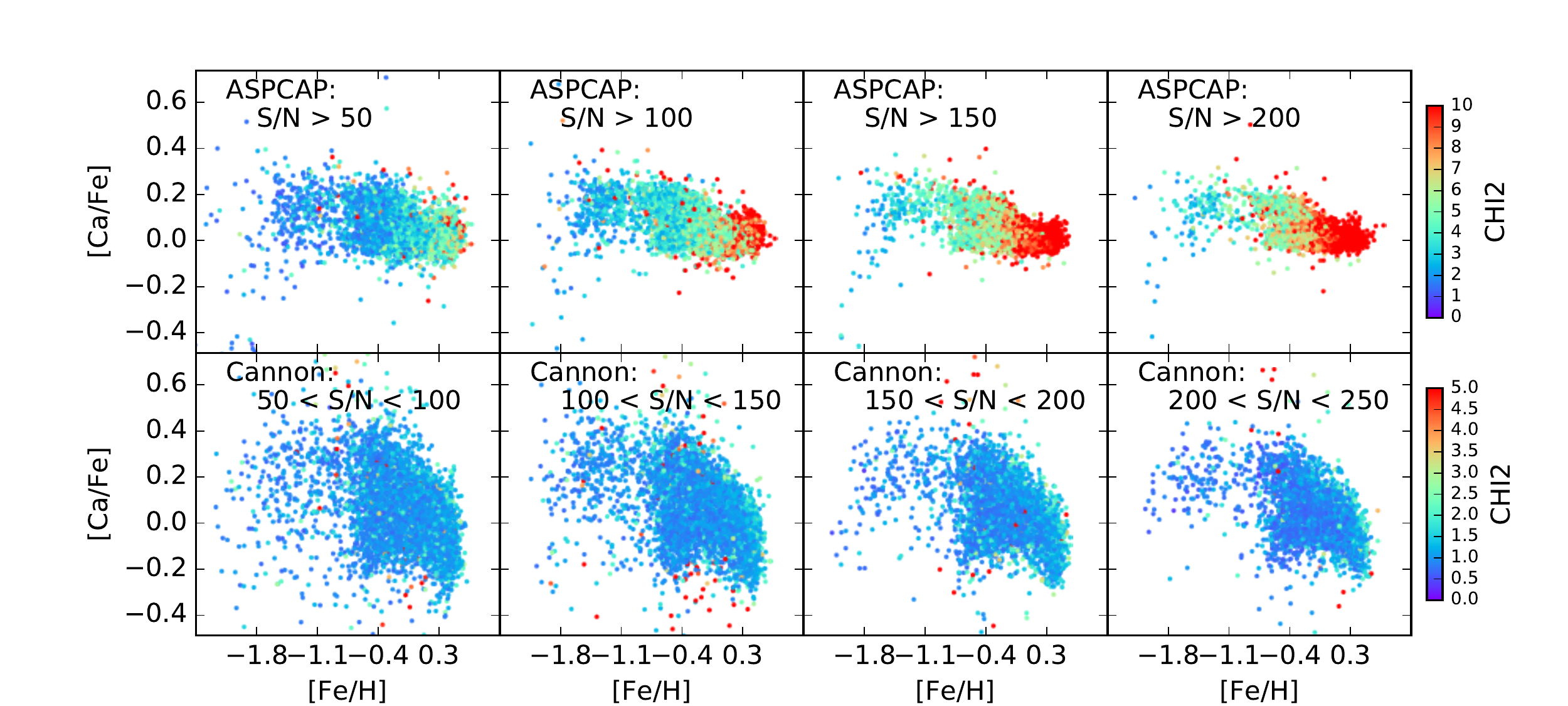}
\caption{Same as Figure \ref{fig:cannon_Ca} but for censored Cannon results.}
\label{fig:cannon_Ca_censor}
\end{figure}

\begin{figure}
\includegraphics[width=0.5 \textwidth]{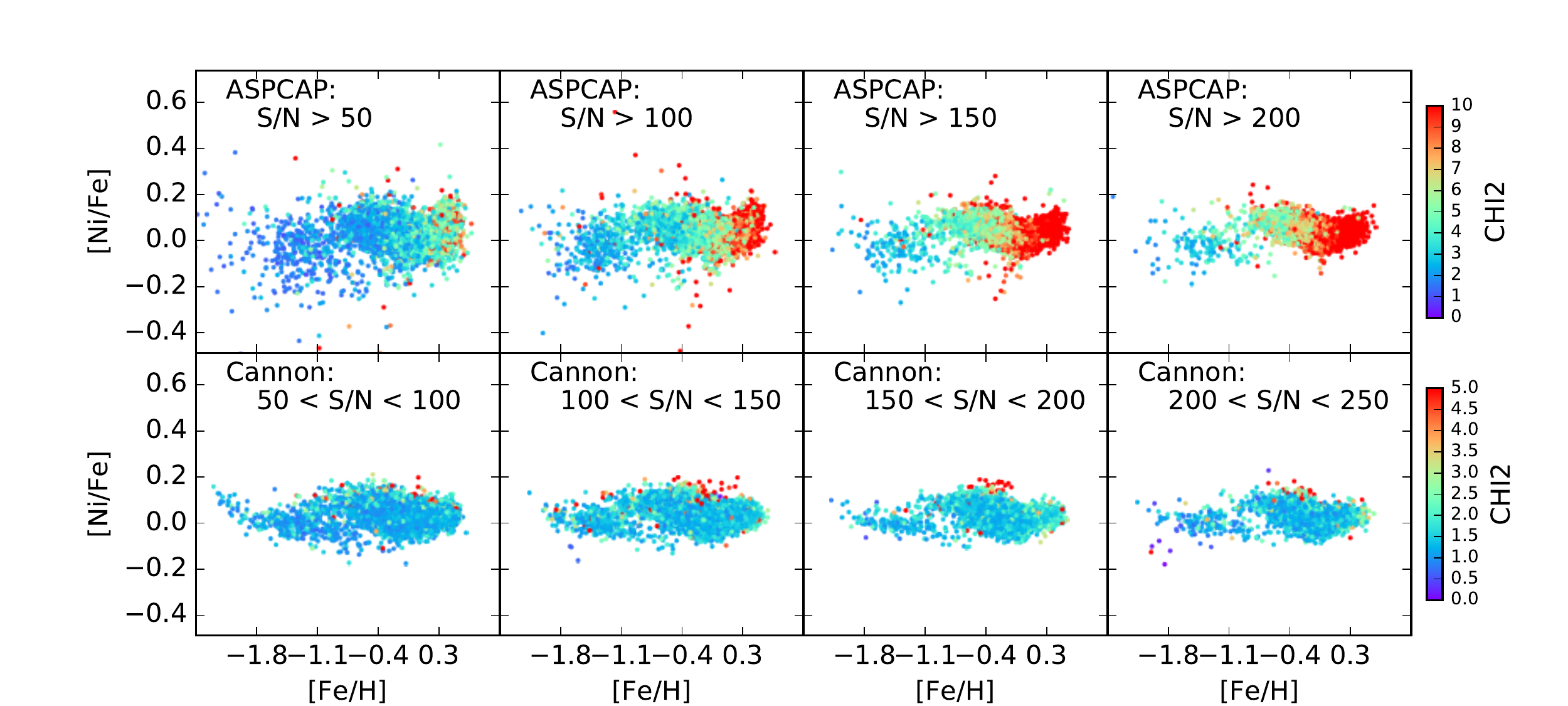}
\caption{Uncensored Cannon results for Ni compared with ASPCAP results,
for stars with 4000 $<$ \teff $<$ 4500 at a range of S/N. Points are
color-coded by the $\chi^2$ of the fit.}
\label{fig:cannon_Ni}
\end{figure}

\begin{figure}
\includegraphics[width=0.5 \textwidth]{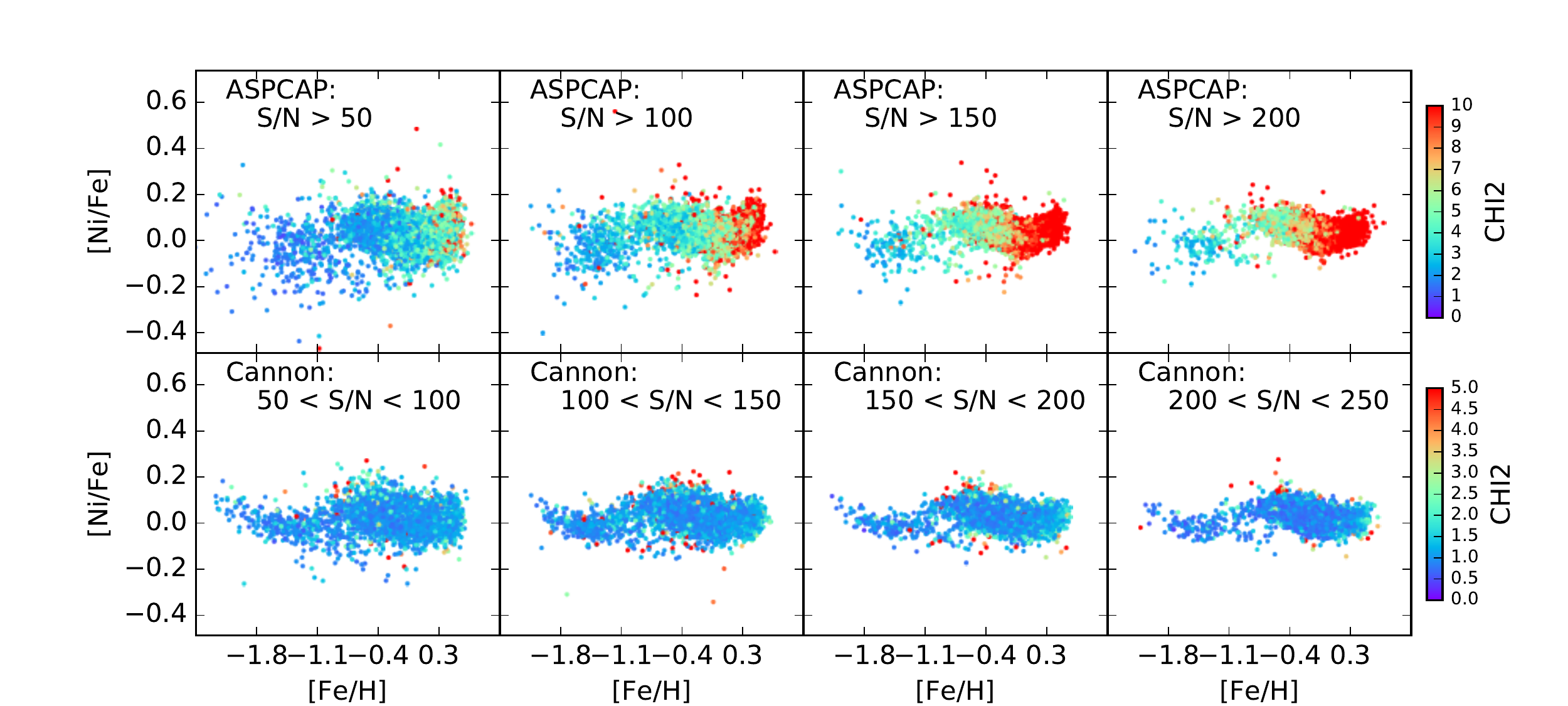}
\caption{Same as Figure \ref{fig:cannon_Ni} but for censored Cannon results}
\label{fig:cannon_Ni_censor}
\end{figure}

Using this wavelength sensoring changes The Cannon results significantly.
An extreme example of the difference between censored and uncensored results can
be seen by comparing Figure \ref{fig:cannon_Ca}, which shows 
Cannon (bottom) and ASPCAP (top) results for \xfe{Ca}, with Figure 
\ref{fig:cannon_Ca_censor}, which shows the same thing for the
censored results. For calcium, while the uncensored Cannon results look tighter
than the ASPCAP results, the censored results look significantly worse.
This is not true in general: for some elements, even the censored
Cannon results look tighter than the ASPCAP results. For example,
Figure \ref{fig:cannon_Ni} and \ref{fig:cannon_Ni_censor} show the same
comparison for \xfe{Ni}, for which even the censored Cannon results look
better.  The reasons for both of these extremes (e.g., significantly worse 
\xfe{Ca} than ASPCAP and significantly better \xfe{Ni} than ASPCAP) are still 
being fully investigated, and these efforts are expected to improve both approaches.

In practice, we apply the following steps to derive Cannon labels:

\begin{enumerate}
\item The combined apStar spectra are normalized, following the
prescription in \citet{Casey2016}.

\item A training set is constructed that attempts to sample a wide range
of stellar parameters: we split the \teff-\logg-\mh\  space into cubes
covering the range 3500$ < $\teff$ < $5500, 0$ < $\logg$ < $3.9, and -2.5$ < $\mh$ < $0.5,
and take the 50 stars with the highest S/N in each cube: this results
in a training set of 1464 stars. For the labels of this subsample, we
adopt the calibrated ASPCAP \teff, \logg, \mh, \am, and \xh{X}
for 20 different elements. Note that we restrict the training set to
giants, both because of increasing uncertainties in  (or lack of) the
ASPCAP calibration relations, and also because with a broader range of
stellar parameter, the quadratic parameterization used by the Cannon-2
is likely to be less accurate.

\item We train The Cannon on this sample, using wavelength censoring
for the labels that refer to individual element abundances. We adopt the
ASPCAP windows for the individual elemental abundances as the wavelength
censors.

\item We apply the derived model to the remainder of the ASPCAP data
set for objects whose ASPCAP parameters fall within the range of the
parameters adopted for the training set.

\end{enumerate}

\subsubsection{Using Cannon labels}

All of the issues/caveats associated with ASPCAP apply to The Cannon results, 
as The Cannon abundances depend on the abundances of the training set
which are derived using ASPCAP. The Cannon provides a $\chi^{2}$ value 
which indicates how well The Cannon model spectrum fits the observed spectrum.
Results with higher $\chi^{2}$ are significantly more uncertain.

%limit to range of training set

%pay attention to chisq!

%all the issues/caveats associate with ASPCAP apply to The Cannon results, since the latter depend on the former.

\section{Datamodel revisions}
\label{sect:data}

As in previous data releases, APOGEE data are released through files
in the science archive server (SAS), including summary data files
(allVisit and allStar) that include the main APOGEE derived quantities
(radial velocities, stellar parameters, and elemental abundances),
and all of the pipeline data products, include the final combined
spectra (apStar) and pseudo-continuum normalized spectra (aspcapStar).
The derived quantities are also available in the Catalog Archive Server
(CAS), an online database.

Several modifications have been made to the derived quantities that are
presented in the SAS summary files and the CAS tables.

\begin{itemize}
\item Since APOGEE-2 data use a different set of targeting flag bits than
APOGEE data (see \citealt{Zasowski2017} for details), DR14 includes the
APOGEE target flags (APOGEE\_TARGET1 and APOGEE\_TARGET2) and the APOGEE-2
target flags (APOGEE2\_TARGET1, APOGEE2\_TARGET2, and APOGEE2\_TARGET3); the
APOGEE values are zero (no bits set) for APOGEE-2 targets, and vice versa

\item Since some stars may have ASPCAP fits from multiple synthetic
grids (those near the grid boundaries), the allStar file and CAS table
include an ASPCAP\_CLASS entry that gives the grid that provided the best
fit. The class names include the temperature code (F, GK, or M), a code
for giant (g) or dwarf (d) grid, and a code for the LSF used (abcd).
In the allStar file, there are arrays, FPARAM\_CLASS and CHI2\_CLASS,
that give the raw parameters and $\chi^2$ values for each fit that was
performed on a given star.

\item In the DR12 allStar file, abundances were provided in arrays
labeled FELEM (uncalibrated) and ELEM (calibrated), as well as in
labels with individual elemental abundance names.  However, the array
values were a bit complicated to interpret, since some of the abundances
were given as \xh{X}, while others were given as \xm{X}, depending on
the grid dimension used for the abundance fit.  In DR13 and DR14, the
FELEM array has been preserved, but the ELEM array has been removed:
calibrated abundances are presented in both X\_M and X\_H arrays (with
abundances relative to M and H, respectively),
as well as in labels with individual element abundance names, relative
to iron (e.g., C\_FE, MG\_FE, NI\_FE).

\end{itemize}

As before, raw parameters are loaded into the FPARAM array, with 
calibrated parameters loaded in a PARAM array and also into named
tags TEFF, LOGG, VICRO, M\_H, ALPHA\_M, C\_M, N\_M, and VSINI (C\_M and N\_M for giants
only, VSINI for dwarfs only). If stars are outside the range for which
calibration objects are available, the calibrated and named quantities
are set to a value of -9999. One important implication of this is that dwarfs do
not have the LOGG tag populated; the raw ASPCAP LOGG is available in
the second element of the FPARAM array.

\subsubsection{Cannon data products}

In the SAS, DR14 includes a summary allStarCannon file, as well as
cannonField and cannonStar files, as described in the following
paragraphs.

The allStarCannon file bundles up all of The Cannon results in a single
file, analogous to the allStar file with ASPCAP results. The allStarCannon
file has been constructed to be a line-for-line match with the allStar
file, to make it simple to use either ASPCAP or Cannon results, or to
compare them. However, the allStarCannon file does not repeat all of the
information contained in the allStar file; it simply supplements it with
Cannon label results. Note that the allStarCannon files do not have Cannon
abundances for many stars, since Cannon results are only provided for stars
that fall within The Cannon training parameter space; we carry along empty
Cannon values for the other stars to preserve the simple line-to-line matching
between the allStar and allStarCannon files. 

The cannonField files bundle up all of the results for stars in a given
field and are analogous to the aspcapField files. These files contain
the spectra and derived best fits, as well as the derived label values,
in a FITS table format file (see the cannonField data model).

The cannonStar files contain the results for individual stars, including
labels, normalized spectra, uncertainties, and best fit spectra, in a
FITS image format (see the cannonStar data model).

For convenience of users who might want to delve
deeper into and experiment with The Cannon results, the
subset of stars used to train The Cannon are saved in a file
(apogee-dr14-giants-xh-censor-training-set.fits) and the model itself is
also saved (as a Python pickle file, apogee-dr14-giants-xh-censor.model).

In the CAS, The Cannon results can be found in the cannonStar table.

\section{Conclusions}
\label{sect:conclusion}

We have described the methodology used for the SDSS/APOGEE
Data Releases 13 and 14, concentrating on the areas in which they differ
from that of DR12 \citep{Holtzman2015}. Improvements
have been made in the data reduction in the areas of telluric correction,
persistence, and radial velocity determination. Methods for determining 
the stellar parameter and abundance determinations were refined.

We describe the calibrations applied to the stellar parameters and abundances
in SDSS/APOGEE DR13 and DR14. We also describe some of the shortcomings
of these calibrations and suggest alternate calibrations for \teff\ 
in DR13, and for \logg\  in both DR13 and DR14. 

Analysis of stars within open clusters suggests
that the precision of the abundances is typically 0.05 dex.  A companion
paper \citep{Jonsson2018} presents a comparison of the calibrated
parameters and abundances with independent optical measurements for an
overlapping sample, and finds that the systematic differences for most
elements in DR14, when compared to the references, are of the order of
0.05 dex. 

We demonstrate that the modifications made in the pipeline to reduce the
effect of persistence seem to be generally effective.

We have also described and presented results using analysis by
the Cannon \citep{Ness2015}.  We find that if we allow The Cannon to train
without any restriction on what part of the spectrum it uses for elemental
abundances, it can use regions where there are not any known features of
the element in question and in many cases, seems to be using features of
other elements. As a result, we run The Cannon in ``censored" mode where
we only allow it to use regions of the spectrum for each element where
lines of that element are known. In this mode, The Cannon results can have
considerably larger scatter than results from uncensored mode, and in some
cases, larger scatter than the ASPCAP results.

Future data releases will likely include further improvements. In particular,
we are working towards stellar parameter and abundance determination with 
a homogeneous grid of model atmospheres across the full temperature range.
We are improving the line list to include hydrides that are important
for cool dwarfs and to include lines of several s-process elements  
(\citealt{Hasselquist2016}, \citealt{Cunha2017}).
We are also working to provide better abundance determinations for elements
whose lines are blended with other elements in the same element group.
\\
\\

Funding for the Sloan Digital Sky Survey IV has been provided by the
Alfred P. Sloan Foundation, the U.S. Department of Energy Office of
Science, and the Participating Institutions. SDSS acknowledges support
and resources from the Center for High-Performance Computing at the
University of Utah. The SDSS web site is www.sdss.org.

SDSS is managed by the Astrophysical Research Consortium for the
Participating Institutions of the SDSS Collaboration including the
Brazilian Participation Group, the Carnegie Institution for Science,
Carnegie Mellon University, the Chilean Participation Group, the French
Participation Group, Harvard-Smithsonian Center for Astrophysics,
Instituto de Astrofísica de Canarias, The Johns Hopkins University,
Kavli Institute for the Physics and Mathematics of the Universe
(IPMU) / University of Tokyo, Lawrence Berkeley National Laboratory,
Leibniz Institut für Astrophysik Potsdam (AIP), Max-Planck-Institut
für Astronomie (MPIA Heidelberg), Max-Planck-Institut für Astrophysik
(MPA Garching), Max-Planck-Institut für Extraterrestrische Physik (MPE),
National Astronomical Observatories of China, New Mexico State University,
New York University, University of Notre Dame, Observatório Nacional /
MCTI, The Ohio State University, Pennsylvania State University, Shanghai
Astronomical Observatory, United Kingdom Participation Group, Universidad
Nacional Autónoma de México, University of Arizona, University of
Colorado Boulder, University of Oxford, University of Portsmouth,
University of Utah, University of Virginia, University of Washington,
University of Wisconsin, Vanderbilt University, and Yale University.

JAH, SRM, and VVS acknowledge support for this research from the National 
Science Foundation (AST-1109178). SRM also acknowledges NSF grant
AST-1616636). JAJ and MP acknowledge support from NSF grant  AST-1211673.
SzM has been supported by the Premium Postdoctoral
Research Program of the Hungarian Academy of Sciences, and by the Hungarian
NKFI Grants K-119517 of the Hungarian National Research, Development and Innovation Office.
HJ acknowledges support from the Birgit and Hellmuth Hertz’ Foundation (via the Royal Physiographic Society of Lund), the Crafoord Foundation, and Stiftelsen Olle Engkvist Byggm\"astare.

%CAP is thankful for support from the Spanish Ministry of Economy and
%Competitiveness (MINECO) through grant AYA2014-56359-P.
%
%DAGH and OZ acknowledge support provided by the Spanish Ministry of Economy and Competitiveness under grant AYA-2011-27754.
%
%TCB acknowledges partial support for this work from grants PHY 08-22648;
%Physics Frontier Center/Joint Institute or Nuclear Astrophysics (JINA),
%and PHY 14-30152; Physics Frontier Center/JINA Center for the Evolution
%of the Elements (JINA-CEE), awarded by the US National Science Foundation.
%
%Szabolcs Meszaros has been supported by the J{\'a}nos Bolyai Research
%Scholarship of the Hungarian Academy of Sciences.

\appendix

\section{DR13 calibrations}

Here we present proposed revisions to the calibrations used for DR13.

\subsection{Effective temperature}
\label{appendix:DR13_teff}

\begin{figure}[t]
\begin{center}
\includegraphics[width=0.45 \textwidth]{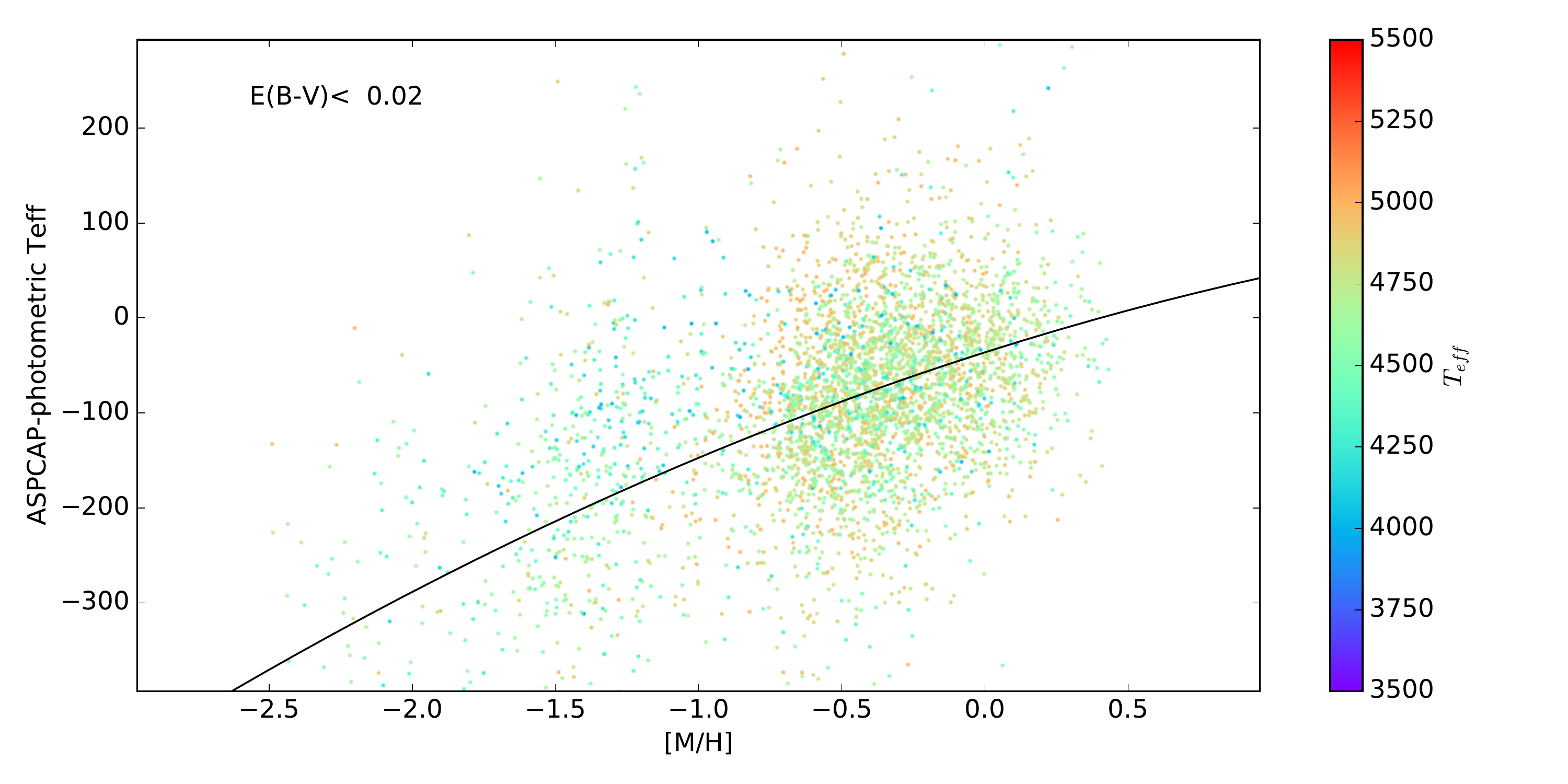}
\end{center}
\caption{Difference between APOGEE DR13 \teff\ and photometric \teff, 
demonstrating the problem with the DR13 \teff, for which no calibration
was applied. Points show mean differences in bins of metallicity, and the 
line gives the derived fit that is recommended to apply to DR13 values.
\referee{Points are color-coded by \teff.}}
\label{fig:dr13_teffcomp}
\end{figure}

No calibration was applied to the spectroscopic \teff\ in DR13. However,
Figure \ref{fig:dr13_teffcomp}
shows the difference between the spectroscopic temperature
and photometric temperatures derived from the
\citet{GHB2009} photometric calibration, as applied to a low-reddening sample
that includes all APOGEE stars with $b>30$ and $E(B-V)<0.02$,
where $E(B-V)$ is the \citet{SFD} reddening in the direction
of the star. It is apparent that there is a large trend with
metallicity, \mh, with a substantial offset for metal-poor
stars. This trend also exists in comparisons with other
spectroscopic samples (e.g., \citealt{Jonsson2018}).

%Similar trends are seen with other samples for which photometric
%temperatures can be derived, and also with other
%spectroscopic samples for stars in common (\eg,
%\citealt{GAIAESO}). \textcolor{red}{Need to clarify this
%last sentence?}

%\textcolor{red}{Katia wonders whether we should even provide
%a post-calibration for DR13.... given the possibility of metallicity-dependence
%in the photometric \teff results. But is there independent
%suggestion of a problem? I think so ... e.g., clusters}

A fit to the low reddening sample yields the relation:
\begin{equation}
\label{eqn:dr13_teff}
T_{eff}(ASPCAP) - T_{eff}(GHB) = - 36.17 + 95.97 [M/H] - 15.09 [M/H]^2
\end{equation}
where \mh\  is the calibrated ASPCAP metallicity, as shown in Figure
\ref{fig:dr13_teffcomp}.
We suggest that if DR13 users are interested in
effective temperatures, equation \ref{eqn:dr13_teff}
should be adopted as a post-calibration to the DR13 release values.

For DR13, all stars were assigned a fixed \teff\ uncertainty, which was
determined from the scatter in the relation between \teff\ and photometric
\teff, reduced by a factor of $\sqrt{2}$ to account for uncertainties in the
photometric effective temperatures: this corresponds to $\sim$ 70 K.
This scatter is dominated by stars near solar metallicity, so is not
strongly affected by the possible systematics at low metallicity discussed
above.

\subsection{Surface gravity}
\label{appendix:DR13_logg}

\begin{figure}[b]
\includegraphics[width=0.5 \textwidth]{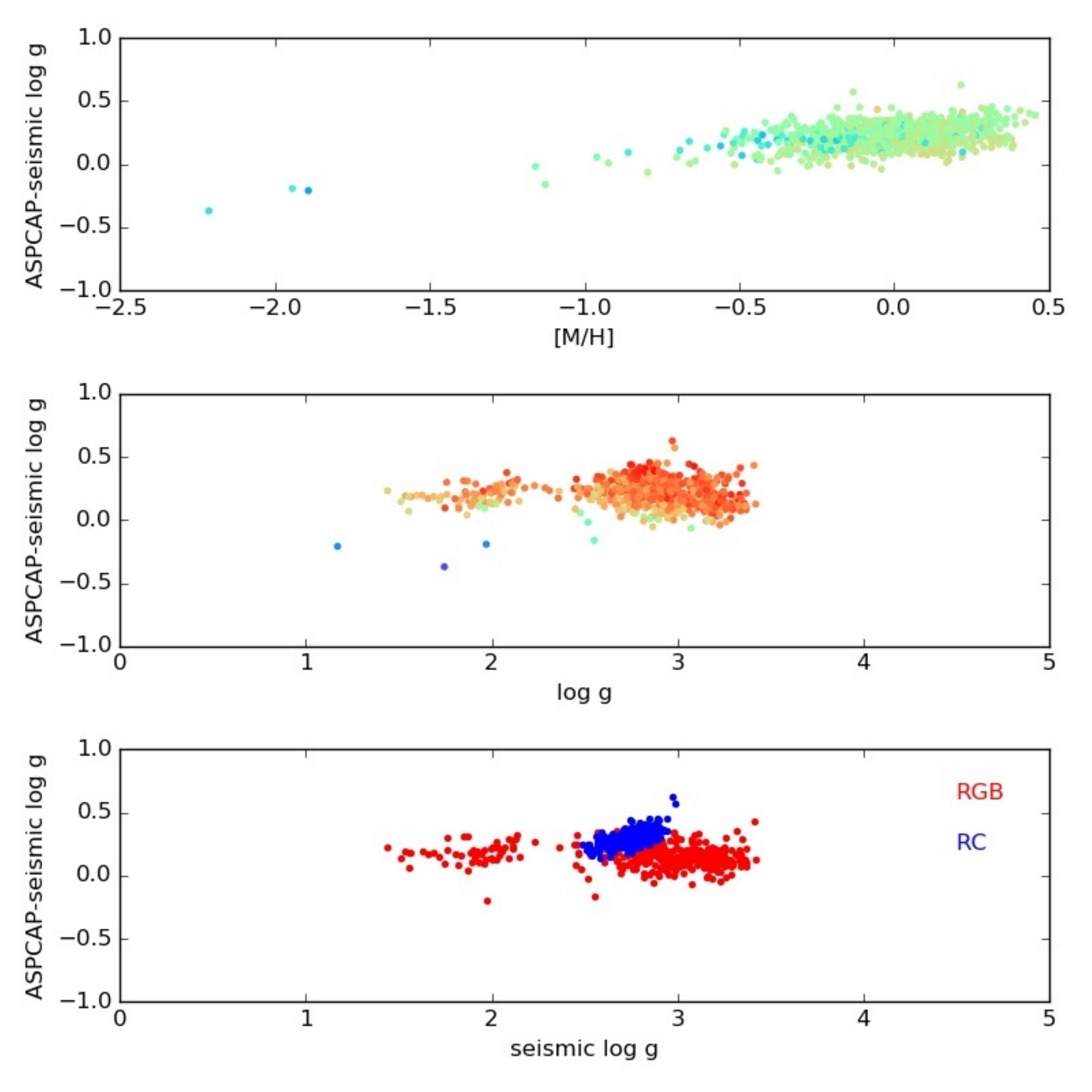}
\includegraphics[width=0.5 \textwidth]{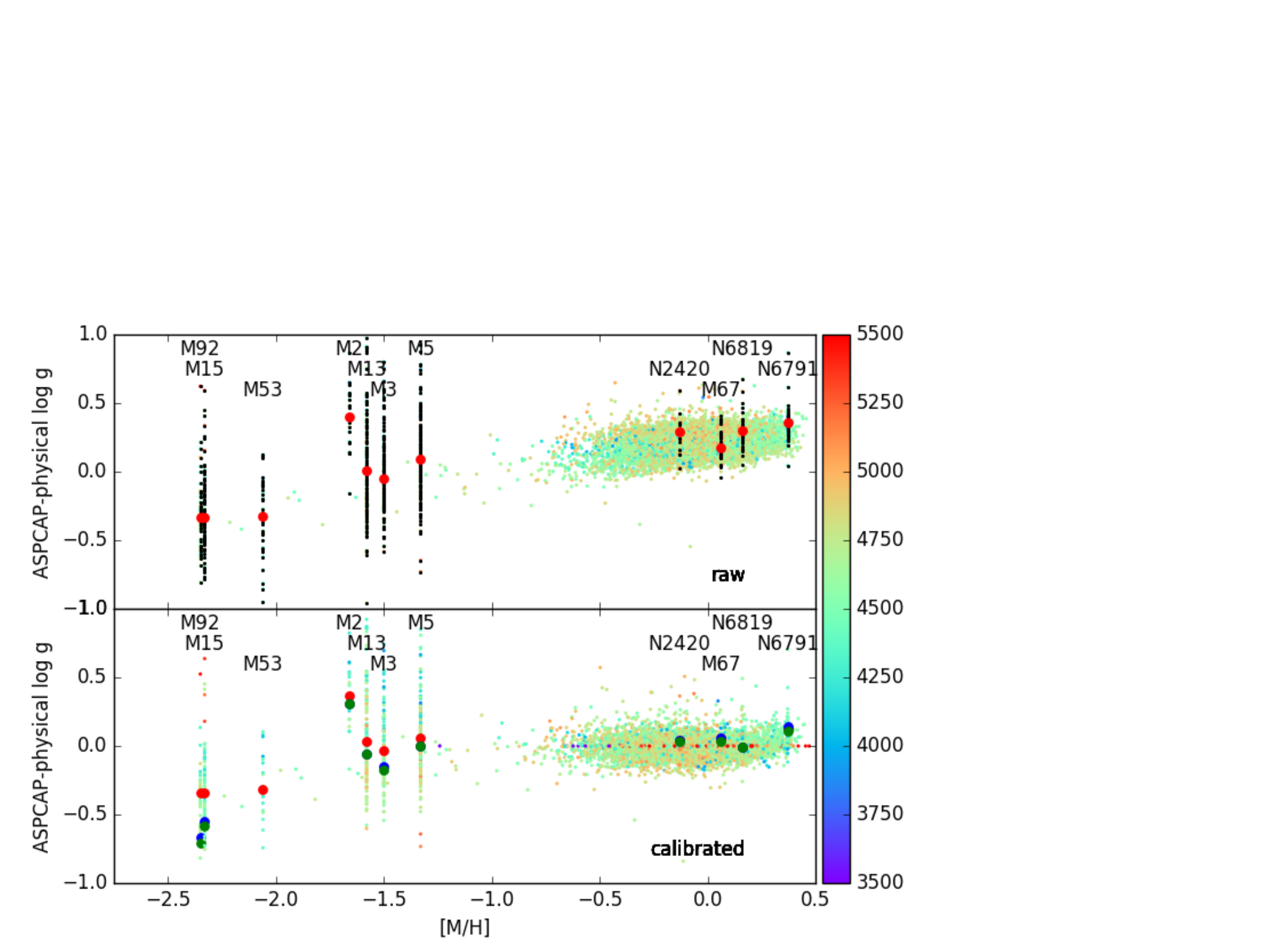}
\caption{Left plot shows comparison of the DR13 surface gravity with asteroseismic
surface gravity. Top panel shows difference as a function of metallicity
with points color-coded by surface gravity, middle panel shows difference
as a function of surface gravity color-coded by metallicity, and bottom
panels shows difference for RGB and RC.
Right panel shows calibration of DR13 ASPCAP surface gravities with asteroseismic and physical (clusters) gravities; points are color-coded by temperature. Large points show the median difference with physical gravities in clusters (red before effective temperature correction, green after correction, blue adopting photometric effective temperature. Top panel shows comparison with raw ASPCAP surface gravities; bottom panel after DR13 surface gravity calibration.}
\label{fig:dr13_loggcomp}
\end{figure}

Figure \ref{fig:dr13_loggcomp}
shows the difference between the ASPCAP DR13 surface gravities
and asteroseismic gravities for stars from the APOKASC 3.6.0 catalog
\citep{Pinsonneault2014}, which includes significantly more stars
with asteroseismic values than were available for DR12.
This comparison suggests that there is a significant trend
with metallicity (top panel), not much trend in surface gravity (middle
panel), and a notable difference between the RGB and RC (lower panel).

For DR13, we adopted a \logg\ correction that depends both on surface
gravity and metallicity.  We note that this differs significantly from
DR12, where only a surface gravity dependence was found and calibrated.

We also derived a separate correction for the ASPCAP \logg\ values
from RGB and RC stars. Based on the asteroseismic sample, we have derived a relation
that allows us to classify the stars correctly at  95\% level,
using \teff, \mh, and \xx{C}{N} (\citealt{Bovy2014},
Pinsonneault, private communication).
For every star, we compute the difference between the ASPCAP raw effective 
temperature and a fiducial metallicity-dependent ridgeline derived by \citet{Bovy2014}:
\begin{equation}
T_{ridge} = 4468+(\log g-2.5)/0.0018 – 382.5 [M/H]
\end{equation}
Stars cooler than the ridgeline temperature are classified as RGB stars, while stars more than 100 K hotter than the ridgeline temperature are classified as RC stars. For stars in the intermediate region, the observed C/N ratio is used to help to discriminate RGB from RC stars.
Stars with
\begin{equation}
[C/N] < -0.113-0.0043 (T-T_{ridge})
\end{equation}
are classified as RGB stars, while stars with
\begin{equation}
[C/N] > -0.088-0.0018 (T-T_{ridge})
\end{equation}
are classified as RC stars. 
For stars in an ambiguous region in C/N space, we interpolate between the RGB and RC corrections.

The adopted surface gravity calibrations are then applied as follows. For
RGB stars:

\begin{equation}
 \log g = \log g (raw) - (0.300 -0.048 \log g + 0.147 [M/H])
\end{equation}

For RC stars:
\begin{equation}
 \log g = \log g (raw) - (-4.442 + 3.326 \log g + 0.147 [M/H] - 0.581 (\log g)^2)
\end{equation}

Due to limited availability of asteroseismic data when the calibration was frozen,
we chose to clip the metallicity correction to \mh$>$-1.5, below
which the sign of the correction appears to reverse (top panel of left figure
in Figure \ref{fig:dr13_loggcomp}.  The inclusion
of additional data suggests that this reversal is, in fact, correct. The
same conclusion is drawn from analysis of clusters for which ``physical"
gravities can be obtained using:
\begin{equation}
g = 4 \pi G M \sigma T_{eff}^4 / L
\end{equation}
where $M$ is the adopted mass of evolved stars in the clusters (from
isochrones, based on cluster metallicity and age), and $L$
is their luminosity, which was derived from the $H$ magnitude,
a bolometric correction taken from the PARSEC isochrones, and
an adopted reddening and distance.  The right panel of 
Figure \ref{fig:dr13_loggcomp} adds the cluster results 
to the surface gravity comparison, and demonstrates that
the surface gravity correction really does change sign at
low metallicity.
%Table \ref{tab:clusters} gives the clusters and adopted 
%reddenings and distances. \textcolor{red}{need table!}

%\begin{figure}
%\includegraphics[width=0.5 \textwidth]{dr13_loggcomp-1.pdf}
%\caption{Calibration of DR13 ASPCAP surface gravities with asteroseismic and physical (clusters) gravities; points are color-coded by temperature. Large points show the median difference with physical gravities in clusters (red before effective temperature correction, green after correction, blue adopting photometric effective temperature. Top panel shows comparison with raw ASPCAP surface gravities; bottom panel after DR13 surface gravity calibration.}
%\label{fig:dr13_logg_clusters}
%\end{figure}

As a result of this additional analysis, we believe that the DR13
calibrated gravities are too low at \mh$<$ -1.5. 
A revision to the DR13 surface gravities is recommended for 
stars with \mh$<$ -1.5:

\begin{equation}
log g (\textrm{post calibration}) = log g (calibrated) – 0.5 ([M/H]+1.5)
\end{equation}

In DR13, quoted uncertainties for \logg\ are the observed scatter in
the relation between raw \logg\ and asteroseismic \logg, which corresponds
to 0.08 for most stars; for stars in the intermediate region where the
classification between RGB and RC was more uncertain, we adopted 0.095.

\subsection{DR13 elemental abundances}

\label{appendix:dr13_abundances}

\begin{figure}
\includegraphics[width=0.45 \textwidth]{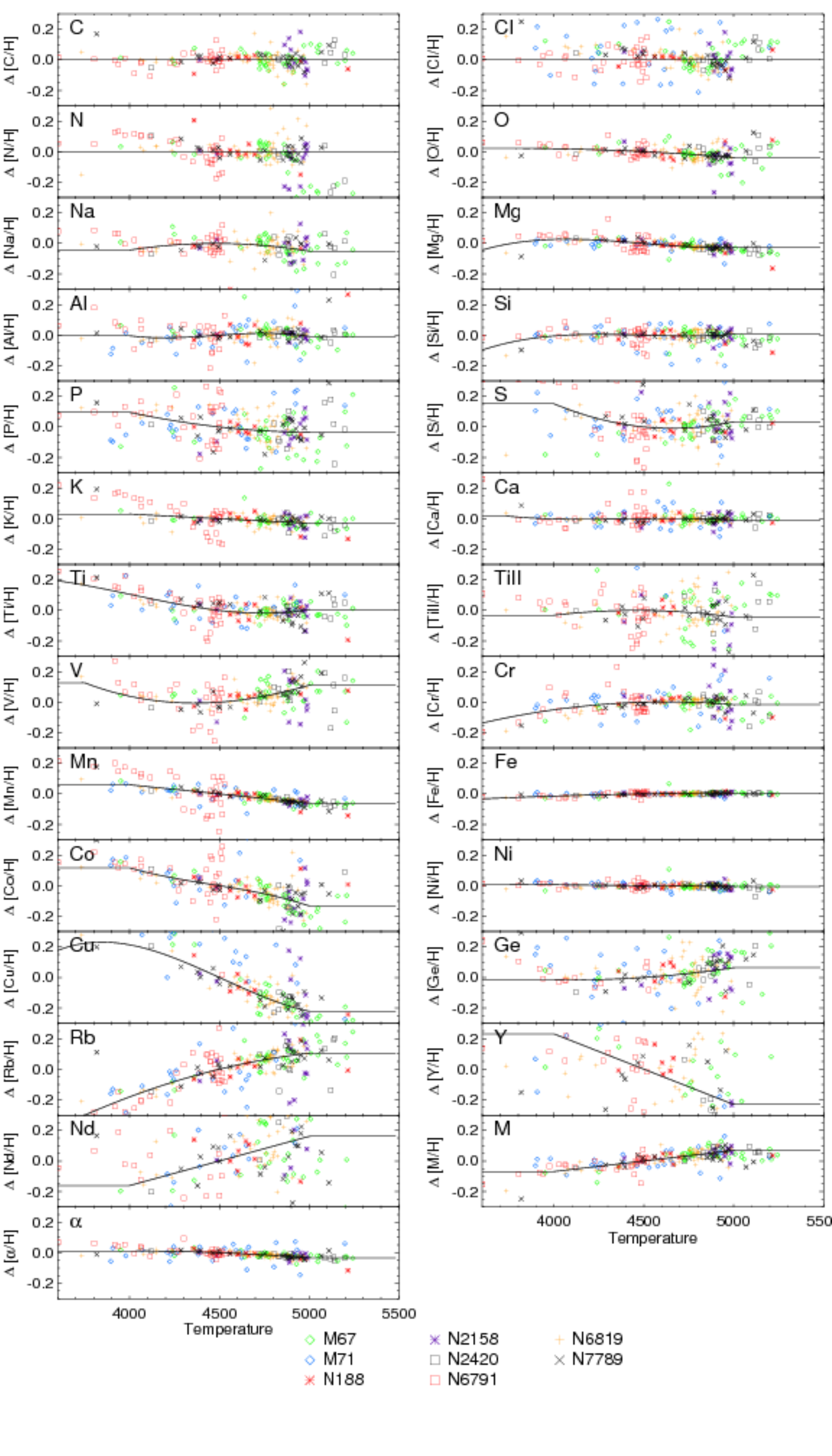}
\includegraphics[width=0.45 \textwidth]{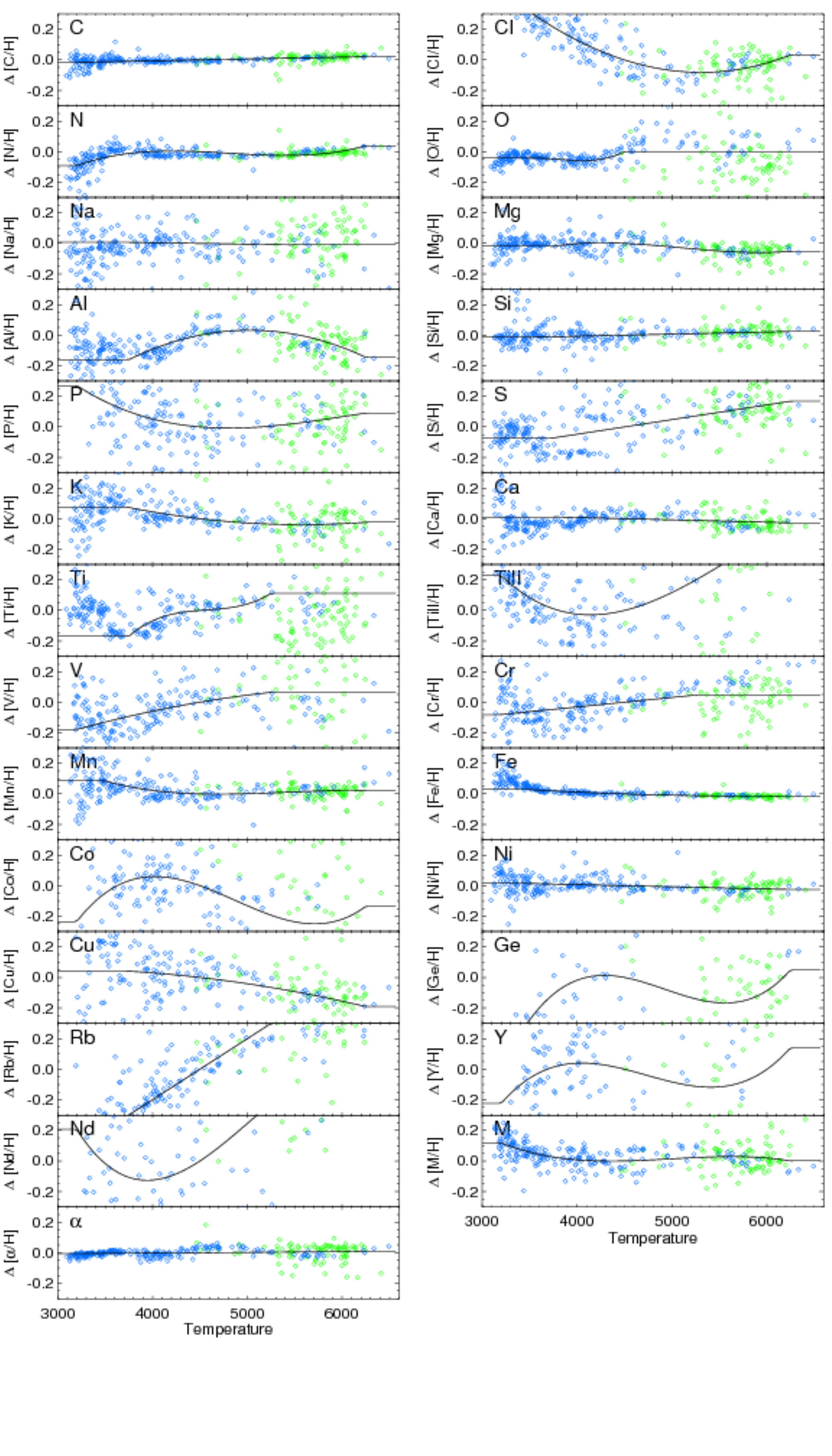}
\caption{DR13 internal abundance calibrations for red giants (left) and
dwarfs (right) applied as a 
function of \teff. \referee{Different point colors and symbols indicate
different clusters.}}
\label{fig:dr13_intcal}
\end{figure}

Figure \ref{fig:dr13_intcal} 
presents the calibration relations that were
adopted for the internal calibration relations for red giants and dwarfs in DR13.

Estimated uncertainties in abundances were derived using the same
methodology as for DR12.  We use the abundance derivations in both
open and globular cluster stars (removing known second generation stars
from the latter) with the underlying assumption that 
individual element abundances are uniform in all cluster members (apart
from C and N, which have mixing effects in giants).  We have chosen to
employ only clusters with metallicity greater than \mh$>$ -1, which
restricts the sample to mostly open clusters. In the selected cluster
sample, we measure the element abundance scatter in bins of temperature,
metallicity, and signal-to-noise (S/N). For each individual element,
we fit these values with a simple functional form:

\begin{equation}
log \sigmaσ = A + B (T_{eff} -4500)/1000. + C [M/H] + D (S/N -100)/100.
\end{equation}

where $\sigma$σ is the scatter among cluster stars relative to the mean derived
abundance.  Note that in the above relation, the fit to log  ensures
that the derived relation will always yield a positive uncertainty. The
values for the coefficients (A, B, C, D) associated with each element for giants
are given in Table \ref{tab:dr13_abun_err}. Also included are the ``Global'' uncertainties,
which represent the scatter around the temperature fits shown in Figure \ref{fig:dr13_intcal}.

\begin{table}
\begin{center}
	\caption{Parameters for DR13 abundance uncertainties.}
	\begin{tabular}{l c c c c c r}
\hline
		\textbf{Element} & \textbf{A} & \textbf{B} & \textbf{C} & \textbf{D} & \textbf{$\sigma$\footnote{\teff = 4500, \mh=0, S/N=100}}	& $\sigma_{global}$\\ 
\hline
			C & -3.243 & 0.608 & -0.757 & -0.257 & 0.039  & ---\\
			C I & -2.804 & 0.403 & -0.743 & -0.319 & 0.061 & ---\\
			N & -2.671 & 0.373 & -0.407 & -0.192 & 0.069 & ---\\
			O & -3.410 & 1.471 & -0.778 & -0.182 & 0.033 & 0.045\\
			Na & -2.389 & 0.140 & -0.926 & -0.323 & 0.092 & 0.059\\
			Mg & -3.980 & 0.284 & -0.949 & -0.115 & 0.019 & 0.029\\
			Al & -2.616 & -0.192 & -0.628 & -0.399 & 0.073 & 0.072\\
			Si & -3.464 & 0.548 & -0.482 & -0.212 & 0.031 & 0.040\\
			P & -1.988 & 0.384 & -0.568 & -0.369 & 0.137 & 0.120\\
			S & -2.199 & -0.030 & -0.402 & -0.295 & 0.111 & 0.122\\
			K & -3.098 & 0.208 & -0.583 & -0.496 & 0.045 & 0.048\\
			Ca & -3.520 & 0.153 & -0.895 & -0.405 & 0.030 & 0.032\\
			Ti & -3.108 & 0.295 & -0.741 & -0.185 & 0.045 & 0.057\\
			Ti II &	-2.192 & 0.328 & -0.538 & -0.267 & 0.112 & 0.116\\
			V & -2.447 & 1.030 & -1.096 & -0.519 & 0.087 & 0.089\\
			Cr & -3.191 & 0.290 & -0.775 & -0.455 & 0.041 & 0.048\\
			Mn & -3.523 & 0.235 & -0.614 & -0.488 & 0.029 & 0.049\\
			Fe & -5.316 & 0.202 & -0.874 & 0.019 & 0.005 & 0.053\\
			Co & -2.062 & 1.064 & -0.656 & -0.523 & 0.127 & 0.095\\
			Ni & -4.067 & 0.442 & -0.816 & -0.395 & 0.017 & 0.015\\
			Cu & -2.140 & -0.096 & -0.559 & -0.426 & 0.118 & 0.083\\
			Ge & -1.893 & 0.258 & -0.665 & -0.395 & 0.151 & 0.107\\
			Rb & -2.325 & 0.466 & -1.117 & -0.360 & 0.098 & 0.082\\
			M & -3.730 & 0.232 & -0.524 & 0.013 & 0.024 & 0.052\\
			$\alpha$ & -4.219 & 0.053 & -0.794 & -0.127 & 0.015 & 0.017\\
\hline
		\end{tabular}
\end{center}
		\label{tab:dr13_abun_err}
	\end{table}

\bibliographystyle{apj}

\bibliography{ref}

\begin{thebibliography}{39}
\expandafter\ifx\csname natexlab\endcsname\relax\def\natexlab#1{#1}\fi

\bibitem[{{Abolfathi} {et~al.}(2017){Abolfathi}, {Aguado}, {Aguilar}, {Allende
  Prieto}, {Almeida}, {Tasnim Ananna}, {Anders}, {Anderson}, {Andrews},
  {Anguiano}, \& et~al.}]{DR14}
{Abolfathi}, B., {Aguado}, D.~S., {Aguilar}, G., {et~al.} 2017, ArXiv e-prints

\bibitem[{{Allende Prieto} {et~al.}(2006){Allende Prieto}, {Beers}, {Wilhelm},
  {Newberg}, {Rockosi}, {Yanny}, \& {Lee}}]{AllendePrieto2006}
{Allende Prieto}, C., {Beers}, T.~C., {Wilhelm}, R., {et~al.} 2006, \apj, 636,
  804

\bibitem[{{Alvarez} \& {Plez}(1998)}]{Alvarez1998}
{Alvarez}, R., \& {Plez}, B. 1998, \aap, 330, 1109

\bibitem[{{Asplund} {et~al.}(2005){Asplund}, {Grevesse}, \&
  {Sauval}}]{Asplund2005}
{Asplund}, M., {Grevesse}, N., \& {Sauval}, A.~J. 2005, in Astronomical Society
  of the Pacific Conference Series, Vol. 336, Cosmic Abundances as Records of
  Stellar Evolution and Nucleosynthesis, ed. T.~G. {Barnes}, III \& F.~N.
  {Bash}, 25

\bibitem[{{Barber} {et~al.}(2006){Barber}, {Tennyson}, {Harris}, \&
  {Tolchenov}}]{Barber2006}
{Barber}, R.~J., {Tennyson}, J., {Harris}, G.~J., \& {Tolchenov}, R.~N. 2006,
  VizieR Online Data Catalog, 6119

\bibitem[{{Bensby} {et~al.}(2014){Bensby}, {Feltzing}, \& {Oey}}]{Bensby2014}
{Bensby}, T., {Feltzing}, S., \& {Oey}, M.~S. 2014, \aap, 562, A71

\bibitem[{{Blanton} {et~al.}(2017){Blanton}, {Bershady}, {Abolfathi},
  {Albareti}, {Allende Prieto}, {Almeida}, {Alonso-Garc{\'{\i}}a}, {Anders},
  {Anderson}, {Andrews}, \& et~al.}]{Blanton2017}
{Blanton}, M.~R., {Bershady}, M.~A., {Abolfathi}, B., {et~al.} 2017, \aj, 154,
  28

\bibitem[{{Bovy}(2016)}]{Bovy2016}
{Bovy}, J. 2016, \apj, 817, 49

\bibitem[{{Bovy} {et~al.}(2011){Bovy}, {Hogg}, \& {Roweis}}]{Bovy11a}
{Bovy}, J., {Hogg}, D.~W., \& {Roweis}, S.~T. 2011, Annals of Applied
  Statistics, 5

\bibitem[{{Bovy} {et~al.}(2014){Bovy}, {Nidever}, {Rix}, {Girardi}, {Zasowski},
  {Chojnowski}, {Holtzman}, {Epstein}, {Frinchaboy}, {Hayden}, {Rodrigues},
  {Majewski}, {Johnson}, {Pinsonneault}, {Stello}, {Allende Prieto}, {Andrews},
  {Basu}, {Beers}, {Bizyaev}, {Burton}, {Chaplin}, {Cunha}, {Elsworth},
  {Garc{\'{\i}}a}, {Garc{\'{\i}}a-Her{\'n}andez}, {Garc{\'{\i}}a P{\'e}rez},
  {Hearty}, {Hekker}, {Kallinger}, {Kinemuchi}, {Koesterke},
  {M{\'e}sz{\'a}ros}, {Mosser}, {O'Connell}, {Oravetz}, {Pan}, {Robin},
  {Schiavon}, {Schneider}, {Schultheis}, {Serenelli}, {Shetrone}, {Silva
  Aguirre}, {Simmons}, {Skrutskie}, {Smith}, {Stassun}, {Weinberg}, {Wilson},
  \& {Zamora}}]{Bovy2014}
{Bovy}, J., {Nidever}, D.~L., {Rix}, H.-W., {et~al.} 2014, \apj, 790, 127

\bibitem[{{Casey} {et~al.}(2016){Casey}, {Hogg}, {Ness}, {Rix}, {Ho}, \&
  {Gilmore}}]{Casey2016}
{Casey}, A.~R., {Hogg}, D.~W., {Ness}, M., {et~al.} 2016, ArXiv e-prints

\bibitem[{{Cunha} {et~al.}(2017){Cunha}, {Smith}, {Hasselquist}, {Souto},
  {Shetrone}, {Allende Prieto}, {Bizyaev}, {Frinchaboy},
  {Garc{\'{\i}}a-Hern{\'a}ndez}, {Holtzman}, {Johnson}, {J{\H o}nsson},
  {Majewski}, {M{\'e}sz{\'a}ros}, {Nidever}, {Pinsonneault}, {Schiavon},
  {Sobeck}, {Skrutskie}, {Zamora}, {Zasowski}, \&
  {Fern{\'a}ndez-Trincado}}]{Cunha2017}
{Cunha}, K., {Smith}, V.~V., {Hasselquist}, S., {et~al.} 2017, \apj, 844, 145

\bibitem[{{De Silva} {et~al.}(2007){De Silva}, {Freeman}, {Asplund},
  {Bland-Hawthorn}, {Bessell}, \& {Collet}}]{DeSilva2007}
{De Silva}, G.~M., {Freeman}, K.~C., {Asplund}, M., {et~al.} 2007, \aj, 133,
  1161

\bibitem[{{De Silva} {et~al.}(2006){De Silva}, {Sneden}, {Paulson}, {Asplund},
  {Bland-Hawthorn}, {Bessell}, \& {Freeman}}]{DeSilva2006}
{De Silva}, G.~M., {Sneden}, C., {Paulson}, D.~B., {et~al.} 2006, \aj, 131, 455

\bibitem[{{Garc{\'{\i}}a P{\'e}rez} {et~al.}(2016){Garc{\'{\i}}a P{\'e}rez},
  {Allende Prieto}, {Holtzman}, {Shetrone}, {M{\'e}sz{\'a}ros}, {Bizyaev},
  {Carrera}, {Cunha}, {Garc{\'{\i}}a-Hern{\'a}ndez}, {Johnson}, {Majewski},
  {Nidever}, {Schiavon}, {Shane}, {Smith}, {Sobeck}, {Troup}, {Zamora},
  {Weinberg}, {Bovy}, {Eisenstein}, {Feuillet}, {Frinchaboy}, {Hayden},
  {Hearty}, {Nguyen}, {O'Connell}, {Pinsonneault}, {Wilson}, \&
  {Zasowski}}]{GarciaPerez2016}
{Garc{\'{\i}}a P{\'e}rez}, A.~E., {Allende Prieto}, C., {Holtzman}, J.~A.,
  {et~al.} 2016, \aj, 151, 144

\bibitem[{{Gonz{\'a}lez Hern{\'a}ndez} \& {Bonifacio}(2009)}]{GHB2009}
{Gonz{\'a}lez Hern{\'a}ndez}, J.~I., \& {Bonifacio}, P. 2009, \aap, 497, 497

\bibitem[{{Gunn} {et~al.}(2006){Gunn}, {Siegmund}, {Mannery}, {Owen}, {Hull},
  {Leger}, {Carey}, {Knapp}, {York}, {Boroski}, {Kent}, {Lupton}, {Rockosi},
  {Evans}, {Waddell}, {Anderson}, {Annis}, {Barentine}, {Bartoszek}, {Bastian},
  {Bracker}, {Brewington}, {Briegel}, {Brinkmann}, {Brown}, {Carr},
  {Czarapata}, {Drennan}, {Dombeck}, {Federwitz}, {Gillespie}, {Gonzales},
  {Hansen}, {Harvanek}, {Hayes}, {Jordan}, {Kinney}, {Klaene}, {Kleinman},
  {Kron}, {Kresinski}, {Lee}, {Limmongkol}, {Lindenmeyer}, {Long}, {Loomis},
  {McGehee}, {Mantsch}, {Neilsen}, {Neswold}, {Newman}, {Nitta}, {Peoples},
  {Pier}, {Prieto}, {Prosapio}, {Rivetta}, {Schneider}, {Snedden}, \&
  {Wang}}]{Gunn2006}
{Gunn}, J.~E., {Siegmund}, W.~A., {Mannery}, E.~J., {et~al.} 2006, \aj, 131,
  2332

\bibitem[{{Gustafsson} {et~al.}(2008){Gustafsson}, {Edvardsson}, {Eriksson},
  {J{\o}rgensen}, {Nordlund}, \& {Plez}}]{Gustafsson2008}
{Gustafsson}, B., {Edvardsson}, B., {Eriksson}, K., {et~al.} 2008, \aap, 486,
  951

\bibitem[{{Hasselquist} {et~al.}(2016){Hasselquist}, {Shetrone}, {Cunha},
  {Smith}, {Holtzman}, {Lawler}, {Allende Prieto}, {Beers}, {Chojnowski},
  {Fern{\'a}ndez-Trincado}, {Garc{\'{\i}}a-Hern{\'a}ndez}, {Hearty},
  {Majewski}, {Pereira}, {Placco}, {Villanova}, \& {Zamora}}]{Hasselquist2016}
{Hasselquist}, S., {Shetrone}, M., {Cunha}, K., {et~al.} 2016, \apj, 833, 81

\bibitem[{{Hawkins} {et~al.}(2016){Hawkins}, {Masseron}, {Jofr{\'e}},
  {Gilmore}, {Elsworth}, \& {Hekker}}]{Hawkins2016}
{Hawkins}, K., {Masseron}, T., {Jofr{\'e}}, P., {et~al.} 2016, \aap, 594, A43

\bibitem[{{Holtzman} {et~al.}(2015){Holtzman}, {Shetrone}, {Johnson}, {Allende
  Prieto}, {Anders}, {Andrews}, {Beers}, {Bizyaev}, {Blanton}, {Bovy},
  {Carrera}, {Chojnowski}, {Cunha}, {Eisenstein}, {Feuillet}, {Frinchaboy},
  {Galbraith-Frew}, {Garc{\'{\i}}a P{\'e}rez}, {Garc{\'{\i}}a-Hern{\'a}ndez},
  {Hasselquist}, {Hayden}, {Hearty}, {Ivans}, {Majewski}, {Martell},
  {Meszaros}, {Muna}, {Nidever}, {Nguyen}, {O'Connell}, {Pan}, {Pinsonneault},
  {Robin}, {Schiavon}, {Shane}, {Sobeck}, {Smith}, {Troup}, {Weinberg},
  {Wilson}, {Wood-Vasey}, {Zamora}, \& {Zasowski}}]{Holtzman2015}
{Holtzman}, J.~A., {Shetrone}, M., {Johnson}, J.~A., {et~al.} 2015, \aj, 150,
  148

\bibitem[{{Jahandar} {et~al.}(2017){Jahandar}, {Venn}, {Shetrone}, {Irwin},
  {Bovy}, {Sakari}, {Kielty}, {Digby}, \& {Frinchaboy}}]{Jahandar2017}
{Jahandar}, F., {Venn}, K.~A., {Shetrone}, M.~D., {et~al.} 2017, ArXiv e-prints

\bibitem[{J\"onsson {et~al.}(2018)J\"onsson, others, {et~al.}}]{Jonsson2018}
J\"onsson, H., others, {et~al.} 2018, in preparation

\bibitem[{{Koesterke}(2009)}]{Koesterke2009}
{Koesterke}, L. 2009, in American Institute of Physics Conference Series, Vol.
  1171, American Institute of Physics Conference Series, ed. I.~{Hubeny}, J.~M.
  {Stone}, K.~{MacGregor}, \& K.~{Werner}, 73--84

\bibitem[{{Majewski} {et~al.}(2017){Majewski}, {Schiavon}, {Frinchaboy},
  {Allende Prieto}, {Barkhouser}, {Bizyaev}, {Blank}, {Brunner}, {Burton},
  {Carrera}, {Chojnowski}, {Cunha}, {Epstein}, {Fitzgerald}, {Garc{\'{\i}}a
  P{\'e}rez}, {Hearty}, {Henderson}, {Holtzman}, {Johnson}, {Lam}, {Lawler},
  {Maseman}, {M{\'e}sz{\'a}ros}, {Nelson}, {Nguyen}, {Nidever}, {Pinsonneault},
  {Shetrone}, {Smee}, {Smith}, {Stolberg}, {Skrutskie}, {Walker}, {Wilson},
  {Zasowski}, {Anders}, {Basu}, {Beland}, {Blanton}, {Bovy}, {Brownstein},
  {Carlberg}, {Chaplin}, {Chiappini}, {Eisenstein}, {Elsworth}, {Feuillet},
  {Fleming}, {Galbraith-Frew}, {Garc{\'{\i}}a}, {Garc{\'{\i}}a-Hern{\'a}ndez},
  {Gillespie}, {Girardi}, {Gunn}, {Hasselquist}, {Hayden}, {Hekker}, {Ivans},
  {Kinemuchi}, {Klaene}, {Mahadevan}, {Mathur}, {Mosser}, {Muna}, {Munn},
  {Nichol}, {O'Connell}, {Parejko}, {Robin}, {Rocha-Pinto}, {Schultheis},
  {Serenelli}, {Shane}, {Silva Aguirre}, {Sobeck}, {Thompson}, {Troup},
  {Weinberg}, \& {Zamora}}]{Majewski2017}
{Majewski}, S.~R., {Schiavon}, R.~P., {Frinchaboy}, P.~M., {et~al.} 2017, \aj,
  154, 94

\bibitem[{{Massarotti} {et~al.}(2008){Massarotti}, {Latham}, {Stefanik}, \&
  {Fogel}}]{Massarotti2008}
{Massarotti}, A., {Latham}, D.~W., {Stefanik}, R.~P., \& {Fogel}, J. 2008, \aj,
  135, 209

\bibitem[{{Masseron} \& {Hawkins}(2017)}]{Masseron2017}
{Masseron}, T., \& {Hawkins}, K. 2017, \aap, 597, L3

\bibitem[{{Ness} {et~al.}(2015){Ness}, {Hogg}, {Rix}, {Ho}, \&
  {Zasowski}}]{Ness2015}
{Ness}, M., {Hogg}, D.~W., {Rix}, H.-W., {Ho}, A.~Y.~Q., \& {Zasowski}, G.
  2015, \apj, 808, 16

\bibitem[{{Ness} {et~al.}(2017){Ness}, {Rix}, {Hogg}, {Casey}, {Holtzman},
  {Fouesneau}, {Zasowski}, {Geisler}, {Shetrone}, {Minniti}, {Frinchaboy}, \&
  {Roman-Lopes}}]{Ness2017}
{Ness}, M., {Rix}, H., {Hogg}, D.~W., {et~al.} 2017, ArXiv e-prints

\bibitem[{{Nidever} {et~al.}(2015){Nidever}, {Holtzman}, {Allende Prieto},
  {Beland}, {Bender}, {Bizyaev}, {Burton}, {Desphande}, {Fleming},
  {Garc{\'{\i}}a P{\'e}rez}, {Hearty}, {Majewski}, {M{\'e}sz{\'a}ros}, {Muna},
  {Nguyen}, {Schiavon}, {Shetrone}, {Skrutskie}, {Sobeck}, \&
  {Wilson}}]{Nidever2015}
{Nidever}, D.~L., {Holtzman}, J.~A., {Allende Prieto}, C., {et~al.} 2015, \aj,
  150, 173

\bibitem[{{Pinsonneault} {et~al.}(2014){Pinsonneault}, {Elsworth}, {Epstein},
  {Hekker}, {M{\'e}sz{\'a}ros}, {Chaplin}, {Johnson}, {Garc{\'{\i}}a},
  {Holtzman}, {Mathur}, {Garc{\'{\i}}a P{\'e}rez}, {Silva Aguirre}, {Girardi},
  {Basu}, {Shetrone}, {Stello}, {Allende Prieto}, {An}, {Beck}, {Beers},
  {Bizyaev}, {Bloemen}, {Bovy}, {Cunha}, {De Ridder}, {Frinchaboy},
  {Garc{\'{\i}}a-Hern{\'a}ndez}, {Gilliland}, {Harding}, {Hearty}, {Huber},
  {Ivans}, {Kallinger}, {Majewski}, {Metcalfe}, {Miglio}, {Mosser}, {Muna},
  {Nidever}, {Schneider}, {Serenelli}, {Smith}, {Tayar}, {Zamora}, \&
  {Zasowski}}]{Pinsonneault2014}
{Pinsonneault}, M.~H., {Elsworth}, Y., {Epstein}, C., {et~al.} 2014, \apjs,
  215, 19

\bibitem[{{Plez}(2012)}]{Plez2012}
{Plez}, B. 2012, {Turbospectrum: Code for spectral synthesis}, Astrophysics
  Source Code Library

\bibitem[{{Ram{\'{\i}}rez} \& {Allende Prieto}(2011)}]{RamirezCAP2011}
{Ram{\'{\i}}rez}, I., \& {Allende Prieto}, C. 2011, \apj, 743, 135

\bibitem[{{Schlegel} {et~al.}(1998){Schlegel}, {Finkbeiner}, \& {Davis}}]{SFD}
{Schlegel}, D.~J., {Finkbeiner}, D.~P., \& {Davis}, M. 1998, \apj, 500, 525

\bibitem[{{SDSS Collaboration} {et~al.}(2016){SDSS Collaboration}, {Albareti},
  {Allende Prieto}, {Almeida}, {Anders}, {Anderson}, {Andrews},
  {Aragon-Salamanca}, {Argudo-Fernandez}, {Armengaud}, \& et~al.}]{DR13}
{SDSS Collaboration}, {Albareti}, F.~D., {Allende Prieto}, C., {et~al.} 2016,
  ArXiv e-prints

\bibitem[{{Shetrone} {et~al.}(2015){Shetrone}, {Bizyaev}, {Lawler}, {Allende
  Prieto}, {Johnson}, {Smith}, {Cunha}, {Holtzman}, {Garc{\'{\i}}a P{\'e}rez},
  {M{\'e}sz{\'a}ros}, {Sobeck}, {Zamora}, {Garc{\'{\i}}a-Hern{\'a}ndez},
  {Souto}, {Chojnowski}, {Koesterke}, {Majewski}, \& {Zasowski}}]{Shetrone2015}
{Shetrone}, M., {Bizyaev}, D., {Lawler}, J.~E., {et~al.} 2015, \apjs, 221, 24

\bibitem[{Wilson {et~al.}(2018)Wilson, others, {et~al.}}]{Wilson2018}
Wilson, J., others, {et~al.} 2018, in preparation

\bibitem[{{Zamora} {et~al.}(2015){Zamora}, {Garc{\'{\i}}a-Hern{\'a}ndez},
  {Allende Prieto}, {Carrera}, {Koesterke}, {Edvardsson}, {Castelli}, {Plez},
  {Bizyaev}, {Cunha}, {Garc{\'{\i}}a P{\'e}rez}, {Gustafsson}, {Holtzman},
  {Lawler}, {Majewski}, {Manchado}, {M{\'e}sz{\'a}ros}, {Shane}, {Shetrone},
  {Smith}, \& {Zasowski}}]{Zamora2015}
{Zamora}, O., {Garc{\'{\i}}a-Hern{\'a}ndez}, D.~A., {Allende Prieto}, C.,
  {et~al.} 2015, \aj, 149, 181

\bibitem[{{Zasowski} {et~al.}(2017){Zasowski}, {Cohen}, {Chojnowski},
  {Santana}, {Oelkers}, {Andrews}, {Beaton}, {Bender}, {Bird}, {Bovy},
  {Carlberg}, {Covey}, {Cunha}, {Dell'Agli}, {Fleming}, {Frinchaboy},
  {Garc{\'{\i}}a-Hern{\'a}ndez}, {Harding}, {Holtzman}, {Johnson}, {Kollmeier},
  {Majewski}, {M{\'e}sz{\'a}ros}, {Munn}, {Mu{\~n}oz}, {Ness}, {Nidever},
  {Poleski}, {Rom{\'a}n-Z{\'u}{\~n}iga}, {Shetrone}, {Simon}, {Smith},
  {Sobeck}, {Stringfellow}, {Szigeti{\'a}ros}, {Tayar}, \&
  {Troup}}]{Zasowski2017}
{Zasowski}, G., {Cohen}, R.~E., {Chojnowski}, S.~D., {et~al.} 2017, \aj, 154,
  198

\end{thebibliography}

\end{document}